\RequirePackage{fix-cm}
\documentclass[twocolumn]{svjour3}          
\smartqed  
\usepackage{rotating,graphicx,color,mathptmx,longtable,amssymb}
\usepackage{latexsym}
\usepackage{hyperref}
\usepackage[normalem]{ulem}

\usepackage{chngcntr}
\counterwithout{equation}{section}
\counterwithout{figure}{section}

\newcommand{\mmax}{M_G^{\mathit{max}}}
  
\journalname{EPJA}
\begin{document}

\title{Equations of state for hot neutron stars}

\titlerunning{EoS for hot neutron stars}        

\author{Adriana R. Raduta         \and
  Flavia Nacu \and 
  Micaela Oertel 
}

\authorrunning{Raduta, Nacu, Oertel} 

\institute{Adriana R. Raduta \at
  National Institute for Physics and Nuclear Engineering (IFIN-HH),
  RO-077125 Bucharest, Romania \\
  \email{araduta@nipne.ro}  
  \and
  Flavia Nacu \at
  National Institute for Physics and Nuclear Engineering (IFIN-HH),
  RO-077125 Bucharest, Romania \\
  \email{flavia.nacu@nipne.ro}           
  \and
  Micaela Oertel \at
  LUTH, Observatoire de Paris, Universit\'e PSL, CNRS,
  Universit\'e de Paris, 92190 Meudon, France \\
  \email{micaela.oertel@obspm.fr}
}

\date{Received: date / Accepted: date}

\maketitle

\begin{abstract}
  We review the equation of state (EoS) models covering a large range
  of temperatures, baryon number densities and electron fractions
  presently available on the \textsc{CompOSE} database. These models
  are intended to be directly usable within numerical simulations of
  core-collapse supernovae, binary neutron star mergers and
  proto-neutron star evolution. We discuss their compliance with
  existing constraints from astrophysical observations and nuclear
  data. For a selection of purely nucleonic models in reasonable
  agreement with the above constraints, after discussing the
  properties of cold matter, we review thermal properties for
  thermodynamic conditions relevant for core-collapse supernovae and
  binary neutron star mergers. We find that the latter are strongly
  influenced by the density dependence of the nucleon effective mass.
  The selected bunch of models is used to investigate the EoS
  dependence of hot star properties, where entropy per baryon and
  electron fraction profiles are inspired from proto-neutron star
  evolution. The $\Gamma$-law analytical thermal EoS used in many
  simulations is found not to describe well these thermal properties
  of the EoS. However, it may offer a fair description of the
  structure of hot stars whenever thermal effects on the baryonic part
  are small, as shown here for proto-neutron stars starting from
  several seconds after bounce.  
\end{abstract}

\section{Introduction}
\label{intro}

The composition and structural properties of cold, mature neutron
stars (NS) in weak $\beta$-equilibrium depend on a one-parameter
equation of state (EoS) that relates the pressure to the energy
density.  On the contrary, dynamics of core-collapse supernovae (CCSN),
evolution of proto-neutron stars (PNS), formation of stellar black holes (BH),
and the post-merger phase of binary neutron star mergers (BNS) require an
EoS depending on three thermodynamic parameters, typically chosen as
temperature $T$, baryon number density $n_B$, and electron fraction
$Y_e$. These need to cover wide domains: $10^{-14}~{\rm fm}^{-3} \leq
n_B \leq 1.5~{\rm fm}^{-3}$, $0 \leq Y_e=n_e/n_B \leq 0.6$ and $0 \leq
T \leq 100~{\rm MeV}$ \cite{Pons_ApJ_1999,Sumiyoshi_2007,Janka_PhysRep_2007,Fischer_2009,Shibata_11,OConnor_2011,hempel12,Mezzacappa2015,Rosswog_15,Baiotti_2017,Connor2018ApJ,Burrows2020MNRAS,Ruiz2020}.
Since a long time, numerous numerical studies of these different phenomena
show a considerable sensitivity to the EoS, see
e.g. \cite{Pons_ApJ_1999,Janka_2012,Bauswein_2012,Koeppel_2019,Bauswein_PRL_2020,Preau_MNRAS_2021}.

In the last decades the EoS of cold NS has been intensively studied
and constrained by experimental nuclear physics data, astrophysical
data from radio/X-ray pulsars, gravitational wave events, and
\textit{ab initio} calculations of (neutron) matter.  The different
data are complementary in the sense that the conditions of each
measurement imply that constraints on the EoS can be obtained within a
particular density and isospin asymmetry domain. In particular,
nuclear experiments are naturally performed for matter with densities
below saturation density
$n_{\mathit{sat}} \approx 0.16~{\rm fm}^{-3} \approx 2 \cdot 10^{14}~{\rm g/cm^3}$
and for a nearly equal number of
protons and neutrons and \textit{ab initio} calculations are the most
reliable for low density ($\lesssim 1-2 n_{\mathit{sat}}$) pure neutron matter.

On the astrophysical side, NS mass measurements provide lower limits
on the maximum gravitational mass of stable configurations
$\mmax$ and, thus, constrain the high density range of
$\beta$ equilibrated EoS.
Recent observations of pulsars with precisely determined masses
around 2 $M_\odot$ correspond to
PSR J1614-2230 ($M=1.908 \pm 0.016 M_{\odot}$, 68.3\% credible interval)
\cite{Demorest_Nature_2010,Arzoumanian_ApJSS_2018};
PSR J0348+0432 ($M=2.01 \pm 0.04 M_{\odot}$, 68.3\% credible interval)
\cite{Antoniadis2013};
MSP J0740+6620 ($M=2.14^{+0.20}_{-0.18} M_{\odot}$ with 95.4\% confidence level
\cite{Cromartie2020} and $M=2.08^{+0.07}_{-0.07} M_{\odot}$ with 68.3\% confidence level
\cite{Fonseca_2021}).
Even more massive NS have been identified in so-called "spider" systems,
\textit{e.g.} PSR J1810+1744 ($M=2.13 \pm 0.04M_{\odot}$ with
68\% confidence level \cite{Romani_ApJL_2021}) but, in contrast to the above
mentioned mass determinations, these mass estimations depend on the companion heating model.
These measurements have triggered a large number of studies on the
appearance of non-nucleonic degrees of freedom at high densities. In particular,
hyperons~\cite{Weissenborn_PRC_2012,Weissenborn_NPA_2012,Bonanno2012,Miyatsu_PRC2013,Colucci_PRC_2013,Oertel_JPG_2015,Chatterjee2015,Fortin_PRC_2016,Fortin_PRC_2017,Li_PLB_2018},
the lowest spin-3/2 baryonic
multiplet~\cite{Drago_PRC_2014,Cai_PRC_2015,Kolomeitsev_NPA_2017,Sahoo_PRC_2018,Li_PLB_2018,Ribes_2019,Raduta_PLB_2021},
meson condensates ~\cite{Malik_EPJA_2021,Thapa_PRD_2021} and a
potential phase transition to quark-matter
~\cite{Weissenborn11,Zdunik_AA_2013,Alford_PRD_2013,DD2F,Bastian_PRC_2017,Montana_PRD_2019,Otto_PRD_2020,Otto_EPJ_2020}
have been considered.  All these works indicate that in contrast to
the first expectations, maximum NS masses of $\approx 2 M_\odot$ do
not exclude these non-nucleonic degrees of freedom to appear in the
most massive neutron stars.

Gravitational waves detected from the late inspiral of the binary
neutron stars merger event
GW170817~\cite{Abbott_PRL119_161101,Abbott_ApJ2017ApJ_L12} have
allowed to determine a combined tidal deformability of the two
NSs~\cite{Abbott_2019}. Due to the estimated masses in GW170817,
$1.36M_{\odot} \leq M_1 \leq 1.60 M_{\odot}$ and $1.16 M_{\odot}\leq
M_2 \leq 1.36M_{\odot}$, this measurement constrains the intermediate
density domain of the EoS and, implicitly, the radii of NSs in this
mass range.

The NICER mission has recently provided two simultaneous mass and
radius measurements for PSR J0030+0451 with
$R(1.44^{0.15}_{-0.14}M_{\odot})=13.02^{+1.24}_{-1.06}~{\rm km}$ \cite{Miller_2019} and
$R(1.34^{+0.15}_{-0.16} M_{\odot})= 12.71^{+1.14}_{-1.19}~{\rm km}$
(68.3\%)~\cite{Riley_2019} and J0740+6620 with $R(2.08 \pm 0.07
M_{\odot})=13.7^{+2.6}_{-1.5}~{\rm km}$ (68\%)~\cite{Miller_may2021}
and $R(2.072^{+0.067}_{-0.066} M_{\odot})=12.39^{+1.30}_{-0.98}~{\rm
  km}$~\cite{Riley_may2021}.  Combining the NICER results with the
limits on the NS maximum mass as well as the tidal deformability from
GW170817~\cite{Abbott_2019} leads to constraints on the
$\beta$-equilibrated EoS for densities in the range
$1.5 n_{\mathit{sat}} \lesssim n_B \lesssim 3n_{\mathit{sat}}$
\cite{Miller_may2021,Raaijmakers_may2021} and which can be translated
into limits for NS radii~\cite{Miller_may2021}.

Altogether, these recent developments have allowed to considerably
narrow down the parameter space for cold $\beta$-equilibrated
EoS. The situation is more complicated for the finite-temperature EoS
potentially out of weak equilibrium where no firm constraint exists so
far. The different analysis performed of GW170817 and its
electromagnetic counterpart (GRB170817A and AT2017gfo) have formidably
confirmed that future multimessenger observations --combining
gravitational wave detection with electromagnetic and neutrino
signals-- from binary neutron star mergers and core-collapse
supernovae with the subsequent PNS evolution have the potential to
sample this complementary domains of the dense matter EoS. Among
others, a combined analysis of the GW and electromagnetic data for
this particular event has been used by several authors to deduce both
a lower~\cite{Bauswein_PRL_2020} and an upper bound on
$\mmax$~\cite{Margalit_17,Ruiz2018,Rezzolla_2018,Shibata_PRD_2019}.
In \cite{Khadkikar_PRC_2021} the importance of thermal effects on the
EoS in this context is demonstrated, relaxing considerably the upper
bound.

In contrast to the cold NS EoS, where systematic studies allow to
relate data to single nuclear matter parameters, see
e.g.~\cite{Fortin_PRC_2016,Margueron_PRC_2018b,Li_PRC_2019,Traversi_APJ_2020},
the dynamic environment
implying an interplay of many different factors, as well as the
thermal and compositional effects in the EoS render this task
difficult for the so-called ``general purpose'' EoS depending on three
parameters. In this context, theoretical developments
\cite{Constantinou_PRC_2014,Constantinou_PRC_2015} as well as
numerical simulations
\cite{Schneider_PRC_2019b,Schneider_ApJ_2020,Yasin_PRL_2020,Raithel2021,Andersen2021}
indicate that properties of hot matter and the corresponding dynamical
evolution of the related astrophysical phenomena are very sensitive to
the value of the nucleon effective mass due to its predominant role in
determining the kinetic energy and thus the strength of thermal
effects.

The main motivation of this work is to try to better understand the
finite temperature EoS. To that end we perform a comparative study of
the general purpose EoS available on \textsc{CompOSE}, focussing on
purely nucleonic ones. The online service
\textsc{CompOSE}~\cite{Typel_2013} provides EoS data in tabular form
and among others most of the currently publicly available general
purpose EoS.  These are delivered as 3D tables as function of $n_B$,
$T$ and $Y_e$. Entries include primary thermodynamic state variables
(energy density, free energy density, chemical potentials, pressure
and entropy) and optionally microscopic quantities (Landau effective
masses, Dirac effective masses, single particle potentials) as well as
relative abundances of different species.  Typical ranges of values
for which this information is offered are $10^{-15} \leq n_B \leq 1-10
~{\rm fm}^{-3}$, $0.1 \leq T \leq 100~{\rm MeV}$ and $0.01 \leq Y_e
\leq 0.6$. The {\tt compose} software allows to interpolate or extract
data for specific thermodynamic conditions, including constant entropy
per baryon and neutrino-less $\beta$-equilibrium and to obtain various
thermodynamic coefficients via numerical differentiation (specific
heats at constant volume and at constant pressure, adiabatic index,
adiabatic compressibility, thermal compressibility, tension
coefficient at constant volume, speed of sound).  In addition to
general purpose tables, EoS for cold $\beta$-equilibrated NS,
symmetric nuclear matter (SNM) and pure neutron matter (PNM) are
available, too.

We start by reviewing in Sec. \ref{sec:Eos} the ensemble of
the general purpose EoS models presently available as well as their
compliance with various constraints from nuclear physics and NS
observations.  We then further analyze the behavior of a selection of
models.  The resulting properties of cold matter are discussed in
Sec. \ref{sec:T=0}, followed by a discussion of hot matter properties
with various degrees of isospin asymmetry and different temperatures
($5 \leq T \leq 50~{\rm MeV}$) in Sec.\ref{sec:T}. In many numerical
approximations, thermal effects are approximately included by adding
to the cold EoS thermal contributions in the form of a
$\Gamma$-law~\cite{Hotokezaka_PRD_2013,Bauswein_PRD_2010,Endrizzi_PRD_2018,Camelio2019}.
Hot matter properties obtained from the full finite-temperature EoS
will be compared with the $\Gamma$-law ones.  The composition
and a number of selected thermodynamic quantities of
hot stellar matter with subsaturation densities are investigated in
Sec. \ref{sec:compo} with special focus on the
model dependence of the results;
the temperature-dependence of the symmetry energy is considered in
Sec. \ref{sec:Esym_T};
the EoS dependence of PNS properties and the performance of the
$\Gamma$-law approximation for thermal effects are studied
Sec. \ref{sec:HotStars}. Conclusions are drawn in
Sec. \ref{sec:Concl}.

Throughout this paper, unless otherwise stated, we use natural units
$k_B=c = \hslash=1$.
The value of the neutron rest mass in vacuum is used to convert number
densities to rest-mass densities.

\section{General purpose EoS models on \textsc{CompOSE}}
\label{sec:Eos}

In this section we provide an overview \footnote{as of August, 30,
  2021}
of models for which general purpose EoS tables are available on the
\textsc{CompOSE} site\footnote{
  \url{https://compose.obspm.fr/}}~\cite{Typel_2013}, before
discussing the thermal properties for some selected models in more
detail. General purpose tables cover a large range of different baryon
number densities $n_B$, temperatures $T$, and electron fractions $Y_e$
with the idea to be applicable to CCSN and
BNS merger simulations. To that end, in general
the contribution of charged leptons (electrons and positrons) as well
as photons are included as ideal Fermi or Bose gases, respectively.
The available models differ only for the hadronic part, there
differences arise from the treatment of inhomogeneous matter
containing nuclear clusters, the interactions between nucleons and
potentially other baryons, and the particle content considered at
suprasaturation densities, \textit{i.e.} which non-nucleonic degrees
of freedom (hyperons, pions, kaons, quarks, \dots) are included.

For reviews about the underlying nuclear physics, see
e.g.~\cite{Oertel_RMP_2017,Burgio_2018}, here we will only briefly
remind the main characteristics of the models on \textsc{CompOSE}.
The standard format for the name of a particular EoS table takes the
form XXX(YYY). XXX thereby indicates the initials of the 
authors in the original publication(s) proposing the corresponding
model. In the case of the general purpose tables this allows to
identify, for models with nucleonic
  particle degrees of freedom,
the treatment of inhomogeneous matter, see
Sec.~\ref{ssec:inhomo}. YYY represents the name of the
interaction, see Sec.~\ref{ssec:effint}, 
and accounts for both nucleonic and exotic sectors.
In tables \ref{tab:SNAmodels},
\ref{tab:HOMOmodels} and \ref{tab:NSEmodels}
we list all EoS tables with information on treatment of inhomogeneous matter,
baryonic interactions and particle content.  


\subsection{Effective interactions between baryons}
\label{ssec:effint}

\subsubsection{Energy density functionals}
Most available tables employ phenomenological approaches and use
effective interactions between nucleons, motivated in terms of energy
density functional theory (DFT). The parameters of the interaction are in
general fitted to different properties of selected nuclei and nuclear
matter properties. Already on the mean-field level these interactions
result in a rather precise description of nuclei and nuclear
matter. An extrapolation to the highest densities and/or to very
neutron rich matter or neutron rich nuclei, \textit{i.e.} far away from the
domain where the interaction actually has been fitted, should be
regarded with some caution. Nevertheless, due to their technical
simplicity, these interactions are the most widely used to construct
general purpose EoS tables. Different classes exist:

\textit{(Generalized) Skyrme effective interactions}: These are based
on non-relativistic density functionals with a large variety of
different parameter sets. The pool of considered data as well as the
relative weight between data that sample isoscalar and isovector
channels explain the large number of existing interactions and their
wide range of behaviors. For a review, see
e.g.~Ref. \cite{Dutra_Skyrme_PRC_2012}. For neutron star applications
it is interesting to note that some parameterisations consider
\textit{ab initio} calculations of neutron matter as data for
adjusting the parameters, e.g. the SLy4 parameterisation~\cite{SLy4}.

\textit{Covariant density-functional models (CDFT)}: These are based on
(special) relativistic functionals, usually written in terms of a
Lagrangian density, where the different interaction channels for the
nucleons are often labeled by the meson with the corresponding quantum
numbers. In general, baryons are considered to interact in the
scalar-isoscalar ($\sigma$), the vector-isoscalar ($\omega$), and the
vector-isovector ($\rho$) channel, sometimes a vector-isoscalar
($\delta$) channel is included and, if hyperons are considered,
different channels coupling baryons with nonzero strangeness. Several
functionals assume non-linear potentials depending on the mean fields;
in other models the interaction parameters are assumed density
dependent.  For a review, see e.g. Ref.~\cite{Dutra_RMF_PRC_2014}.

The SU(3) Chiral Mean Field (CMF) model
\cite{DS_ApJ_2008,Dexheimer_PASA_2017} falls in this category,
too. However, the structure of the interaction is taken from a
non-linear realization of the sigma model incorporating chiral
symmetry. In particular, at high densities and/or temperatures, chiral
symmetry is restored, which can be seen in a reduction of the
effective baryon masses. The model is in good agreement with nuclear
physics data and at low densities and/or temperatures, the nuclear
liquid-gas first-order phase transition is reproduced. 

\subsubsection{Ab initio calculations} Due to the computational expenses
needed to determine the dense matter EoS starting from the basic few
body interactions between baryons, no tables from genuinely
microscopic calculations exist. However, tables with effective
interactions fitted to the results of microscopic calculations
wherever they exist, are available. One example is the APR
interaction~\cite{APR_PRC_1998}, which is a fit to the variational
calculations by \cite{AP_PRC_1997}. The latter employ realistic
two-body (Argonne V18~\cite{ArgonneV18}) and the three-body (Urbana
UIX~\cite{Carlson_NPA_1983,Pudliner_PRL_1995}) nuclear potentials
derived from fits of scattering data and that were shown to reproduce
a series of atomic nuclei properties.  The APR interaction includes a
phenomenological extension to correctly describe the effect of
three-body forces in dense matter.  Due to the three-nucleon
interaction, energies of both symmetric nuclear matter (SNM) and pure
neutron matter (PNM) manifest a discontinuity in slope such that
different analytic forms are required to fit the low and high density
domains~\cite{APR_PRC_1998}.  This discontinuity is interpreted as a
transition to a spin-isospin ordered phase, \textit{i.e.} a neutral
pion condensate, and the density where it occurs depends on the
isospin asymmetry of the system.  The extension to finite temperatures
of the APR EoS in \cite{Constantinou_PRC_2014} was performed assuming
that the transition density does not depend on temperature. Note
nevertheless that the borders of the phase coexistence region manifest
a slight temperature-dependence.

The calculation of Ref.~\cite{Kanzawa_NPA_2007} (``KOST'') relies on
the same basic two-body (Argonne V18~\cite{ArgonneV18}) and the
three-body (Urbana UIX~\cite{Carlson_NPA_1983,Pudliner_PRL_1995})
nuclear potentials as APR~\cite{APR_PRC_1998}.  Differences to the APR
EoS arise from the simplified two-body cluster variational approach
and the fact that relativistic boost corrections are not included in
\cite{Kanzawa_NPA_2007}.  They result in slightly different energies
of symmetric nuclear matter and neutron matter at zero temperature.
More importantly, at variance with APR, the KOST EoS
does not manifest any discontinuity in
the energy and, thus, no transition to a spin-isospin ordered phase at
high densities.  A detailed comparison between these two microscopic
EoS was performed in \cite{Constantinou_PRC_2014}.  The description of
homogeneous matter in Ref.~\cite{Togashi_NPA_2017} (``KOST2'') is
based on the KOST EoS, additionally
corrected for spurious deuteron-like correlations due to the two-body
cluster approximation which were shown to introduce nonzero
contributions to the energy per nucleon in the low density limit.

\subsection{Treatment of inhomogeneous matter}
\label{ssec:inhomo}
At low densities and temperatures, matter becomes clusterised. For the
considered conditions, matter is in equilibrium with respect to strong
and electromagnetic interaction, determining the abundances of the
different nuclei.  However, since in particular for the thermodynamic
properties of the EoS, the detailed distribution of nuclei has not a
large influence~\cite{Burrows_ApJ_1984}, many models approximate the
full nuclear distribution by one light ($^4$He)-nucleus and a
representative heavy nucleus within the so-called ``single nucleus
approximation'' (SNA). More recently, among others due to the
importance of the detailed distribution for weak interaction
rates~\cite{Hix_PRL_2003,Sullivan_ApJ_2016,Pascal_PRC_2020}, extended
``nuclear statistical equilibrium'' (NSE) models have been developed,
including an interaction between the nuclear clusters and the
surrounding gas of unbound nucleons, too. Models not only differ by
the considered pool of nuclei, but by the considered nuclear masses,
the treatment of thermally excited states, the treatment of the
transition to homogeneous matter, and the interaction with nucleons,
too. Let us now give a brief overview:

\subsubsection{SNA models}

The \textit{Lattimer and Swesty (LS) model}~\cite{LS_NPA_1991} is the
first general purpose EoS which has been made publicly available for
use in numerical simulations. For each $T, n_B, Y_e$ the equilibrium
configuration is determined by minimizing the total Helmholtz free
energy, considering one representative heavy nucleus,
$\alpha$-particles and unbound nucleons. The unique heavy nucleus is
treated within the finite-temperature compressible liquid-drop model,
$\alpha$-particles are considered to form an ideal Boltzmann gas and
unbound nucleons interact via a Skyrme like momentum independent
interaction. The energy functional accounts for shape-dependent
surface, Coulomb and translational energy corrections.  An excluded
volume correction mimics interactions between unbound nucleons,
$\alpha$-particles and the heavy nucleus. The transition from
inhomogeneous to homogeneous nuclear matter is performed by a Maxwell
construction.

Later on, the model has been extended to very low densities within a
noninteracting NSE approach~\cite{Oertel_PRC_2012} and to include the
contribution of $\Lambda$-hyperons at high densities, see
\cite{Oertel_PRC_2012,Peres_PRD_2014}.

The SNA models by \textit{Schneider, Roberts and Ott
  (SRO)}~\cite{SRO_PRC_2017,Schneider_PRC_2019} employ essentially the
approach by Lattimer \& Swesty for inhomogeneous matter. The two main
differences are that they consider a variety of Skyrme interactions
between unbound nucleons which offer more flexibility, especially at
high densities, and that the transition from inhomogeneous to
homogeneous nuclear matter is simplified. It is performed at the
density where the competing structures provide equal values for the
free energy. This obviously implies that there is no phase coexistence
region, at variance with the original LS model \cite{LS_NPA_1991}.
Technical improvements with respect to the original LS model include
accurate values of $\alpha$-particle binding energy, neutron and
proton masses and treatment of kinetic translational energy. In
Refs.~\cite{SRO_PRC_2017,Schneider_PRC_2019} tables have been computed
using a NSE treatment for inhomogeneous matter with 3335 different
nuclei. Apart from the NRAPR model, these are, however, not available
on \textsc{CompOSE} for the moment.

The \textit{model by H. Shen {\em et
    al.}}~\cite{STOS_PTP_1998,STOS_NPA_1998} represents the second
``standard'' general purpose EoS besides the LS~\cite{LS_NPA_1991} one
since tables have been made publicly available more than twenty years
ago. As LS, they rely on SNA and consider nucleons, $\alpha$-particles
and one heavy nucleus in the inhomogeneous phase. A Thomas-Fermi
approximation with parameterized density distributions in spherical
Wigner-Seitz cells is used for determining the thermodynamically
favorable state by minimizing the free energy density with respect to
the cell volume, central densities, spatial extensions and surface
diffuseness of protons and neutrons in the nucleus as well as
densities of unbound neutrons and protons.  The surface energy term is
determined by performing Thomas-Fermi calculations of finite nuclei so
as to reproduce some gross features of atomic nuclei.  Translational
degrees of freedom of the unique nucleus are not considered.
$\alpha$-particles are considered to form an ideal Boltzmann gas.
Excluded volume corrections for the $\alpha$-particles, the nucleon
gas and the heavy nucleus are implemented. The transition from
inhomogeneous to homogeneous nuclear matter is realized by minimizing
the free energy density at a given density.  In contrast to
LS~\cite{LS_NPA_1991}, they employ a CDFT effective interaction for
unbound nucleons.  The original version uses the TM1
parameterisation~\cite{TM1}, a more recent version (SNH(TM1e)) is
based on TM1e~\cite{Shen_ApJ_2020,SNSHS_2019}, allowing for a
different symmetry energy. The TNTYST(KOST2)
EoS~\cite{Togashi_NPA_2017} shares the same Thomas-Fermi theoretical
framework for describing inhomogeneous matter but employs a
microscopic interaction, KOST2
\cite{Kanzawa_NPA_2007,Togashi_NPA_2017}.  This is coherently used for
treating unbound nucleons in subsaturated matter, suprasaturation
homogeneous matter and the unique heavy nucleus, whose density
function is supplemented with a gradient term to account for surface
effects.

The original STOS(TM1) model has been extended to include the
contributions of hyperons and pions at high densities with different
hyperonic interactions~\cite{Ishizuka_JPG_2008} or a transition to
quark matter modeled with in the MIT bag
model~\cite{Sagert_PRL_2009,Sagert_JPG_2010,Fischer_ApJ_2011}.  The
transition from the hadronic to the quark phase is thereby obtained
via a Gibbs construction.

\subsubsection{Extended NSE models}
In the \textit{Furusawa {\em et al.}
  models }~\cite{Furusawa_ApJ_2011,Furusawa_ApJ_2013,Furusawa_NPA_2017,Furusawa_JPG_2017}
inhomogeneous matter is considered to be made of unbound nucleons and
a wide distribution of light (here defined as nuclei with $Z\leq 5$)
and heavy ($Z \geq 6$) nuclei.  Heavy nuclei are considered up to $Z
\leq 1000$ and $N \leq 1000$.  The binding energies of light nuclei
are computed by modifying the experimental data \cite{Audi_2012} with
phenomenological terms accounting for Coulomb screening by the uniform
electron gas, Pauli energy shift \cite{DD2} by other baryons and, for
$^2$H, $^3$H, $^3$He and $^4$He, interactions with
unbound nucleons~\cite{Furusawa_ApJ_2013}.

Heavy nuclei are computed within the liquid drop approximation with
various corrections taking into account dense matter effects: electron
screening is implemented in the Coulomb term, and phenomenological
temperature and density dependent corrections are added to shell and
surface energies~\cite{Furusawa_NPA_2017}.  Translational energy of
nuclei is calculated as in the ideal Boltzmann gas with a geometrical
estimation of the free volume.

For unbound nucleons, the FYS(TM1) model uses CDFT with the TM1
parameterisation, and the FT(KOST2) uses the microscopic
variational calculations from Refs.~\cite{Togashi_NPA_2017}.

The \textit{models by G. Shen {\em et
    al.}}~\cite{GShen_PRC_2010,GShen_PRC_2010p,GShen_PRC_2011p,GShen_PRC_2011}
consider a gas of neutrons, protons, $\alpha$-particles, and thousands
of nuclei with $A \geq 12$.  At subnuclear densities nonideal gas
behavior is incorporated via the virial expansion. At very low
density, this description reduces to noninteracting nuclear
statistical equilibrium, whereas at higher densities the results match
smoothly a CDFT description of homogeneous matter. Tables with
different parameterisations of the CDFT interaction exist:
NL3~\cite{NL3}, FSUgold~\cite{FSUGold} and
FSUgold2.1~\cite{GShen_PRC_2011}.

The models by \textit{Hempel and Schaffner-Bielich} rely on the
extended NSE model developed in
Ref.~\cite{Hempel_NPA_2010}. Inhomogeneous matter at low densities is
considered to be made of a gas of unbound self-interacting nucleons
and an ensemble of light and heavy nuclei.  The gas of unbound
nucleons is treated within the CDFT approach, as is also the case of
homogeneous matter at high densities.  The distribution of nuclei at
subsaturation densities is treated within NSE and for the free energy
density the Maxwell-Boltzmann expression of a classical ideal gas is
used.  For the nuclear masses different choices are considered that
automatically define the pool of nuclei, for details see
Table~\ref{tab:NSEmodels}.  Coulomb screening by the electron gas,
which modifies the nuclear binding energy, is implemented in the
Wigner-Seitz cell approximation. A phenomenological temperature
dependent description of excited nuclear states is included.
Interactions between nuclei and unbound nucleons are
mimicked via the excluded volume correction. The transition from
inhomogeneous to homogeneous matter is realized under the constraints
of local charge neutrality and equal values of proton fraction for the
two phases. Pressure equality is determined only in an approximative
way: pressures are only compared at the density grid-points of the
final EoS table.  The two phases are then connected by a thermodynamic
consistent interpolation. Tables with a dozen of different
parameterisations for the CDFT interaction of unbound nucleons exist,
see Table~\ref{tab:NSEmodels}.


The original model has been extended to account for
$\Lambda$-hyperons~\cite{BHB_2014} or the full baryonic
octet~\cite{Marques_PRC_2017,FOP_2017} at high densities or a
condensate of $K^-$-mesons~\cite{Malik_EPJA_2021}. A transition to quark
matter is constructed in the class of BBKF
models~\cite{Bastian_PRC_2017,Bastian_PRD_2021}.

\textit{Typel models} The generalized relativistic density functional
(GRDF) model was developed in Ref. \cite{DD2}.  At subsaturation
densities it accounts for a wide collection of light and heavy nuclei
and effective two-nucleon states in the continuum with
medium-dependent energies.  At variance with other NSE models, where
the transition to homogeneous matter takes place because of the
excluded volume, in GRDF this transition occurs because nuclei
dissolve in the surrounding gas of unbound nucleons.  This process is
realized via a density and temperature dependent shift of atomic
masses accounting for the Pauli principle. Its parametrized form is
obtained by fits to predictions from microscopic quantum statistical
calculations~\cite{Roepke_PRC_2009}.  GRDF1 and GRDF2 differ in the
parametrization of the mass shifts of heavy nuclei as well as the
values of the employed nuclear masses.  For the level density,
temperature dependent expressions are considered.  The parametrized
form of the mass shift is correlated with the density where the
dissociation of massive nuclei occurs; in GRDF1 (GRDF2) this happens
at $n_{\mathit{sat}}/3$ ($n_{\mathit{sat}}/2$).  Both GRDF1 and GRDF2 employ the DD2
\cite{DD2} parameterisation for the interaction of unbound nucleons.
For more details, see Ref. \cite{Typel_JPG_2018}.


The extended NSE model of \textit{Gulminelli and
  Raduta}~\cite{Gulminelli_PRC_2015,Raduta_NPA_2019} takes into
account a large set of heavy and light nuclei, including
$^{2,...,6}$H, $^{3,...,10}$He, $^{3,...,13}$Li, $^{7,...,18}$Be.  At
variance with other NSE models, this model also accounts for nuclei
beyond drip lines.  Their nuclear binding energies are described
according to LDM(SLy4)~\cite{Danielewicz_NPA_2009} modified in two
respects.  First, a phenomenological pairing term is added.  Then, two
correction terms are included such as to smoothly match, for each
isotopic chain, the liquid-drop predictions with the limiting values
of the DZ10~\cite{DZ_PRC_1995} mass model, employed whenever
experimental data are not available.  The allowed mass range of
clusters is $2 < A < 300$.  The full list of low-lying resonances for
nuclei with $4 \leq A \leq 10$ has been considered.  For the level
density a realistic expression fitted on experimental data is used.
Unbound nucleons are modeled with the SLy4 Skyrme interaction~\cite{SLy4}.
Excluded volume effects between nuclear clusters and unbound nucleons
are implemented via energy shifts of clusters binding energies.

In order to describe the transition from inhomogeneous to homogeneous
matter, for fixed values of $Y_q$ and $T$ the clusterized phase is
computed within the NSE approximation up to a maximum density where the
NSE procedure converges, typically $4 \cdot 10^{-2} - 9 \cdot 10^{-2}$
fm$^{-3}$.  Homogeneous matter is supposed to appear, independently of
temperature and proton fraction, at $n_t = 10^{-1}$ fm$^{-3}$. For
intermediate values of density, chemical composition and thermodynamic
observables are computed by linear interpolation between the boundary
values.

\begin{table*}
  \caption{List of general purpose EoS tables based on SNA, available
    on \textsc{Compose}.  For each EoS model we provide information
    on: nuclear effective interaction; low density extension by NSE;
    considered degrees of freedom.  Also given are the maximum mass of
    cold $\beta$-equilibrated NS ($\mmax$); radius of canonical $1.4M_{\odot}$
    NS ($R_{14}$); radius of a $2.072M_{\odot}$ NS ($R_{2.072}$);
    limits of combined tidal deformability $\tilde \Lambda=16\left[
      \left( M_1+12 M_2\right) M_1^4 \Lambda_1 + \left( M_2+12
      M_1\right) M_2^4 \Lambda_2 \right]/13 \left(M_1 +M_2\right)^5$
    corresponding to the GW170817 event with an estimated total mass
    $M_T=2.73^{+0.04}_{-0.01} M_{\odot}$ and a mass ratio range $0.73
    \leq q=M_2/M_1 \leq 1$.  The last but one column specifies whether
    tables exist also for purely baryonic matter, in addition to those
    corresponding to the whole mixture.  Present astrophysical
    constraints on EoS regard: i) the lower limit of maximum
    gravitational mass; ii) radii of canonical mass NS; iii) radii of
    a massive NS; iv) range for the tidal deformability obtained from
    the GW170817 event, see text for details.  Models in tension with
    constraint i) $M \geq 2.01 - 0.04 M_{\odot}$ \cite{Antoniadis2013}
    are marked in bold.  
    Values outside the ranges
    ii) $11.80 \leq R(1.4M_{\odot}) \leq 13.10$ km \cite{Miller_may2021},
    iii) $11.41~{\rm km} \leq R(2.072M_{\odot})\leq 13.69~{\rm km}$ \cite{Riley_may2021},
    iv) $110 \leq \tilde \Lambda \leq 800$ \cite{Abbott_2019}
    are also marked in bold.
    For $\mmax$ and $R_{14}$ provided are the values on \textsc{CompOSE}
    or the original publications.  $R_{2.072}$, $\tilde
    \Lambda(q=0.73)$ and $\tilde \Lambda(q=1)$ are calculated by using
    for the crust the EoS models by \cite{NV_1973} and
    \cite{HDZ_1989}.  Other notations are: $q$ stands for $u$, $d$,
    $s$ quarks; $\Lambda$ denotes the $\Lambda$-hyperon; $Y$
    generically denotes the $\Lambda$, $\Sigma$ and
    $\Xi$ hyperons; $\pi$ stands for pions.
  } 
\label{tab:SNAmodels}       
\begin{tabular}{llcccccccl}
\hline\noalign{\smallskip}
class/ & low dens. & d.o.f & $\mmax$ & $R_{14}$ & $R_{2.072}$ & $\tilde \Lambda$  & $\tilde \Lambda$ & baryonic & Ref.  \\
 model & extension & exotica & ($M_{\odot}$) & (km)  & (km) & q=0.73 & q=1 & tables \\
\noalign{\smallskip}\hline\noalign{\smallskip}
Lattimer and Swesty \\
LS(LS220) &  n &- & 2.06 & 12.7 & - & 664 & 596 & - & \cite{LS_NPA_1991} \\
no low dens.\\
LS(LS220) &  y &- &  2.06 & 12.7 & - & 664 & 596 & - & \cite{LS_NPA_1991,Oertel_PRC_2012}\\
with low dens. \\
{\bf GROM(LS220L)} &  n & $\Lambda$ & {\bf 1.91} & 12.4 & - & 576 & 498 & - & \cite{LS_NPA_1991,Oertel_PRC_2012}\\
  no low dens.\\
  {\bf GROM(LS220L)} &  y & $\Lambda$ & {\bf 1.91} & 12.4 & - & 576 & 498 & - & \cite{LS_NPA_1991,Oertel_PRC_2012}\\
  with low dens.\\
\noalign{\smallskip}\hline\noalign{\smallskip}
Schneider \em{et al.}\\
SRO(APR) & n & $\pi$ condens. & 2.16 & {\bf 11.3} & {\bf 10.9} & 299 & 272 & - & \cite{Schneider_PRC_2019} \\
{\bf SRO(NRAPR)} & y  & - & {\bf 1.94} & 11.9 & - & 385 & 340 & - & \cite{SRO_PRC_2017,Schneider_PRC_2019} \\
SRO(SLy4)  & n & - & 2.05 & {\bf 11.7} & - & 369 & 334 & - & \cite{SRO_PRC_2017} \\
SRO(SkAPR) & n & - & 1.97 & 12.0 & - & 535 & 490 & - & \cite{SRO_PRC_2017} \\
SRO(LS220) & n & - & 2.04 & 12.7 & - & 658 & 593 & - & \cite{SRO_PRC_2017} \\
SRO(KDE0v1)& n & - & 1.97 & {\bf 11.7} & - & 340 & 303 & - & \cite{SRO_PRC_2017} \\
{\bf SRO(LNS)}& n & - & {\bf 1.72} & {\bf 11.0} & - & 235 & 196 & - & \cite{SRO_PRC_2017} \\
\noalign{\smallskip}\hline\noalign{\smallskip}
HShen \em{et al.} \\
STOS(TM1)          &  n & - & 2.23 & {\bf 14.5} & {\bf 13.7} & {\bf 1376} & {\bf 1279} & y & \cite{STOS_PTP_1998,STOS_NPA_1998}  \\
{\bf STOS(TM1L)}   & n & $\Lambda$ & {\bf 1.90} & {\bf 14.4} & - & {\bf 1366} & {\bf 1283} & y & \cite{STOS_PTP_1998,STOS_NPA_1998}  \\
{\bf IOTSY(TM1Y-30)} & n & Y & {\bf 1.63} & {\bf 14.3} & - & - & {\bf 1258}  & - & \cite{Ishizuka_JPG_2008} \\
{\bf IOTSY(TM1Y0)} & n & Y & {\bf 1.64} & {\bf 14.3} & - & {\bf 1361} & {\bf 1286} &  - & \cite{Ishizuka_JPG_2008} \\
{\bf IOTSY(TM1Y30)} & n & Y & {\bf 1.64} & {\bf 14.3} & - & {\bf 1362} & {\bf 1286} & - & \cite{Ishizuka_JPG_2008} \\
{\bf IOTSY(TM1Y90)} & n & Y & {\bf 1.64} & {\bf 14.3} & - & {\bf 1362} & {\bf 1286} & - & \cite{Ishizuka_JPG_2008} \\
{\bf IOTSY(TM1Y-30pi)} & n & $Y, \pi$ & {\bf 1.66} & {\bf 13.6} & - & {\bf 844} & 781 & - & \cite{Ishizuka_JPG_2008} \\
{\bf IOTSY(TM1Y0pi)} & n & $Y, \pi$ & {\bf 1.66} & {\bf 13.6} & - & {\bf 858} & 778 & - & \cite{Ishizuka_JPG_2008} \\
{\bf IOTSY(TM1Y30pi)} & n & $Y, \pi$ & {\bf 1.66} & {\bf 13.6} & - & {\bf 858} & 778 & - & \cite{Ishizuka_JPG_2008} \\
{\bf IOTSY(TM1Y90pi)} & n & $Y, \pi$ & {\bf 1.66} & {\bf 13.6} & - & {\bf 858} & 778 & - & \cite{Ishizuka_JPG_2008} \\
SFHPST(TM1B139) & n & $q$ & 2.08 & 12.6 & - & n.a. & n.a. & - & \cite{STOS_PTP_1998,STOS_NPA_1998,Sagert_PRL_2009,Sagert_JPG_2010,Fischer_ApJ_2011}\\
SFHPST(TM1B145) & n & $q$ & 2.01 & 13.0 & - & n.a. & n.a. & - & \cite{STOS_PTP_1998,STOS_NPA_1998,Sagert_PRL_2009,Sagert_JPG_2010,Fischer_ApJ_2011}\\
{\bf SFHPST(TM1B155)} & n & $q$ & {\bf 1.70} & {\bf 10.7} & - & n.a. & n.a. & - & \cite{STOS_PTP_1998,STOS_NPA_1998,Sagert_PRL_2009,Sagert_JPG_2010,Fischer_ApJ_2011}\\
{\bf SFHPST(TM1B165)} & n & $q$ & {\bf 1.51} & {\bf 9.1} & - & n.a. & n.a. & - & \cite{STOS_PTP_1998,STOS_NPA_1998,Sagert_PRL_2009,Sagert_JPG_2010,Fischer_ApJ_2011}\\
SNSH(TM1e) & n & - & 2.12 & 13.1 & 12.5& 793 & 756 & - & \cite{Shen_ApJ_2020,SNSHS_2019} \\
\noalign{\smallskip}\hline\noalign{\smallskip}
Togashi {\em et al.} \\
TNTYST(KOST2) & n & - & 2.21 & {\bf 11.5} & {\bf 11.1} & 358  & 332 &  y & \cite{Togashi_NPA_2017} \\
\noalign{\smallskip}\hline\noalign{\smallskip}
\end{tabular}
\end{table*}

\begin{table*}
  \caption{The same as in Table \ref{tab:SNAmodels} for EoS models
    accounting only for homogeneous matter.  In addition to
    $R_{2.072}$, $\tilde \Lambda(q=0.73)$ and $\tilde \Lambda(q=1)$
    $R_{14}$ is calculated here by using the NS crust models by
    \cite{NV_1973} and \cite{HDZ_1989}.  The matching with the core
    EoS is performed at $n_{cc}=0.07~{\rm fm^{-3}}$.}
  \label{tab:HOMOmodels}       
  \begin{tabular}{llccccccl}
\hline\noalign{\smallskip}
class / & d.o.f & $\mmax$ & $R_{14}$ & $R_{2.072}$ & $\tilde \Lambda$ & $\tilde \Lambda$ & baryonic & Ref.  \\
model  & exotica & ($M_{\odot}$) & (km) & (km) &q=0.73 &q=1 & tables \\
\noalign{\smallskip}\hline\noalign{\smallskip}
Dexheimer {\em et al.} \\
DNS(CMF) & Y, q & 2.1 & {\bf 14.0} & 12.6 & {\bf 1114} & {\bf 1043} & y & \cite{DNS_PRC_2015,Dexheimer_PASA_2017} \\
\noalign{\smallskip}\hline\noalign{\smallskip}
\end{tabular}
\end{table*}

\begin{table*}
  \caption{The same as in Table \ref{tab:SNAmodels} for EoS models
    based on NSE.  Whenever more than one reference is provided for
    mass tables, these are used over complementary $(A,Z)$ domains.
    The mass data are employed using the following hierarchy:
    experimental data $>$ predictions of mass models (Finite Range
    Droplet Model (FRDM) \cite{Moller_ADNDT_1995}, Duflo-Zuker (DZ)
    \cite{DZ_PRC_1995}) $>$ estimations based on the liquid drop model
    (LDM).  Other notations are: $nn$, $np$
    and $pp$ stand for $^1S_0$ continuum correlations of the mentioned
    particles.  * means that for MBB(DD2K) only tables corresponding
    to purely baryonic matter are available.  }
  \label{tab:NSEmodels}       
  \begin{tabular}{llcccccccl}
\hline\noalign{\smallskip}
class / & mass & d.o.f & $\mmax$ & $R_{14}$ & $R_{2.072}$ & $\tilde \Lambda$ & $\tilde \Lambda$ & baryonic & Ref.  \\
model & table  & exotica & ($M_{\odot}$) & (km) & (km) & q=0.73 &q=1 & tables \\
\noalign{\smallskip}\hline\noalign{\smallskip}
GShen \em{et al.} \\
{\bf SHO(FSU1)}  & \cite{Moller_ADNDT_1995} & - & {\bf 1.75} & 12.8 & - & {\bf 1182} & {\bf 932} &  y & \cite{GShen_PRC_2011} \\
SHO(FSU2) & \cite{Moller_ADNDT_1995} & - & 2.12 & {\bf 13.6}  & {\bf 14.2} & {\bf 2604} & {\bf 2307} & y & \cite{GShen_PRC_2011} \\
SHT(NL3)        & \cite{Moller_ADNDT_1995} & -  & 2.78 & {\bf 14.9} & {\bf 14.9} &  {\bf 1639} & {\bf 1555} & y & \cite{GShen_PRC_2011p} \\
\noalign{\smallskip}\hline\noalign{\smallskip}
Furusawa {\em et al.} \\
FYSS(TM1)   & LDM & - & 2.22 & {\bf 14.4} & {\bf 13.7} & {\bf 1376} & {\bf 1279} & y & \cite{Furusawa_ApJ_2011,Furusawa_ApJ_2013,Furusawa_NPA_2017}\\
FTNS(KOST2) & LDM & - & 2.25 & {\bf 11.5} & {\bf 11.1} & 360 & 334 & y & \cite{Furusawa_ApJ_2011,Furusawa_ApJ_2013,Furusawa_NPA_2017,Furusawa_JPG_2017}\\
\noalign{\smallskip}\hline\noalign{\smallskip}
Hempel {\em et al.} \\
Steiner {\em et al.}; Banik {\em et al.}  \\
 HS(NL3) & \cite{Lalazissis_ADNDT_1999} & - & 2.79 & {\bf 14.8} & {\bf 14.9} & {\bf 1571} & {\bf 1483} & y & \cite{Hempel_NPA_2010} \\
 HS(TM1) & \cite{Geng_PTP_2005} & -  & 2.21 & {\bf 14.5} & 13.7 & {\bf 1351} & {\bf 1255} & y & \cite{Hempel_NPA_2010} \\
 HS(TMA) & \cite{Geng_PTP_2005} & -  & 2.02 & {\bf 13.8} & - & {\bf 1128} & {\bf 1052} & y & \cite{Hempel_NPA_2010} \\
 HS(DD2) & \cite{Audi_NPA_2003},\cite{Moller_ADNDT_1995} & - & 2.42 & {\bf 13.1} & 13.1 & 799 & 758 & y & \cite{Hempel_NPA_2010} \\
{\bf HS(FSG)} & \cite{Roca-Maza_PRC_2008} & -  & {\bf 1.74} & 12.6 & - & 539 & 439  & y & \cite{Hempel_NPA_2010} \\
{\bf HS(IUF)} & \cite{Roca-Maza_PRC_2008} & -  & {\bf 1.95} & 12.7 & - & 608 & 570 & y & \cite{Hempel_NPA_2010}\\
{\bf BHB(DD2L)} & \cite{Audi_NPA_2003}, \cite{Moller_ADNDT_1995} & $\Lambda$ & {\bf 1.95} & {\bf 13.2} & - & 787 & 757 & y & \cite{Hempel_NPA_2010,BHB_2014} \\
 BHB(DD2Lphi) & \cite{Audi_NPA_2003}, \cite{Moller_ADNDT_1995} & $\Lambda$ & 2.10 & {\bf 13.2} & 12.2 & 790 & 757 & y &\cite{Hempel_NPA_2010,BHB_2014} \\
 SFH(SFHo)  &\cite{Audi_NPA_2003}, \cite{Moller_ADNDT_1995} & - & 2.06 & 11.9 & - & 401 & 366 & y & \cite{Hempel_NPA_2010,SHF_ApJ_2013} \\
 FOP(SFHoY) & \cite{Audi_NPA_2003},\cite{Moller_ADNDT_1995} & Y & 1.99 & 11.9 & - & 401 & 366 & y & \cite{Hempel_NPA_2010,SHF_ApJ_2013,FOP_2017} \\
 SFH(SFHx)  & \cite{Audi_NPA_2003},\cite{Moller_ADNDT_1995} & - & 2.13 & 12.0 & {\bf 11.3} & 466 & 428 & y & \cite{Hempel_NPA_2010,SHF_ApJ_2013} \\
 OMHN(DD2Y) & \cite{Audi_NPA_2003},\cite{Moller_ADNDT_1995} & Y & 2.03 & {\bf 13.2} & - & 787 & 756 & - & \cite{Hempel_NPA_2010,Marques_PRC_2017} \\
 MBB(DD2K)  & \cite{Audi_NPA_2003},\cite{Moller_ADNDT_1995} & $K^-$ & 2.19 & {\bf 13.2} & - & n.a. & n.a. & * & \cite{Hempel_NPA_2010,Malik_EPJA_2021}\\
\noalign{\smallskip}\hline\noalign{\smallskip}
Typel {\em et al.} & {\bf GRDFN\_DD2} \\
 PT(GRDF1\_DD2) & \cite{Audi_2012}, DZ10\cite{DZ_PRC_1995} & - & 2.42 & {\bf 13.2} & 13.1 & 781 & 742 & - & \cite{Typel_JPG_2018} \\
 PT(GRDF2\_DD2) & \cite{Audi_2017}, DZ31\cite{DZ_PRC_1995} & - & 2.42 & {\bf 13.2} & 13.2 & 663 & 636 & - & \cite{Typel_JPG_2018} \\
\noalign{\smallskip}\hline\noalign{\smallskip}
Bastian {\em et al.}\\
 BBKF(DD2F-SF)1.1 & \cite{Audi_NPA_2003}, \cite{Moller_ADNDT_1995} & q & 2.13 & 12.2 & {\bf 10.7} & 507 & 467 & y & \cite{Bastian_PRC_2017,Bauswein_PRL_2019,Bastian_PRD_2021} \\
BBKF(DD2F-SF)1.2 & \cite{Audi_NPA_2003}, \cite{Moller_ADNDT_1995} & q & 2.15 & 12.2 & 11.4 & 501 & 473 & y & \cite{Bastian_PRC_2017,Bauswein_PRL_2019,Bastian_PRD_2021} \\
BBKF(DD2F-SF)1.3 & \cite{Audi_NPA_2003}, \cite{Moller_ADNDT_1995} & q & 2.02 & 12.2 & - & 512 & 467 & y & \cite{Bastian_PRC_2017,Bauswein_PRL_2019,Bastian_PRD_2021} \\
BBKF(DD2F-SF)1.4 & \cite{Audi_NPA_2003}, \cite{Moller_ADNDT_1995} & q & 2.02 & 12.2 & - & 516 & 467 & y & \cite{Bastian_PRC_2017,Bauswein_PRL_2019,Bastian_PRD_2021} \\
BBKF(DD2F-SF)1.5 & \cite{Audi_NPA_2003}, \cite{Moller_ADNDT_1995} & q & 2.03 & 12.2 & - &  488 & 467 & y & \cite{Bastian_PRC_2017,Bauswein_PRL_2019,Bastian_PRD_2021} \\
BBKF(DD2F-SF)1.6 & \cite{Audi_NPA_2003}, \cite{Moller_ADNDT_1995} & q & 2.00 & 12.2 & - & 513 & 467 & y & \cite{Bastian_PRC_2017,Bauswein_PRL_2019,Bastian_PRD_2021} \\
 BBKF(DD2F-SF)1.7 & \cite{Audi_NPA_2003}, \cite{Moller_ADNDT_1995} & q & 2.11 & 12.2 & {\bf 11.2} & 514 & 467 & y & \cite{Bastian_PRC_2017,Bastian_PRD_2021} \\ 
 BBKF(DD2-SF)1.8 & \cite{Audi_NPA_2003}, \cite{Moller_ADNDT_1995} & q & 2.06 & {\bf 11.0} & - & 218 & 180 & y & \cite{Bastian_PRC_2017,Bastian_PRD_2021} \\
 BBKF(DD2-SF)1.9 & \cite{Audi_NPA_2003}, \cite{Moller_ADNDT_1995} & q & 2.17 & {\bf 11.3} & {\bf 11.2} & 228 & 196 & y & \cite{Bastian_PRC_2017,Bastian_PRD_2021} \\
\noalign{\smallskip}\hline\noalign{\smallskip}
        Gulminelli \\
        and Raduta\\
        RG(SLy4) & \cite{Audi_2012}, DZ10 \cite{DZ_PRC_1995}, & - & 2.07 & 11.9 & - & 369 & 334 & - & \cite{Gulminelli_PRC_2015,Raduta_NPA_2019} \\
                 &      & LDM(SLy4) \cite{Danielewicz_NPA_2009} & \\
\noalign{\smallskip}\hline
\end{tabular}
\end{table*}

\section{EoS of cold matter}
\label{sec:T=0}
\begin{table*}
  \caption{List of nucleonic effective interactions on which general
    purpose EoS tables on \textsc{CompOSE} are built.  For each
    interaction we list the properties of symmetric nuclear matter at
    saturation density ($n_{\mathit{sat}}$): energy per nucleon ($E_{\mathit{sat}}$);
    compression modulus ($K_{\mathit{sat}}$); skewness ($Q_{\mathit{sat}}$); symmetry
    energy ($J_{sym}$); slope ($L_{sym}$) and curvature ($K_{sym}$) of
    the symmetry energy; Landau effective mass of nucleons
    ($m_{eff}$).  We list in addition the neutron-proton effective
    mass splitting in neutron matter at $n_{\mathit{sat}}$ ($\Delta m_{eff}$).
    Presently accepted constraints are as follows: $E_{\mathit{sat}} =15.8
    \pm 0.3$ MeV \cite{Margueron_PRC_2018a}; $n_{\mathit{sat}}=0.155 \pm 0.005$
    fm$^{-3}$ \cite{Margueron_PRC_2018a}; $K_{\mathit{sat}}=230 \pm 40$ MeV
    \cite{Khan_PRL_2012}; $E_{sym}=31.7 \pm 3.2$ MeV
    \cite{Oertel_RMP_2017}; $L_{sym}=58.7 \pm 28.1$ MeV
    \cite{Oertel_RMP_2017}.
    Note nevertheless that the model-dependence inherent to the analyses
    for obtaining these values explains that other favored
    domains have been obtained in other works; for $L_{sym}$ see for instance
    \cite{Lattimer_EPJA_2014,BaoAnLi_Universe_2021,Reed_PRL_2021};
    for $K_{sat}$ see \cite{Shlomo_EPJA_2006}, too.
  }
\label{tab:effint}       
\begin{tabular}{lcccccccccccl}
\hline\noalign{\smallskip}
 int. & class & $n_{\mathit{sat}}$ & $E_{\mathit{sat}}$ & $K_{\mathit{sat}}$ & $Q_{\mathit{sat}}$ & $J_{sym}$ & $L_{sym}$ & $K_{sym}$ & $Q_{sym}$ & $m_{eff}$ &
$\Delta m_{eff}$ & Ref.  \\
& & (fm$^{-3}$) & (MeV) & (MeV) & (MeV) & (MeV) & (MeV) & (MeV) & (MeV) & ($m_n$) & ($m_n$)  \\
 \noalign{\smallskip}\hline\noalign{\smallskip}
 LS220 & Skyrme & 0.155 & -16.64 & 219.85 & -410.80 & 28.61 & 73.81 & -24.04 & 96.17 & 1.00 & 0.00 &  \cite{LS_NPA_1991} \\
 SLy4 & Skyrme & 0.159 & -15.97 & 229.91 & -363.11 & 32.00 & 45.94 & -119.73 & 521.53 & 0.695 & -0.184 &  \cite{SLy4} \\
 KDE0v1 & Skyrme & 0.165 & -16.88 & 227.53 & -384.83 & 34.58 & 54.70 & -127.12 & 484.44 & 0.744 & -0.128 &  \cite{Agrawal_PRC_2005} \\
 LNS & Skyrme & 0.175 & -15.96 & 210.76 & -382.50 & 33.43 & 61.45 & -127.35 & 302.52 & 0.826 & 0.228 &  \cite{LNS} \\
 NRAPR & Skyrme & 0.161 & -16.50 & 225.64 & -362.51 & 32.78 & 59.64 & -123.32 & 311.60 & 0.694 & 0.214 & \cite{Steiner_PhysRep_2005} \\
 SkAPR & Skyrme & 0.160 & -16.00 & 266.0 & -348.3 & 32.59 & 58.47 & -102.63 & 420.02 & 0.698 & 0.211 & \cite{Schneider_PRC_2019} \\
FSUgold & CDFT & 0.148      & -16.28 & 229.54 & -523.93  & 32.56 & 60.44 & -51.40 & 425.72 & 0.668 & 0.089 & \cite{FSUGold} \\
FSUgold2.1& CDFT & 0.148      & -16.28 & 229.54 & -523.93& 32.60 & 60.50  &  n.a.   & n.a.   & 0.668 & 0.089 & \cite{GShen_PRC_2011} \\
NL3     & CDFT & 0.148      & -16.24 & 271.53 & 202.91   & 37.40 & 118.53 & 100.88 & 181.31 & 0.659 & 0.090 & \cite{NL3} \\
TM1     & CDFT & 0.145      & -16.26 & 281.16 &-285.22   & 36.89 & 110.79 &  33.63 & -66.54 & 0.689 & 0.085 & \cite{TM1} \\
TM1e    & CDFT & 0.145      & -16.3  & 281    &-285      & 31.38 &  40    &   3.57 & n.a.   & 0.647 & 0.086 & \cite{TM1e} \\
TMA     & CDFT & 0.147      & -16.02 & 318.15 &-572.12   & 30.66 &  90.14 & 10.75 & -108.74 & 0.686 & 0.086 & \cite{TMA} \\
DD2     & CDFT & 0.149      & -16.02 & 242.72 & 168.65   & 31.67 &  55.04 & -93.23 & 598.14 & 0.626 & 0.097 & \cite{DD2} \\
DD2F    & CDFT & 0.149      & -16.02 & 242.72 & 168.65   & 31.67 &  55.04 & -93.23 & 598.14 & 0.626 & 0.097 & \cite{DD2F} \\
IUF     & CDFT & 0.155      & -16.40 & 231.33 &-290.28   & 31.30 &  47.21 &  28.53 & 370.71 & 0.670 & 0.093 & \cite{IUF} \\
SFHo    & CDFT & 0.158      & -16.19 & 245.4  &-467.8    & 31.57 &  47.10 & -205.5 & n.a.   & 0.811 & 0.078 & \cite{SHF_ApJ_2013} \\
SFHx    & CDFT & 0.160      & -16.16 & 238.8  &-457.2    & 28.67 &  23.18 & -40.00 & n.a.   & 0.771 & 0.083 & \cite{SHF_ApJ_2013} \\
CMF      & CDFT & 0.15       & -16.0  & 300    & 281      & 30    &  88   &   27   & n.a.   & n.a.  & n.a. &  \cite{DS_ApJ_2008,DNS_PRC_2015} \\
KOST2  & micro & 0.160    & -16.09 & 245    & n.a.  & 30.0  &  35    & n.a.   & n.a.   & n.a   & n.a. & \cite{Kanzawa_NPA_2007,Togashi_NPA_2017} \\ 
APR     & micro & 0.160    & -16.00 & 266.0  & -1054.2  & 32.59 &  58.47 & -102.63& 1217.0 & 0.698 & 0.211 & \cite{APR_PRC_1998} \\
\noalign{\smallskip}\hline
\end{tabular}
\end{table*}

Before addressing thermal effects, let us discuss some properties of
the different EoS models in cold matter. Actually, most of the tables
do not have a zero temperature entry, but the lowest available
temperature is in general $T = 0.1$ MeV. Except for very low densities
$n_B \lesssim 10^{-9} \mathrm{fm}^{-3}$, the EoS at this low temperature is
almost identical to the zero temperature one since the other energies in the
system are much higher than $T$ and in particular the chemical
potentials $\mu_i \gg T$.

Table \ref{tab:effint} summarizes nuclear matter parameters obtained
employing the different interactions. These parameters enter a Taylor
expansion of the energy per baryon of homogeneous nuclear
matter in terms of baryon number density and isospin, taken around
saturation density and isospin symmetry (\textit{i.e.} same number of
protons and neutrons). Mathematically this expansion is expressed
with the variables $\delta = \left(n_n-n_p\right)/n_B=1-2 Y_p$, where
$Y_p=n_p/n_B$ denotes the proton fraction, and
$\chi=\left(n_B-n_{\mathit{sat}}\right)/3 n_{\mathit{sat}}$,
respectively.
The first step in this approach consists in expressing
the energy per nucleon at arbitrary values of density $n_B$ and
isospin asymmetry $\delta$ as a sum between an isoscalar term
and an isovector term
\begin{equation}
  E(n_B,\delta)/A=E_0(n_B,0)+\delta^2 E_{sym}(n_B,0)~,
  \label{eq:EofnB}
\end{equation}
which implies a parabolic approximation for the isospin dependence.
The first term in Eq. (\ref{eq:EofnB}) represents
the energy per baryon of symmetric matter while
and the second represents the symmetry energy.
Note that other works in the literature
\cite{Piekarewicz_PRC_2009,Chen_PRC_2009}
account for higher order terms in isospin asymmetry. In this case,
the symmetry energy is defined as the lowest order coefficient in
the expansion of $E(n_B,\delta)/A$ in powers of $\delta$, 
\begin{eqnarray}
  E_{sym}(n_B,0)=\frac12 \frac{\partial^2 \left(E/A \right)}{\partial \delta^2}|_{n_B,\delta=0}~.
  \label{eq:Esym_curv}
\end{eqnarray}

In turn the isoscalar and isovector terms in Eq. (\ref{eq:EofnB})
are expanded in terms of baryon number density as
\begin{eqnarray}
  E_0(n_B,0)&=&E_{\mathit{sat}}(n_{\mathit{sat}},0)+\frac1{2!} K_{\mathit{sat}}(n_{\mathit{sat}},0) \chi^2\nonumber \\ &&
  + \frac1{3!} Q_{\mathit{sat}}(n_{\mathit{sat}},0) \chi^3 + {\cal O}(\chi^4),
  \label{eq:E0} 
\end{eqnarray}
and, respectively,
\begin{eqnarray}
  E_{sym}(n_B,0)&=&J_{sym}(n_{\mathit{sat}},0)+L_{sym}(n_{\mathit{sat}},0) \chi \nonumber \\ &&
  +\frac1{2!} K_{sym}(n_{\mathit{sat}},0) \chi^2+ {\cal O}(\chi^3).
  \label{eq:Esym}
\end{eqnarray}

We note that the explanation of odd terms missing in the development of
$E(n_B,\delta)/A$ in powers of $\delta$ resides in the
assumed QCD isospin invariance, which is an approximation, and the fact
that Coulomb interaction is disregarded.

We also note that the approximation in Eq. (\ref{eq:EofnB}) suggests a second definition for
the symmetry energy, as the per-nucleon cost of converting SNM ($\delta=0$) in PNM ($\delta = 1$)
\begin{equation}
  E_{sym} (n_B)=\left(E/A\right)_{PNM}(n_B)-\left(E/A\right)_{SNM}(n_B)~.
  \label{eq:Esym_diff}
\end{equation}
It is obvious that the definitions in Eqs. (\ref{eq:EofnB}) and (\ref{eq:Esym_diff})
only agree if higher order terms in $\delta$ are small,
thus typically close to saturation density \cite{Chen_PRC_2009}.


The caption of Table \ref{tab:effint} lists bounds for these nuclear matter
parameters obtained from a variety of nuclear experiments.

In the following we analyze the predictions of the different models in
the limiting cases of symmetric matter (SM) and neutron matter (NM).
Only purely nucleonic models which comply with the constraints on
$\mmax$, $L_{sym}$ and $K_{\mathit{sat}}$ specified in the captions of Tables
\ref{tab:SNAmodels} and \ref{tab:effint} are considered. These are
LS220, SRO(APR), SRO(SLy4), SRO(SkAPR), SRO(KDE0v1), TNTYST,
SHO(FSUgold2.1), HS(DD2), HS(IUF), SFHo.  In order to enlarge slightly
the pool of models for better comparison, we additionally include
SRO(NRAPR), SFHx and SNSH(TM1e). The first one thereby slightly
underestimates $\mmax$, the second $L_{sym}$ and the last one slightly
overestimates $K_{\mathit{sat}}$. All results reported for cold matter
correspond to the lowest available temperature value, typically
$T=0.1$ MeV. Similarly, results reported as neutron matter correspond
to the lowest available $Y_q$ value, typically $Y_q=0.01$, which
should be very close to the pure neutron matter case.

\subsection{Symmetric matter}
\label{ssec:SNM_T=0}

\begin{figure}
\includegraphics[width=0.99\columnwidth]{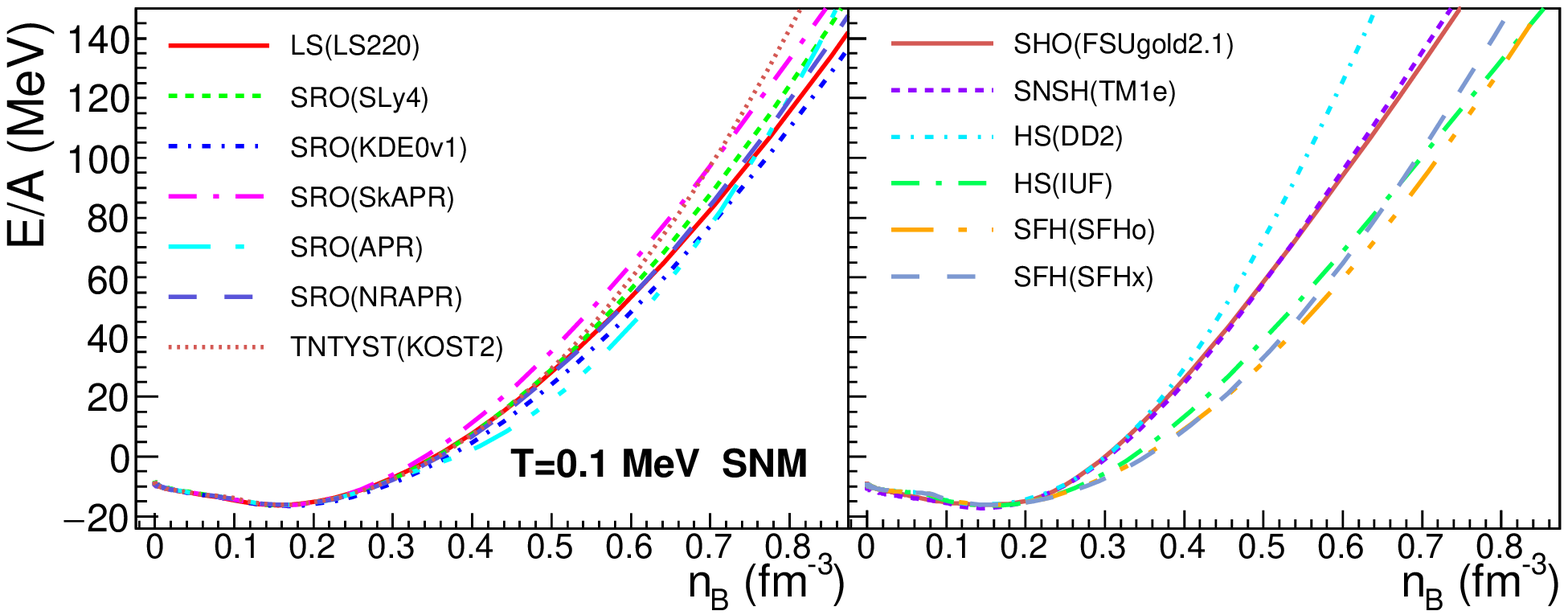}
\includegraphics[width=0.99\columnwidth]{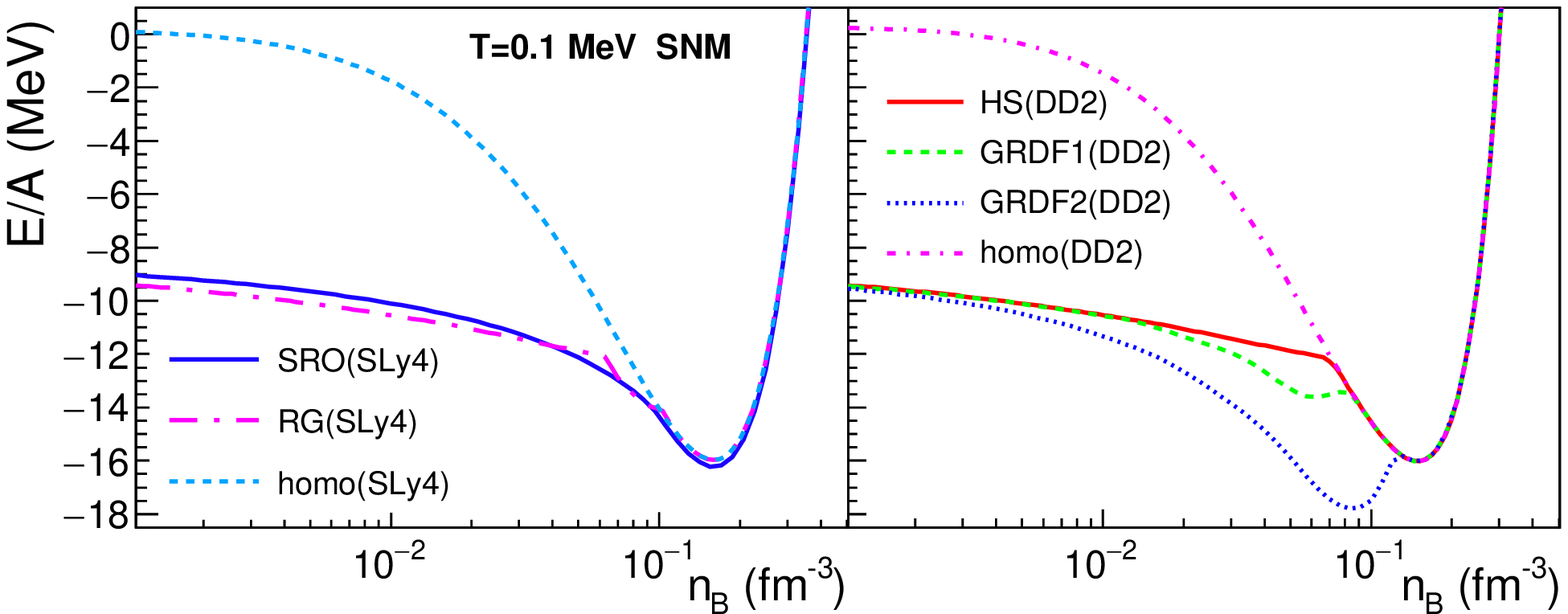}
\caption{Energy per nucleon as a function of baryon number density in
  cold baryonic SM. The bottom panels illustrate the low density behavior for a
  selection of models which employ the same nucleonic interaction but
  rely on different modelling of the inhomogeneous phase.  The various
  EoS models are mentioned in the legend.
}
\label{fig:EperA_SNM_T=0} 
\end{figure}

The energy per nucleon of cold baryonic SM predicted by the models
introduced above is illustrated in the top panels of
fig.~\ref{fig:EperA_SNM_T=0} as function of baryon number density.
The contribution of the baryonic sector is obtained by subtracting
from the values provided in the tables the respective contributions of
leptons and photons as ideal gases.  A residual effect of the electron
gas nevertheless persists in the cluster binding energies which have
been modified in order to account for Coulomb screening.  One first
notes that for $n_B \lesssim 2 n_{\mathit{sat}}$ the predictions of the various
models differ only slightly.  There are two reasons for that.

First, because of the competing effects of Coulomb and surface
energies which get minimized when matter is clusterised or uniform,
respectively, subsaturation matter possesses an inhomogeneous
structure.  This structure basically consists of a dense component
with a low isospin asymmetry (a nuclear cluster) and a dilute
component with high isospin asymmetry (unbound nucleons). All EoS
models discussed in this work account for inhomogeneous structures at
subsaturation densities. For symmetric matter, the main components
contributing to the energetics in this density domain, the nuclear
clusters, are very well constrained by nuclear data.  This explains,
too, that $\lim_{n_B \to 0} E/A \to -8.5$ MeV (see the bottom panels).
This last value corresponds to the maximum binding energy of atomic nuclei,
well reproduced by all models. For comparison, the bottom panels show the
behavior of homogeneous symmetric nuclear matter, too. It is clear that the
energetics are very different, in particular $\lim_{n_B \to 0} E/A\to 0$
for homogeneous matter. 

Although the predictions are rather similar, there remains some
dependence on the treatment of inhomogeneous matter, see the bottom
panels of fig.~\ref{fig:EperA_SNM_T=0}, where different EoS models are
compared which use the same interaction for unbound nucleons, SLy4 on
the left and DD2 on the right.  Discrepancies between SRO(SLy4)
vs. RG(SLy4) at subsaturation densities are attributable to the use of
SNA in SRO(SLy4) and NSE in RG(SLy4).  Discrepancies at $\approx
n_{\mathit{sat}}$, which make SRO(SLy4) deviate from the values in
Table \ref{tab:effint}, are probably artifacts of the procedure
used to describe the transition from inhomogeneous to homogeneous
matter \cite{SRO_PRC_2017}. Discrepancies between the GRDF-models and
HS(DD2) are due to the fact that the former models implement a
phenomenological shift of atomic masses that accounts for the
antisymmetrization correlations between the free nucleons and the
cluster single particle states and include nucleonic resonances. In
addition, the transition from inhomogeneous to homogeneous matter is
not yet implemented in a fully consistent way in the GRDF models which
might explain the displaced minimum of $E/A$. An improved version is
under construction and should soon be available~\cite{Stefanprivate}.
Discrepancies between GRDF1(DD2) and GRF2(DD2) are, in turn, due to
different parameterisations of the mass shifts.

Second, for higher densities, when nuclear clusters dissolve and
matter becomes homogeneous, the energy per baryon of symmetric matter
is still well constrained in the vicinity of saturation density by the
low order nuclear matter parameters. Indeed, as can be seen from
Table~\ref{tab:effint}, the employed interactions show only a small
spread for $n_{\mathit{sat}}, E_{\mathit{sat}},$ and
$K_{\mathit{sat}}$ and the differences in the poorly constrained
higher order parameters only show up above roughly
$2n_{\mathit{sat}}$. Discrepancies between SRO(APR) and TNTYST(KOST2), both
employing the same two- and three-body potentials, are due to a
slightly different many-body treatment resulting among others in the
pion condensed state typical to APR which is absent in KOST2, see
Sec.~\ref{ssec:effint}.  For $n_B \gtrsim 0.32$~fm$^{-3}$ KOST2 is
stiffer than APR; it actually corresponds to the extension of the low
density phase of APR beyond the density at which the transition to the
pion condensate takes place.  Incidentally, the two microscopic EoS
models and those based on Skyrme-like interactions (top left panel),
explore a more narrow range in $E/A$ than the EoS models based on CDFT
(top right panel).

\subsection{Neutron matter}
\label{ssec:PNM_T=0}

\begin{figure}
\includegraphics[width=0.99\columnwidth]{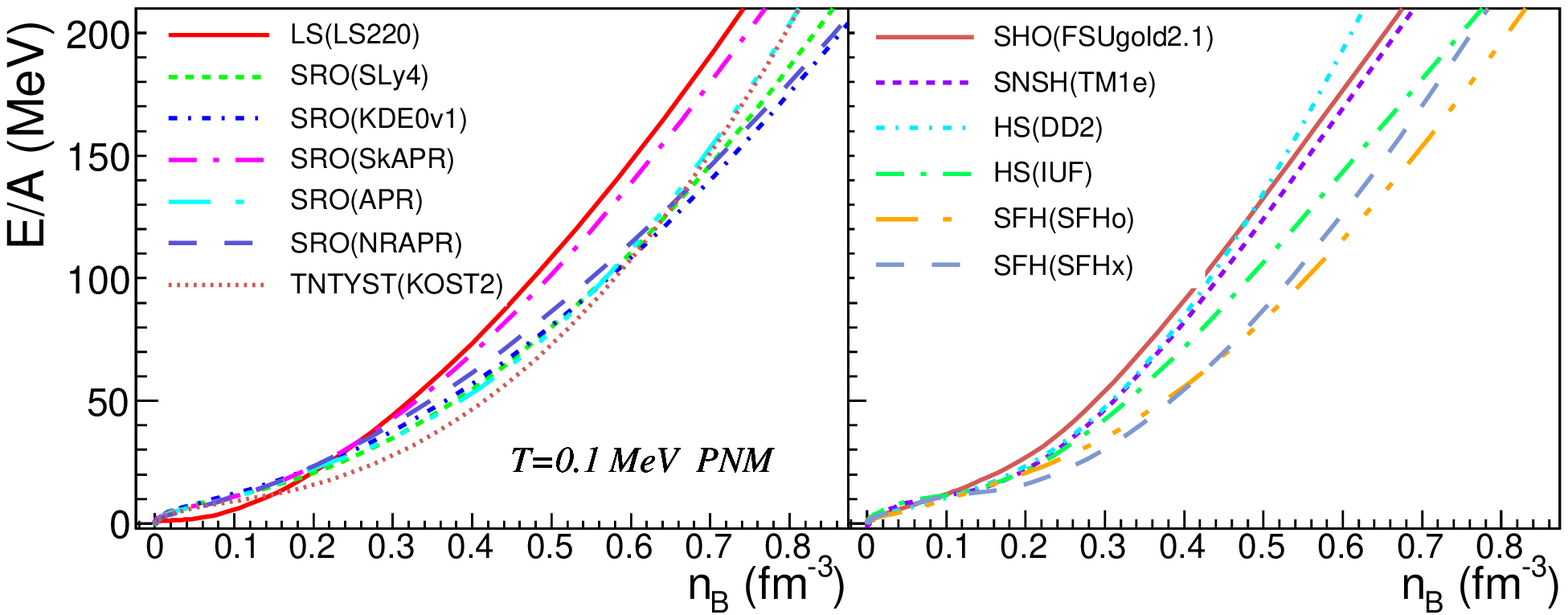}
\includegraphics[width=0.99\columnwidth]{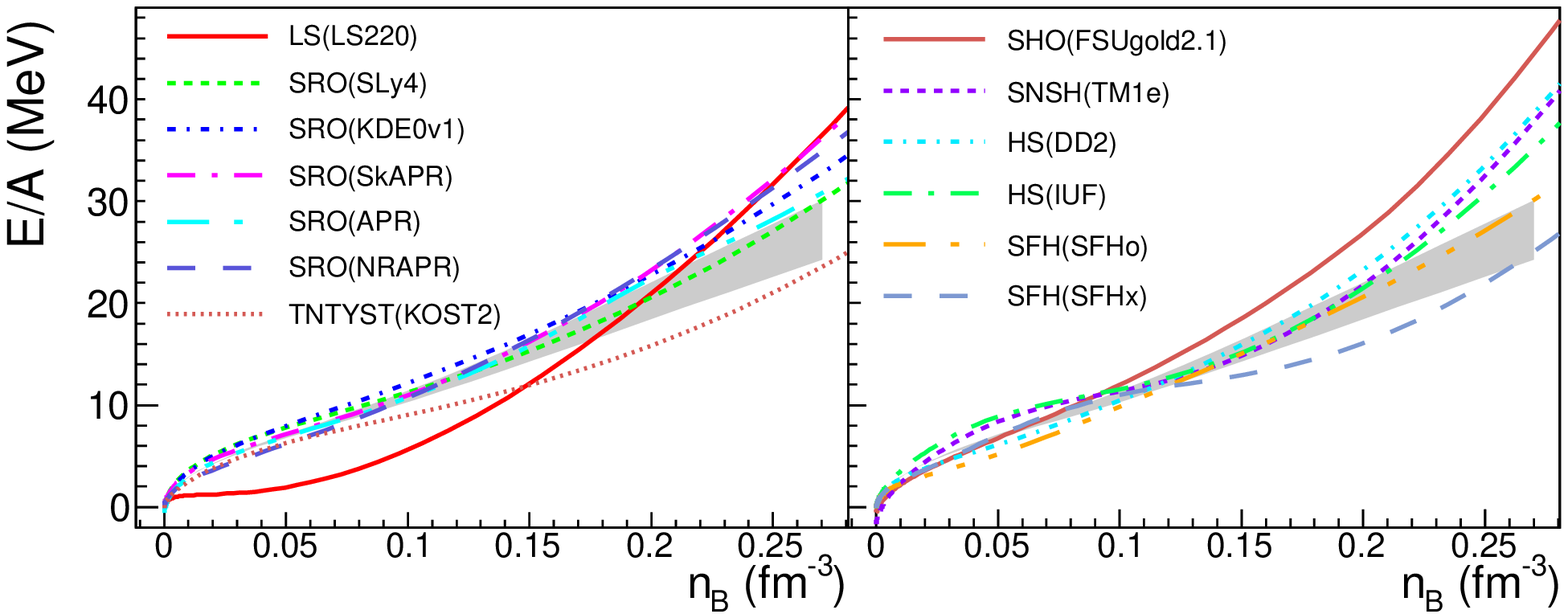}
\caption{Energy per neutron as a function of baryon number
  density in cold NM.  The bottom panels represent a zoom in the low
  density region. The gray area thereby corresponds to predictions of
  many-body perturbation theory~\cite{Drischler_PRC_2016}.  Various
  EoS models are considered as mentioned in the legend.  }
\label{fig:EperA_PNM_T=0} 
\end{figure}

The energy per neutron of cold NM predicted by various models
introduced in Sec.~\ref{sec:Eos} is illustrated in
fig. \ref{fig:EperA_PNM_T=0} as function of baryon number density. The
bottom panels focus on the low density region, where the model results
are compared with predictions of many-body perturbation
theory~\cite{Drischler_PRC_2016}, based on a set of Hamiltonians with
NN and 3N interactions from chiral EFT.  At variance with what happens
in matter with $Y_p\neq 0$, NM at subsaturation densities is
homogeneous. The fact that the effective interactions in the isovector
channel are only weakly constrained by nuclear data is reflected in a
significant dispersion among the predictions of the different models
over the whole density range. As for SM, this dispersion augments with
increasing density. The available models based on Skyrme-like
interactions span an uncertainty range comparable to that spun by the
available models based on CDFT interactions.
For $0.1 \lesssim n_B \lesssim 0.5$~fm$^{-3}$
KOST2 provides lower values than APR, as one would expect based on
lower values of $J_{sym}$ and $L_{sym}$ in the first model with respect
to the latter; for higher densities, the two EoS models merge.
This behavior has been previously commented on in
\cite{Kanzawa_NPA_2007,Constantinou_PRC_2014}.
SLy4 and APR agree best with the microscopic calculations in
\cite{Drischler_PRC_2016}.

\subsection{Symmetry energy}
\label{ssec:Esym_T=0}
\begin{figure}
\includegraphics[width=0.99\columnwidth]{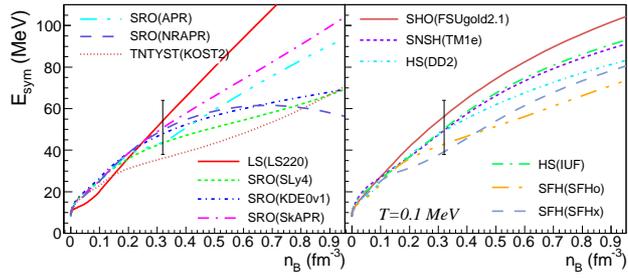}
\caption{Symmetry energy per nucleon, defined as
  $E_{sym}(n_B)=E_{NM}(n_B)-E_{SM}(n_B)$, as function of baryon
  number density at T=0.1 MeV.  The vertical error bar corresponds to
  the constraint $E_{sym}(2n_{\mathit{sat}})=51 \pm 13$ MeV
  \cite{BaoAnLi_Universe_2021}.  }
\label{fig:EsymperA_T=0} 
\end{figure}

The uncertainties in the isovector channel of the effective
interactions show up in the symmetry energy, eq.~(\ref{eq:Esym_diff}),
too.  It is illustrated in fig.~\ref{fig:EsymperA_T=0} as function of
baryon number density for the models introduced in Sec. \ref{sec:Eos}.
The constraint at twice saturation density of nuclear matter,
$E_{sym}(2n_{\mathit{sat}})=51 \pm 13$ MeV, obtained in
Ref. \cite{BaoAnLi_Universe_2021} from a compilation of recent
analyses of NS and heavy ion collision data is shown, too.  We note
that, in contrast to what happens for uncharged uniform matter where
$\lim_{n_B \to 0} E_{sym} \to 0$, clusterisation at subsaturation
densities for symmetric matter translates into non-vanishing symmetry
energies in the limit $n_B\to 0$ \cite{DD2,Raduta_EPJA_2014}.  These
dishomogenities at subsaturation densities break the approximately
quadratic dependence of the energy per baryon on $\delta$ in
eq.~(\ref{eq:E0}), too, such that eqs. (\ref{eq:Esym_curv}) and
(\ref{eq:Esym_diff}) for determining the symmetry energy are no longer
equivalent~\cite{Raduta_EPJA_2014,Typel_EPJA_2014}.
We also note that, similarly to what we have seen for SM, the
occurrence at subsaturation densities of a dense phase with a density
close to saturation, results in a reduced sensitivity to the
underlying nucleon effective interaction of the symmetry energy
\cite{Raduta_EPJA_2014}.

Except SRO(NRAPR), which provides a non-monotonic $E_{sym}(n_B)$, the
qualitative behavior of the other EoS models considered here is the
same at suprasaturation densities, \textit{i.e.} $E_{sym}$ increases
with $n_B$.  Quantitative differences nevertheless exist: LS220
provides the stiffest evolution, while KDE0v1, KOST2, SLy4 and SFHo
provide much softer ones.
For $n_B \gtrsim 0.1$~fm$^{-1}$ KOST2 and APR differ also in what regards
the symmetry energy. At densities smaller than 0.2~fm$^{-3}$, which corresponds to
the density where PNM computed by APR~\cite{APR_PRC_1998} exhibits the
transition to pion condensed state, this is due to the differences
between the predictions of the two variational calculations.  At
higher densities the discrepancies are more important and they are
mainly due to the pion-condensed state in APR.  All models are
compatible with the constraint on $E_{sym}(2 n_{\mathit{sat}})$ derived in
Ref.~\cite{BaoAnLi_Universe_2021}. KOST2 thereby predicts values
slightly below the given range, but the deviation remains well
compatible within errors.

The influence of the density dependence of the symmetry energy,
expressed in terms of $L_{sym}$ and $K_{sym}$, has been intensively
discussed in the context of cold NSs. The most noteworthy correlations
have been found between radii and tidal deformabilities of
intermediate-mass NS
\cite{Lattimer_ApJ_2001,Steiner_PhysRep_2005,Fortin_PRC_2016,Margueron_PRC_2018b,Schneider_PRC_2019,Hu_PTEP_2020,Malik_PRC_2020,Li_PRC_2021}
and $L_{sym}$. Since the proton fraction $Y_p$ in $\beta$-equilibrated
matter strongly depends on the symmetry energy, the density threshold
for nucleonic dUrca process \cite{Fortin_PRC_2016}, too, shows a
strong correlation with $L_{sym}$.  By controlling the radial density
profiles of different species, $L_{sym}$ and $K_{sym}$ have been also
shown to determine the magnitude and density dependence of the pairing
gaps of fermions and, thus, modify the neutrino emission from compact
star interiors \cite{Fortin_PRD_2021}.  This observation equally
applies to purely nucleonic stars as hyperonic stars.

\subsection{Effective masses in isospin asymmetric matter}
\label{ssec:meff_T=0}

\begin{figure}
\includegraphics[width=0.99\columnwidth]{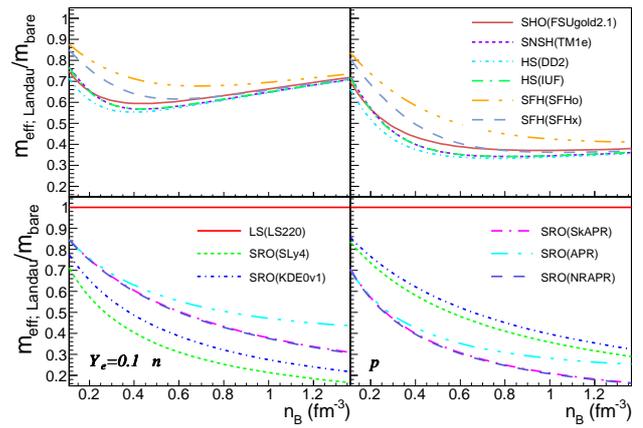}
\caption{Normalized neutron (left) and proton (right) Landau effective masses
  in cold asymmetric matter with $Y_e=0.1$ as a function of baryon number density.
  Only considered are densities above the transition from inhomogeneous
  to homogeneous matter.
  Predictions of different models as mentioned in the legend.
}
\label{fig:meff_T=0} 
\end{figure}

\begin{figure}
  \begin{center}
    \includegraphics[width=0.79\columnwidth]{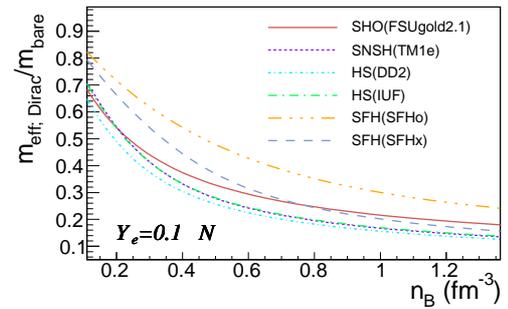}
    \end{center}
\caption{Same as fig.~\ref{fig:meff_T=0} for the Dirac effective
  masses within the different CDFT models.}
\label{fig:meffD_T=0} 
\end{figure}

\begin{figure}
  \begin{center}
    \includegraphics[width=0.99\columnwidth]{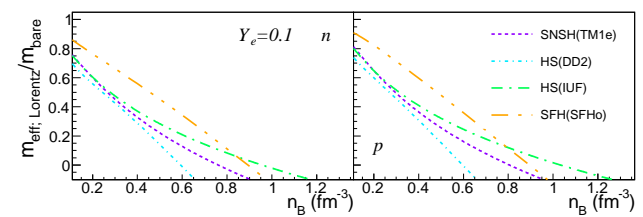}
    \end{center}
\caption{Same as fig.~\ref{fig:meff_T=0} for the Lorentz effective
  masses within a selection of CDFT models.}
\label{fig:meffLo_T=0} 
\end{figure}

An important ingredient of the nuclear EoS are the nucleonic effective
masses. The concept of effective mass was introduced in order to
describe the motion of particles in a momentum dependent potential
similarly with the motion of quasi-particles in a momentum-independent
potential. Later on it has been generalized to account for
momentum and/or energy dependence of the single particle potential
\cite{Li_PhysRep_2018}. In the most general case effective masses
reflect the space and time non-locality of the interaction and the
Pauli exclusion principle.

There are three types of effective masses considered usually for the models
under consideration in this paper: the Dirac mass, the Landau mass
and the Lorentz mass.  The Dirac mass is a genuine relativistic
quantity. It is defined through the scalar self-energy $\Sigma_i^S$
in the Dirac equation,
\begin{equation}
  m_{\mathit{eff;}~{\rm Dirac}; ~i}=m_i+\Sigma_i^S~,
  \label{eq:massD}
\end{equation}
where $i=n,p$ denotes neutrons and protons and $m_i$ stands for the bare mass
of the particle $i$.

The Landau mass is defined in terms of the single particle density
of states
$d\epsilon_i/dp_i$ at the Fermi surface at zero temperature,
\begin{equation}
  m_{\mathit{eff;}~\rm{Landau}; ~i}=p_i \left[
    \left( \frac{d \epsilon_i}{d p_i} \right)_{p = p_F;i}
    \right]^{-1}~,
  \label{eq:massL}
\end{equation}
and characterizes the momentum dependence of the single-particle potential.
In eq. (\ref{eq:massL}) $p_i$ and $\epsilon_i$ denote particle's
momentum and single particle energy and $p_{F;i}$ stands for the Fermi momentum.
In Skyrme-like non-relativistic models as well as in APR, the Landau effective
mass allows one to cast the single particle energies,
\begin{equation}
  \epsilon_i=\frac{p^2}{2m_i}+U_i\left(n_n, n_p, \tau_n, \tau_p \right)~,
\end{equation}
where $\tau_i$ and
$U_i$ are the kinetic energy densities and single-particle
momentum-dependent potentials in
\begin{equation}
  \epsilon_i=\frac{p^2}{2m_{\mathit{eff;}~\rm{Landau}; ~i}}+V_i\left(n_n, n_p\right)~,
\end{equation}
with the potentials $V_i$ depending only on particle densities.

In CDFT models at zero temperature the Landau effective mass
  relates to the Dirac mass via
\begin{equation}
   m_{\mathit{eff;}~\rm{Landau}; ~i}=\sqrt{p_{F;i}^2+m_{\mathit{eff;}~\rm{Dirac}; ~i}^2}~.
\end{equation}

The Lorentz mass characterizes the energy dependence of the
Schr\"odinger-equivalent
potential and at $T=0$ is defined as \cite{Jaminon_PRC_1989}
\begin{equation}
  m_{\mathit{eff;}~\rm{Lorentz}; ~i}=m_{\mathit{eff;}~\rm{Landau}; ~i}+m_i-\epsilon_i\left( p_{F;i}\right)~, 
  \label{eq:massLo}
\end{equation}
If the nucleon self-energies are independent of momentum, as it is the
case of the models considered hereby, the single-particle energies read
\begin{equation}
  \epsilon_i= \Sigma_i^V +\sqrt{p^2+m_{\mathit{eff;}~\rm{Dirac}; ~i}^2}~,
  \label{eq:spe}
\end{equation}  
where $\Sigma_i^V$ stands for the nucleon vector self-energy.
As such eq. (\ref{eq:massLo}) becomes
\begin{equation}
  m_{\mathit{eff;}~\rm{Lorentz}; ~i}= m_i -\Sigma_i^V~.
\end{equation}
According to \cite{Jaminon_PRC_1989}
it is the Lorentz mass that should be compared with the nonrelativistic
effective mass.

The thermal effects we are discussing here are determined by the mass
entering the kinetic energy term, that is the Dirac mass in CDFT
models and the Landau mass in non-relativistic models.

Naturally, the effective masses are correlated with the nuclear matter
parameters characterizing the EoS around saturation density, the
symmetry energy playing a dominant role in the effective mass
difference of protons and neutrons. For a recent review on
the interplay between nucleon effective mass and the symmetry energy
as well as experimental constraints, compliance with \textit{ab initio}
calculations and the role on thermal and transport properties of
neutron rich matter, see \cite{Li_PhysRep_2018}.

Fig. \ref{fig:meff_T=0} illustrates the density dependence of neutron
(left) and proton (right panels) Landau effective masses
(eq.~(\ref{eq:massL})) as function of density at $T=0.1$ MeV and
$Y_e=0.1$ for the different interactions\footnote{The TNTYST table on
  \textsc{CompOSE} does not provide that data, see fig. 9 in
  \cite{Kanzawa_NPA_2007} for that model.}.  With the exception of
LS220, which considers a momentum-independent interaction, all models
predict a complex density dependence of the effective masses.
Models with Skyrme-like effective interactions and the microscopic APR
model provide for both neutrons and protons effective masses that
decrease with density. Note that, despite the fact that NRAPR
\cite{Steiner_PhysRep_2005} was designed to reproduce the effective
masses in APR along with some nuclear data, notable differences exist
between the predictions of these models in neutron-rich matter.  The
reason is that the complex density dependence of the effective masses
in APR cannot be reproduced with the functional form of Skyrme-like
interactions~\cite{Schneider_PRC_2019}. In this context let us mention
that the density dependence of effective masses in KOST/TNTYST
qualitatively resembles those of Skyrme models, see fig. 9 in
Ref. \cite{Kanzawa_NPA_2007}.  As it was the case with energies,
discrepancies among APR and KOST/TNTYST arise from differences in the
variational procedures.  In contrast, the predictions of NRAPR are
nearly identical to those of SkAPR, which might serve to highlight the
effects of other EoS ingredients, without interference from
$m_{\mathit{eff};~\rm{Landau}}$ comparing these two EoS.  For the CDFT models the
Landau effective masses first decrease and then starting from roughly $2
n_{\mathit{sat}}$ increase again. This feature is a relativistic
effect: at high densities momentum dominates over mass.  SLy4 (SFHo)
provides the strongest (most limited) density dependence of
$m_{\mathit{eff};~\rm{Landau};i}(n_B)$ as well as the lowest (highest) values at
high densities. 

With the exception of SRO(SLy4) and SRO(KDE0v1) that lead to
$m_{\mathit{eff;}~\rm{Landau};p} > m_{\mathit{eff;};~\rm{Landau};n}$, all models predict
$m_{\mathit{eff;};~\rm{Landau};n} > m_{\mathit{eff};~\rm{Landau};p}$ in neutron-rich matter.
We recall that the magnitude of the neutron-proton mass splitting is
proportional to the isospin asymmetry and that its sign depends upon
isoscalar and isovector contributions to the effective masses, with
the latter loosely constrained by experimental data on isovector giant
dipole resonances~\cite{Li_PhysRep_2018}.  Microscopic calculations
that include three body forces converge in predicting
$m_{\mathit{eff};~\rm{Landau};n} > m_{\mathit{eff};~\rm{Landau};p}$ in neutron-rich matter
\cite{Sammarruca_PRC_2005,Dalen_PRC_2005,Baldo_PRC_2014,Catania_PRC_2020}.

The Dirac effective masses, eq. (\ref{eq:massD}), typical to CDFT models
are shown in fig.~\ref{fig:meffD_T=0} for zero temperature and $Y_e=0.1$.
They monotonically decrease with density. Please note that there is no difference
in proton and neutron effective masses here, since the considered CDFT models have
no isovector contribution ($\delta$-channel) to the effective masses.

The density dependence of the Lorentz mass, eq. (\ref{eq:massLo}),
is depicted for a selection of CDFT models in Fig. \ref{fig:meffLo_T=0}.
As before, $T=0$ and $Y_e=0.1$. It comes out that the density-dependence of
this quantity is stronger than those of Landau and Dirac effective masses.
Moreover, for $n_B \gtrsim 4-7 n_{sat}$ it vanishes. These results agree
with those in \cite{Chen_PRC_2007}, where a much larger collection of
CDFT models has been considered.

Finally we note that the density-dependence of nucleon effective
masses may be constrained by the results of {\em ab initio}
calculations. For instance, such calculations, based on a set of
commonly used Hamiltonians including two- and three-nucleon forces
derived from chiral effective-field theory, are available in
\cite{Somasundaram_PRC_2021}. They show that the density-decrease of
the nucleon (neutron) Landau effective mass in SM (NM) saturates at
$n_B\approx 0.2~{\rm fm}^{-3}$ ($n_B \approx 0.10-0.17~{\rm
  fm}^{-3}$), where $m_{\mathit{eff;}~{\rm Landau}}/ m_{bare}
\approx 0.65$ (0.85).  It is clear that none of the models
considered here agrees with this behavior.  

The density dependence of nucleon effective masses
at temperatures of the order $T \ll 1$ MeV has been shown
to significantly impact the
thermal evolution of isolated and accreting NS through the reduction
by a factor of up to two or three of the superfluid pairing gaps in
the $^1S_0$ and $^3PF_2$ channels \cite{Chen_NPA_1993}; the
emissivities of different neutrino emitting
processes~\cite{Yakovlev_PhysRep_2001,Baldo_PRC_2014}; magnitude of
thermal conductivity~\cite{Baiko_AA_2001} and specific
heats~\cite{Yakovlev_2004}.

The role of nucleonic effective masses in finite temperature
environments has been discussed in connection with the dynamics of
CCSN, contraction of the PNS and the neutrino and
GW signals \cite{Schneider_PRC_2019b,Yasin_PRL_2020,Andersen2021};
evolution of PNS in failed CCSN and subsequent formation of BH \cite{Schneider_ApJ_2020};
post-merger strain of GW~\cite{Raithel2021}.
All these works employed parametrized EoS
models which allow independent tuning of various NM parameters. In
some cases also phenomenological EoS models widely used in the
community have been employed.  Finite temperature effects have been
implemented consistently~\cite{SRO_PRC_2017} or via the so-called
$M^*$-approximation \cite{Raithel_ApJ_2019}, see Sec. \ref{ssec:Xth}.
Ref. \cite{Schneider_PRC_2019b} shows that large
effective masses result in higher temperatures of the neutrinospheres
as well as lower values of the neutrinosphere density and
radius. These effects are visible for all flavor neutrinos but the
largest effect is obtained for electron neutrinos and
antineutrinos. Modifications of the neutrinosphere obviously impact
the electron neutrino and antineutrino energies at a few hundreds ms
after bounce and the neutrino luminosity: high effective masses lead
to larger neutrino energies and luminosities.  For postbounce times
between roughly $100~{\rm ms}$ and $1~{\rm s}$ the central density,
temperature
and PNS radius manifest strong dependence on $m_{\mathit{eff}}$,
too~\cite{Schneider_PRC_2019b,Yasin_PRL_2020},  larger
$m_{\mathit{eff}}$ values leading to higher values of $n_{B;c}$ and lower
values of $T_c$ and $R_{PNS}$. While
\cite{Schneider_PRC_2019b,Yasin_PRL_2020,Andersen2021} agree that
$m_{\mathit{eff}}$ also impacts the evolution of the shock radius,
they put forward opposite correlations.  Ref. \cite{Andersen2021}
shows that shock revival is stronger and occurs at earlier times after
bounce when $m_{\mathit{eff}}$ is large, in agreement with what one
would expect from an increased neutrino heating. Due on its influence
on turbulence and convection in the gain region, these impact the
gravitational wave emission, too.
Higher effective masses lead thereby to larger
amplitude gravitational waves for postbounce times $ t\gtrsim 0.1
s$~\cite{Andersen2021}.  By performing numerical simulations
\cite{Schneider_ApJ_2020} show that $m_{\mathit{eff}}$ also determines
the onset of collapse into a BH for failed CCSN.  According to these
simulations, the collapse starts immediately after the mass of PNS
exceeds the maximum gravitational mass of hot NS, with the
latter being strongly dependent on $m_{\mathit{eff}}$.  Large values of the
effective masses lead to low values of $\mmax(PNS)$ and, thus, fast
collapse.  The effect of $m_{\mathit{eff}}$ in hot stellar
environments is remarkable as it largely dominates the
effects of all other nuclear matter parameters
\cite{Schneider_PRC_2019b,Schneider_ApJ_2020,Andersen2021}.

\section{Thermal quantities}
\label{sec:T}

In the following section we shall analyze a series of thermal quantities,
including thermal energy and pressure and thermodynamic coefficients.
Thermal effects on thermodynamic state variables will be estimated
by looking at the difference between the finite temperature
and the zero temperature quantity
\begin{equation}
  X_{\mathit{th}}=X(n_B, Y_e, T)-X(n_B, Y_e, 0)~.
  \label{eq:Xth}
\end{equation}

The values reported for baryons alone are obtained by subtracting
from the values provided by the {\tt compose} code, and which correspond
to the mixture baryons+leptons+photons, the corresponding leptonic and
photonic contributions.

\subsection{Energy density and pressure}
\label{ssec:Xth}

\begin{figure}
  \includegraphics[width=0.99\columnwidth]{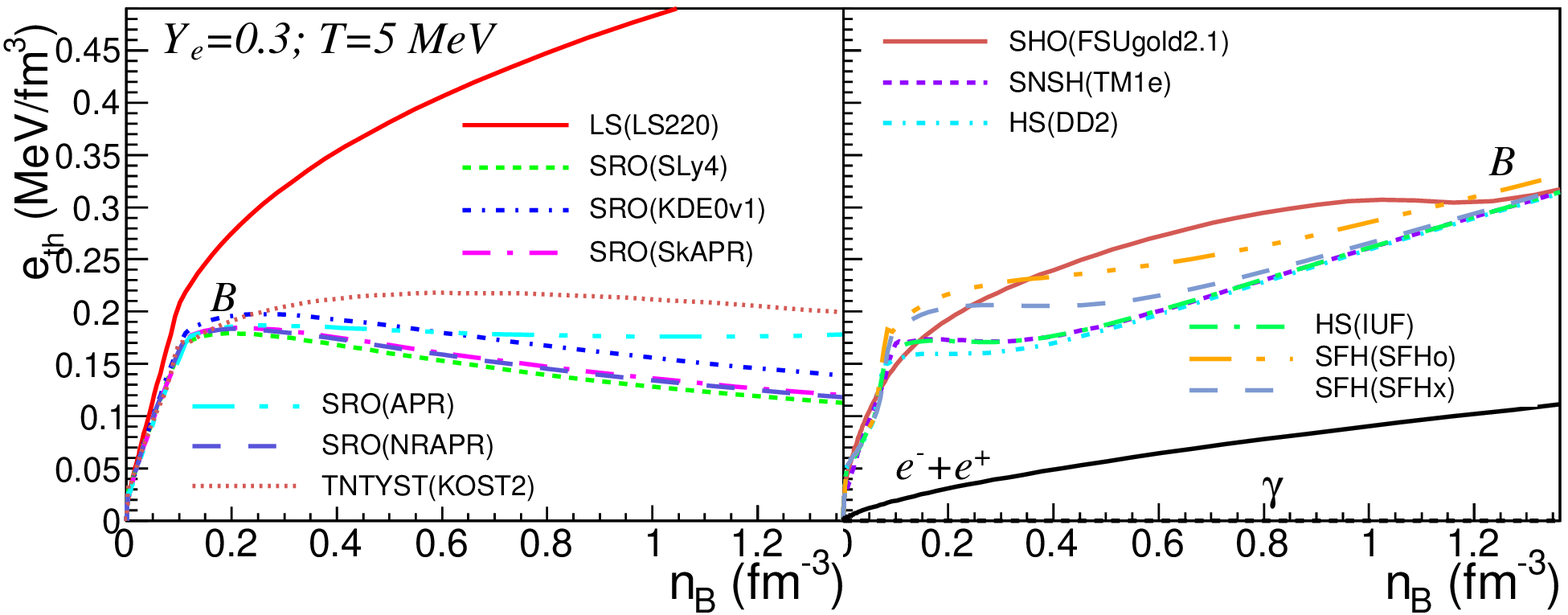}
\includegraphics[width=0.99\columnwidth]{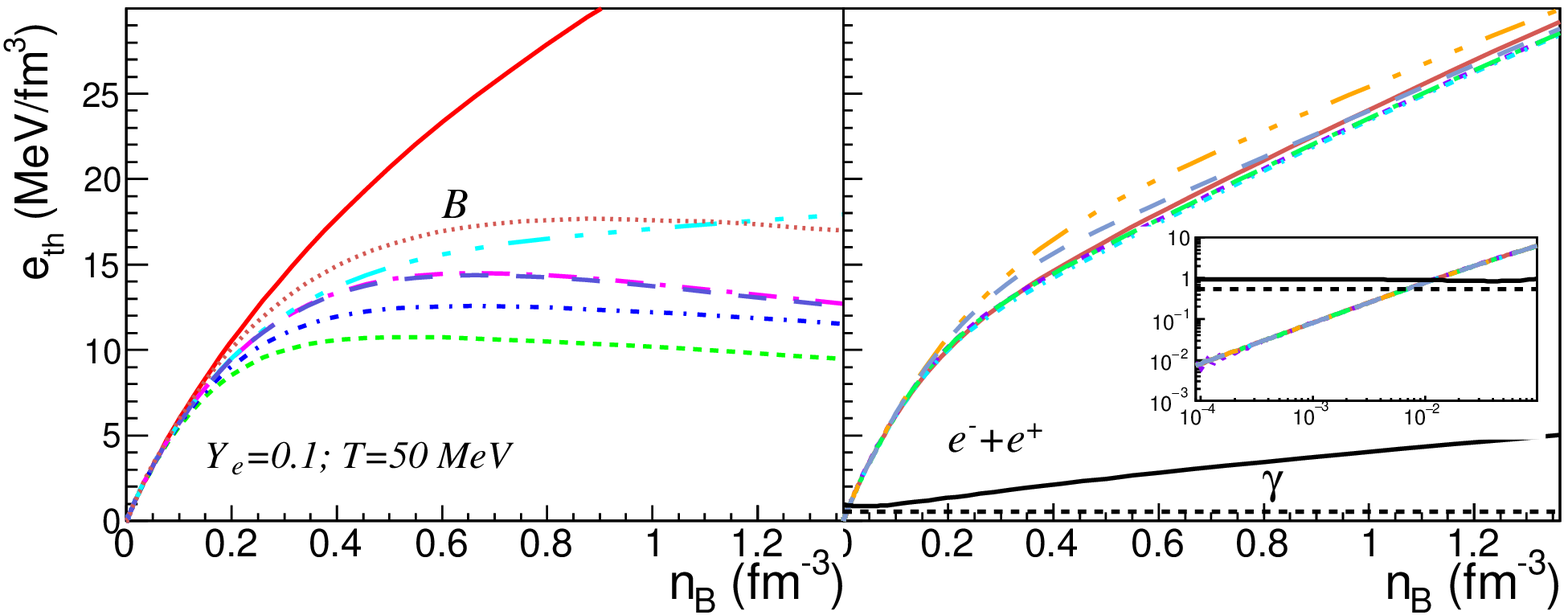}
\caption{Baryonic contribution to the thermal energy density $e^B_{\mathit{th}}$,
  eq.~(\ref{eq:Xth}), as function of baryon number density for
  ($Y_e=0.3$, $T$=5 MeV) (top) and ($Y_e=0.1$, $T$=50 MeV) (bottom).
  The results are depicted for various EoS models.
  In the right panels the contributions of leptons
  (black solid line) and photons (black dashed line) are shown, too.
  The insert in the right bottom panel shows the results at low densities.}
\label{fig:eth} 
\end{figure}

\begin{figure}
  \includegraphics[width=0.99\columnwidth]{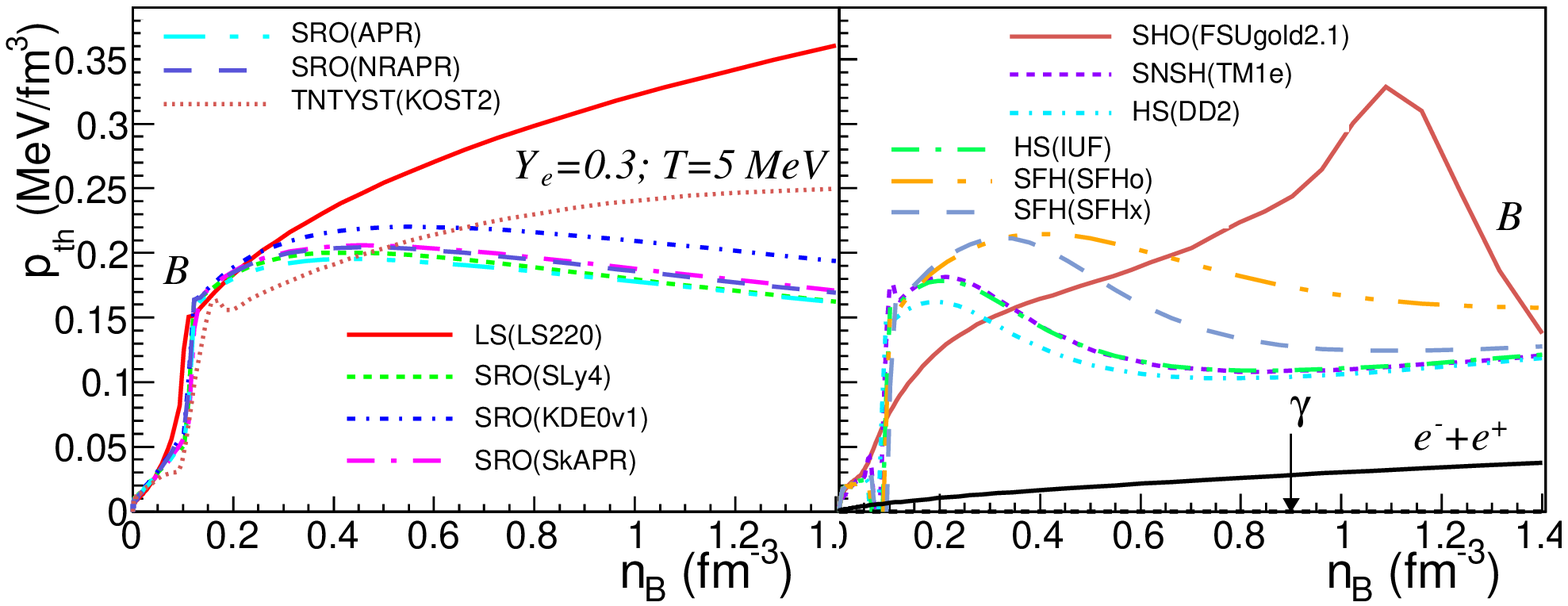}
  \includegraphics[width=0.99\columnwidth]{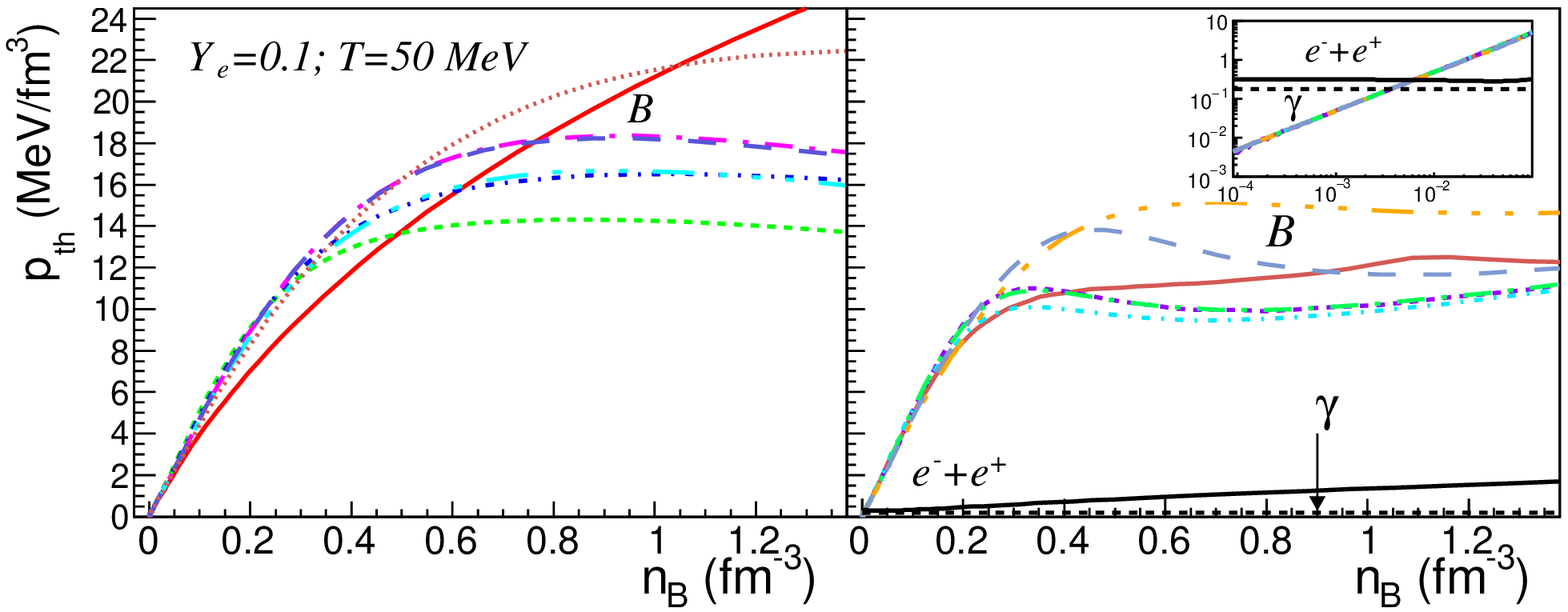}
  \caption{Same as fig. \ref{fig:eth} for the thermal pressure.
  }
\label{fig:pth} 
\end{figure}

\begin{figure}
  \includegraphics[width=0.99\columnwidth]{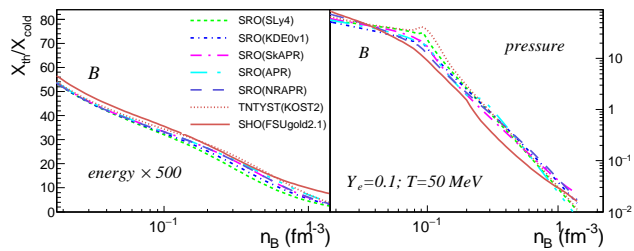}
  \caption{Ratio of thermal ($T = 50$ MeV) to cold baryonic
    energy density (left) and pressure (right) as function of baryon
    number density for a selection of EoS models. An electron fraction
    of $Y_e=0.1$ is considered.  In the left panel $e_{\mathit{th}}/e_{cold}$
    is multiplied by 500.  }
\label{fig:Xratio} 
\end{figure}

\begin{figure}
  \includegraphics[width=0.99\columnwidth]{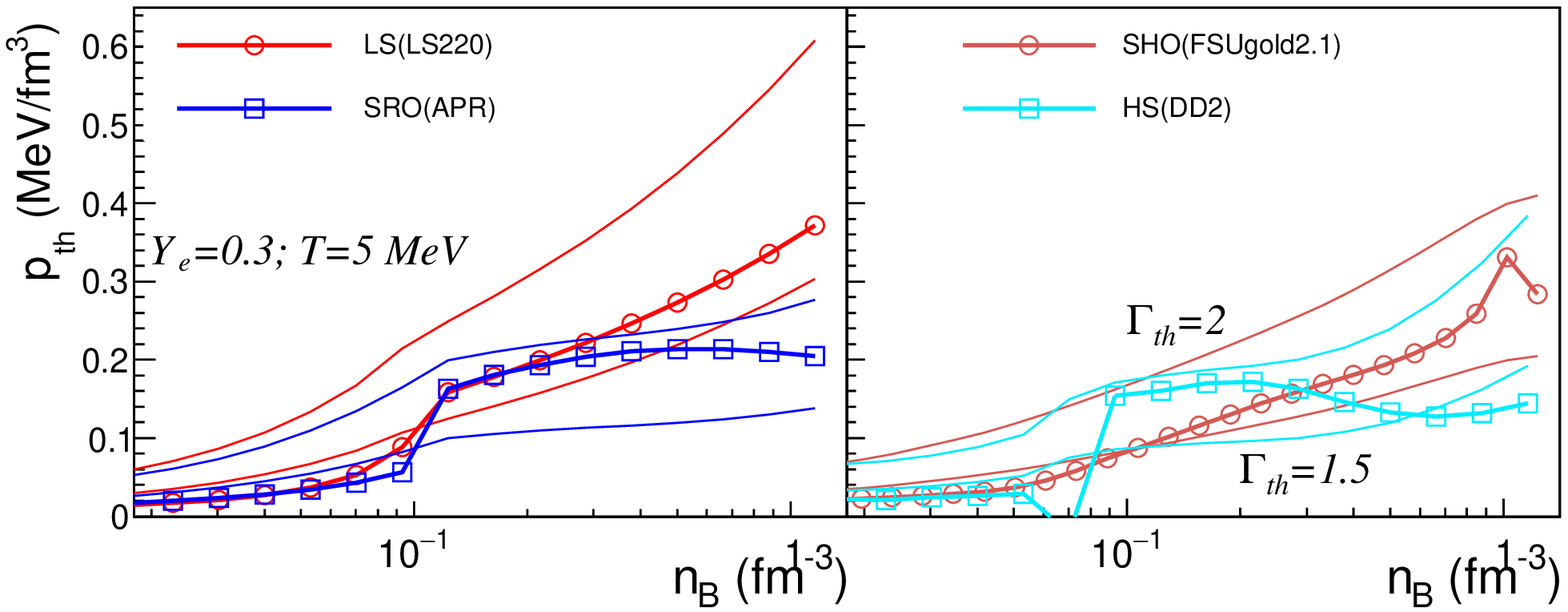}
  \includegraphics[width=0.99\columnwidth]{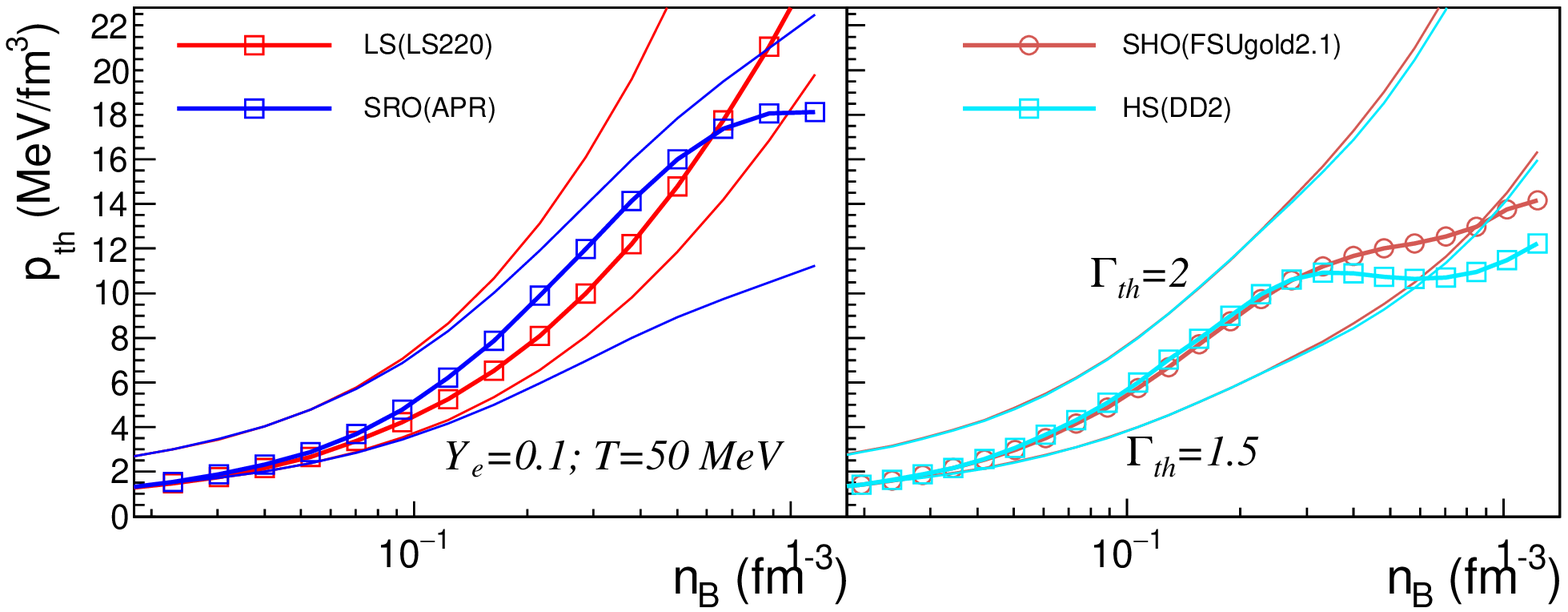}
  \caption{Predictions of eq. (\ref{eq:Xth}) for total thermal pressure
    (thick lines and symbols) are confronted with predictions of
    eq. (\ref{eq:GammaEoS}) with $\Gamma_{\mathit{th}}$=1.5 (lower thin lines)
    and 2 (upper thin lines).
    The considered models are specified in the legend.
  }
\label{fig:pth_Gammalaw} 
\end{figure}

Fig. \ref{fig:eth} and fig. \ref{fig:pth} illustrate the density
dependence of the baryonic thermal energy density and pressure,
respectively. Temperature and charge fraction have been fixed to
($T$=5 MeV, $Y_e=0.3$) (top panels) and ($T$=50 MeV, $Y_e=0.1$)
(bottom panels). For low densities, $n_B \lesssim
0.1~\mathrm{fm}^{-3}$, and high temperatures both
$e_{\mathit{th}}$ and $p_{\mathit{th}}$ behave linearly, have no EoS
model dependence and approach zero for vanishing density.  These
features are characteristic of ideal gases and are obtained
whenever the interactions are weak enough. Though not visible
in the figures, at low temperatures the behavior depends on the
effective interaction as well as on the approach adopted to model
inhomogeneous matter and, for NSE based models, depart from
linearity, too. These obviously means that the ideal gas limit is
violated.
$p_{\mathit{th}}(n_B)$ predicted by HS(DD2), HS(IUF), SFHo, SFHx
and SNSH(TM1e) for ($T=5$ MeV, $Y_e$=0.3) manifest discontinuities at
$n_B \approx n_{\mathit{sat}}/2$.  These are due to the way in which the
transition from inhomogeneous to homogeneous matter is dealt with.
For $0.1 \lesssim n_B \lesssim 0.2-0.5~\mathrm{fm}^{-3}$, $e_{\mathit{th}}$
still increases with density while $p_{\mathit{th}}$ manifests a complex
behavior.  For both quantities the values become
EoS-dependent. Correlations between the magnitude of thermal effects
and values of the nucleonic Landau effective masses are straightforward
to establish for non-relativistic models. We remind that in APR and Skyrme-line
models the Landau effective mass does not depend on
temperature. Indeed, in these models $m_{\mathit{eff};{\rm Landau};i}$
exclusively depend on particle densities.
Fig. \ref{fig:eth} and fig. \ref{fig:pth} show that the largest values
of $e_{\mathit{th}}$ correspond to models with large nucleonic Landau
effective masses, and the largest values for $p_{\mathit{th}}$ are
obtained with small $m_{\mathit{eff};{\rm Landau}}$.
The first correlation is easy to understand:
large values of $m_{\mathit{eff};{\rm Landau}}$ result in low single particle
energies,
easy to populate by thermal excitation.
The latter correlation is less trivial. Though, based on the thermodynamic
relation $p=n^2 \partial \left(e/n \right)/\partial n|_T$, it is
clear that, in addition to $m_{\mathit{eff};{\rm Landau};i}$ the pressure will
depend on $\partial m_{\mathit{eff};{\rm Landau};i}/\partial n_j$, with $i,j=n,p$,
too. This issue was previously discussed in
Refs. \cite{Constantinou_PRC_2014} where analytical expressions have
been worked out for Hamiltonian models.  For higher densities a
diversity of behaviors is obtained.  LS220, for which effective masses
are equal to bare masses, predicts that both $e_{\mathit{th}}$ and $p_{\mathit{th}}$
increase with $n_B$; this result is qualitatively similar to what
happens for free gases.  TNTYST and non-relativistic models, all
showing monotonic density decreases of $m_{\mathit{eff};{\rm Landau};n/p}(n_B)$,
predict thermal effects that have a maximum.  For $T=5$ MeV ($T=50$
MeV) the maxima occur at $n_B \approx 1-5 n_{\mathit{sat}}$
($n_B \approx 4-6n_{\mathit{sat}}$).  For APR, the flattening of
$m_{\mathit{eff};{\rm Landau};n/p}(n_B)$ for
$n_B \gtrsim 1~{\rm fm}^{-3}$ leads to thermal effects intermediate to
those predicted by LS220 and non-relativistic models.  More precisely,
$p_{\mathit{th}}(n_B)$ and $e_{\mathit{th}}(n_B,T=5~\rm{MeV})$ have a maximum while
$e_{\mathit{th}}(n_B, T=50~\rm{MeV})$ increases.  Except
$e_{\mathit{th}}(n_B)$ at $T=50$ MeV, for which the results resemble those of
LS220, CDFT models predict curves that show a complex behavior and a
strong EoS dependence at high densities.  Correlations with nucleonic
Dirac effective masses are nevertheless difficult to extract,
since within CDFT the $m_{\mathit{eff};{\rm Dirac}}$-dependence of
single particle energies is rather weak, see eq. (\ref{eq:spe}).

The significance of thermal effects can be judged considering the
modifications brought to "cold" quantities. Fig. \ref{fig:Xratio}
illustrates the ratios of thermal energy density and pressure relative
to their cold counterparts for $T=50$ MeV and $Y_e=0.1$.  For both
quantities naturally the higher the density the smaller the thermal
effects.  Modifications induced on the pressure are important and
depend on the EoS model, those on energy density are insignificant.
The ratio of thermal to cold pressure can reach about 100 at a tenth
of saturation density, whereas it decreases to a few 1/10 for
densities above 2 $n_{\mathit{sat}}$.  Thus the outer shells of a NS
will be more affected than the inner ones and finite temperature will
modify more NS radii than masses. Similarly thermal effects will be
stronger in low mass NS.

The right panels in Figs. \ref{fig:eth} and \ref{fig:pth} show
contributions of lepton and photon gases, too. As expected, the latter
are dominant only at high temperatures and low densities.

Due to the limited number of available finite temperature EoS models,
many numerical simulations have been performed with phenomenological
extensions of cold EoS, generally using a $\Gamma$-law for the thermal
part~\cite{Hotokezaka_PRD_2013,Bauswein_PRD_2010,Endrizzi_PRD_2018,Camelio2019}.
The total pressure is then written as
\begin{equation}
  P(n_B,e)=P_{cold}(n_B)+ \left(\Gamma_{\mathit{th}} -1 \right) e_{\mathit{th}}(n_B),
  \label{eq:GammaEoS}
\end{equation}
where $\Gamma$ is a constant factor with values taken in the range
$1.5 \leq \Gamma \leq 2$. No dependence on temperature,
density or chemical composition of the $\Gamma$-factor is considered.

Fig. \ref{fig:pth_Gammalaw} compares the total thermal pressure
calculated by eq. (\ref{eq:Xth}) with the estimation provided by the
$\Gamma$-law, eq. (\ref{eq:GammaEoS}).  It is obvious that the
monotonic behavior of eq. (\ref{eq:GammaEoS}) is far from accounting
for the complex density dependence of $p_{\mathit{th}}(n_B)$, due to the
complex density dependence of the Landau (for non-relativistic models)
and Dirac (for CDFT models) effective masses.
Moreover at
extreme density values the predictions of eq. (\ref{eq:Xth}) are
situated out of the domain limited by eq. (\ref{eq:GammaEoS}) with
$\Gamma_{\mathit{th}}$=1.5 and 2.  The inability of the $\Gamma$-law to
correctly account for the thermal pressure was recently noticed in
Ref.~\cite{Raithel_ApJ_2019}, who proposed an alternative
solution. The so-called $M^*$-approximation allows to calculate the
pressure of stellar matter at arbitrary values of proton fraction and
temperature using analytic expressions whose parameters account for
the symmetry energy and the density dependence of the effective mass.
The functional dependence proposed in \cite{Raithel_ApJ_2019} based on
CDFT calculations is able to account within a few percent for the
thermal pressure of the respective CDFT EoS.  Nevertheless it performs
modestly for non-relativistic EoS models.
This suggests that different density dependence of the
effective mass have to be employed in the different formalisms
to reproduce the same finite temperature phenomenology, 
because the quantities are not the same and are not expected to
behave in the same way as a function of density.

\begin{figure*}
  \includegraphics[width=0.49\linewidth]{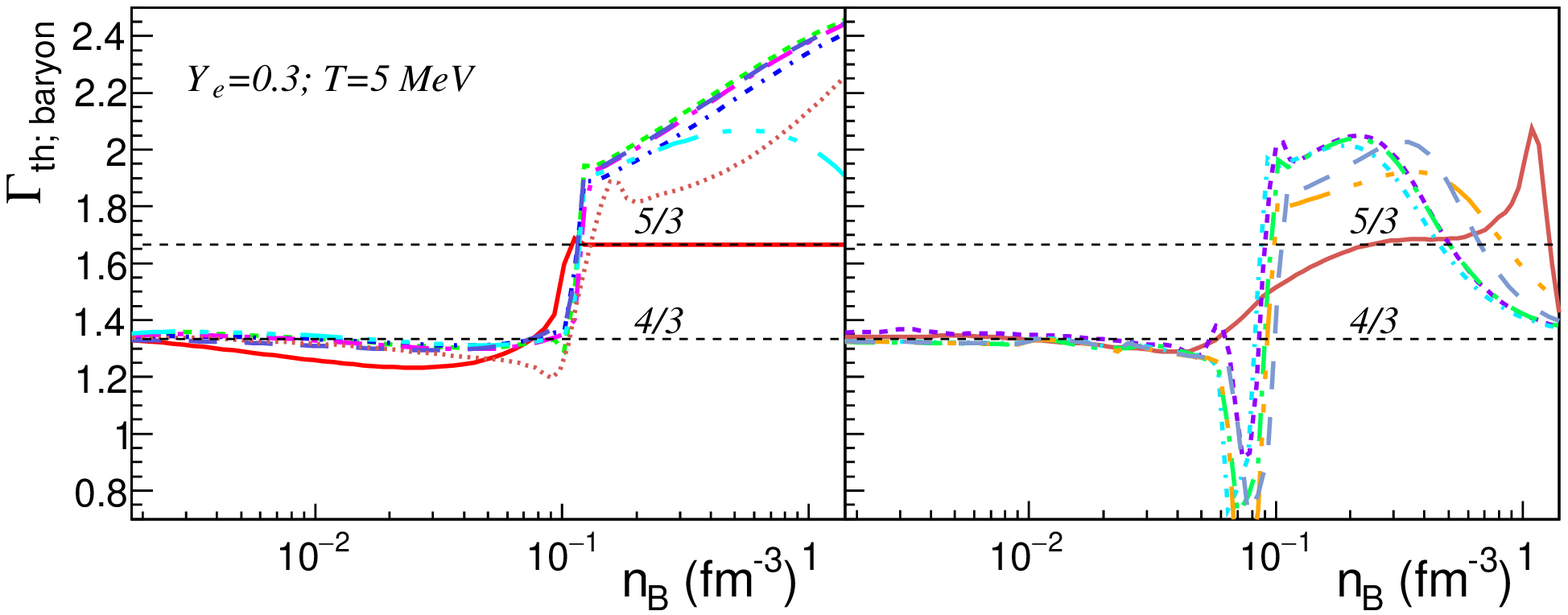}
  \includegraphics[width=0.49\linewidth]{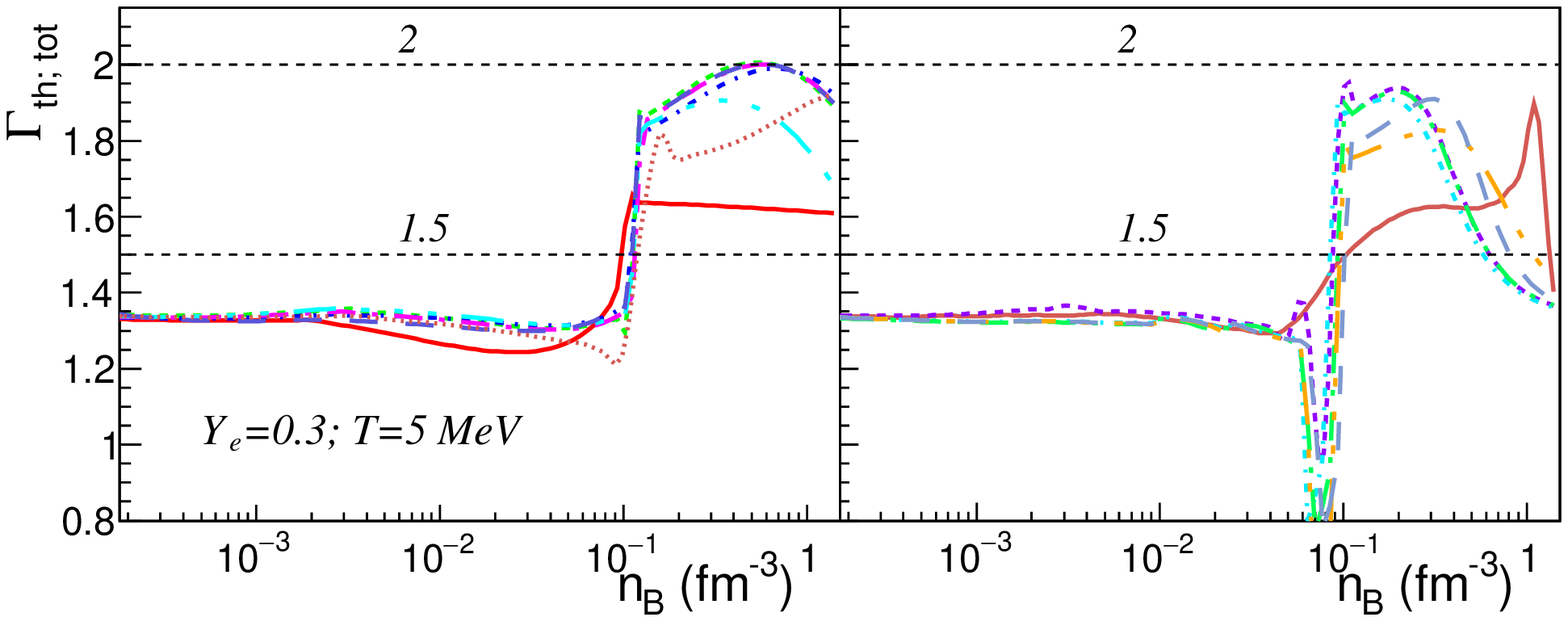}
  \includegraphics[width=0.49\linewidth]{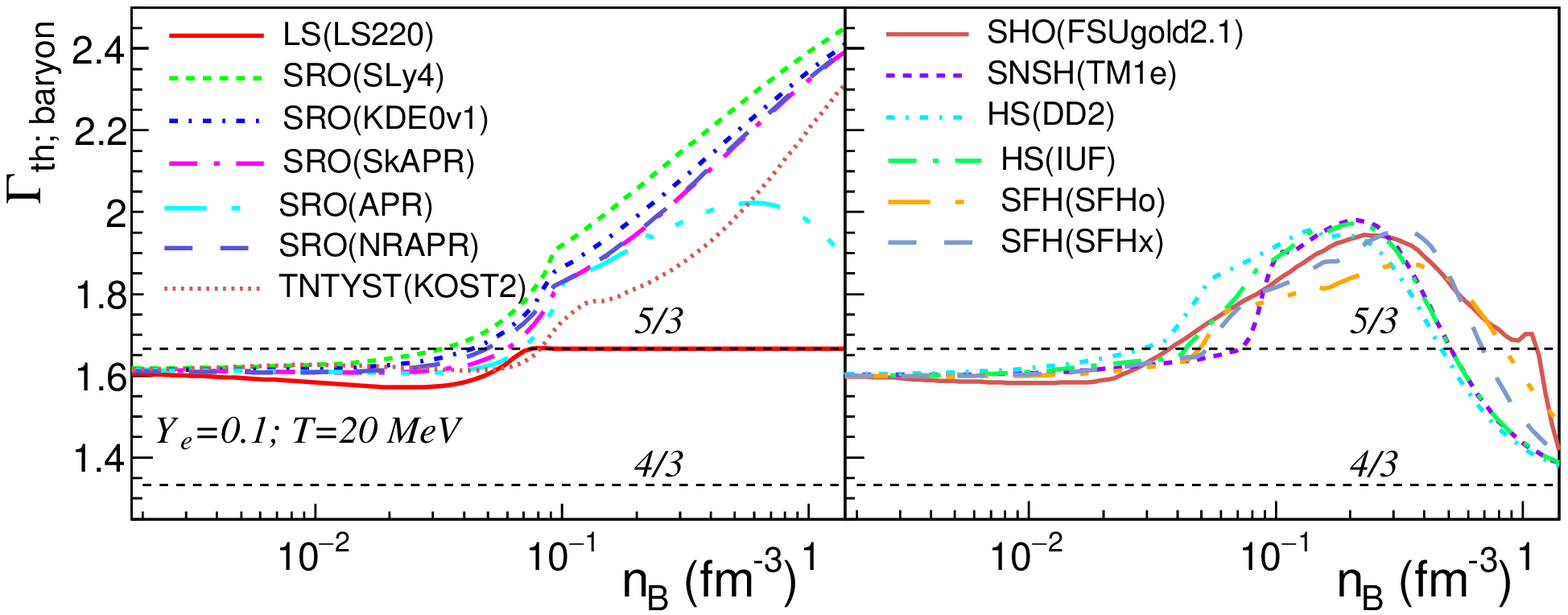}
  \includegraphics[width=0.49\linewidth]{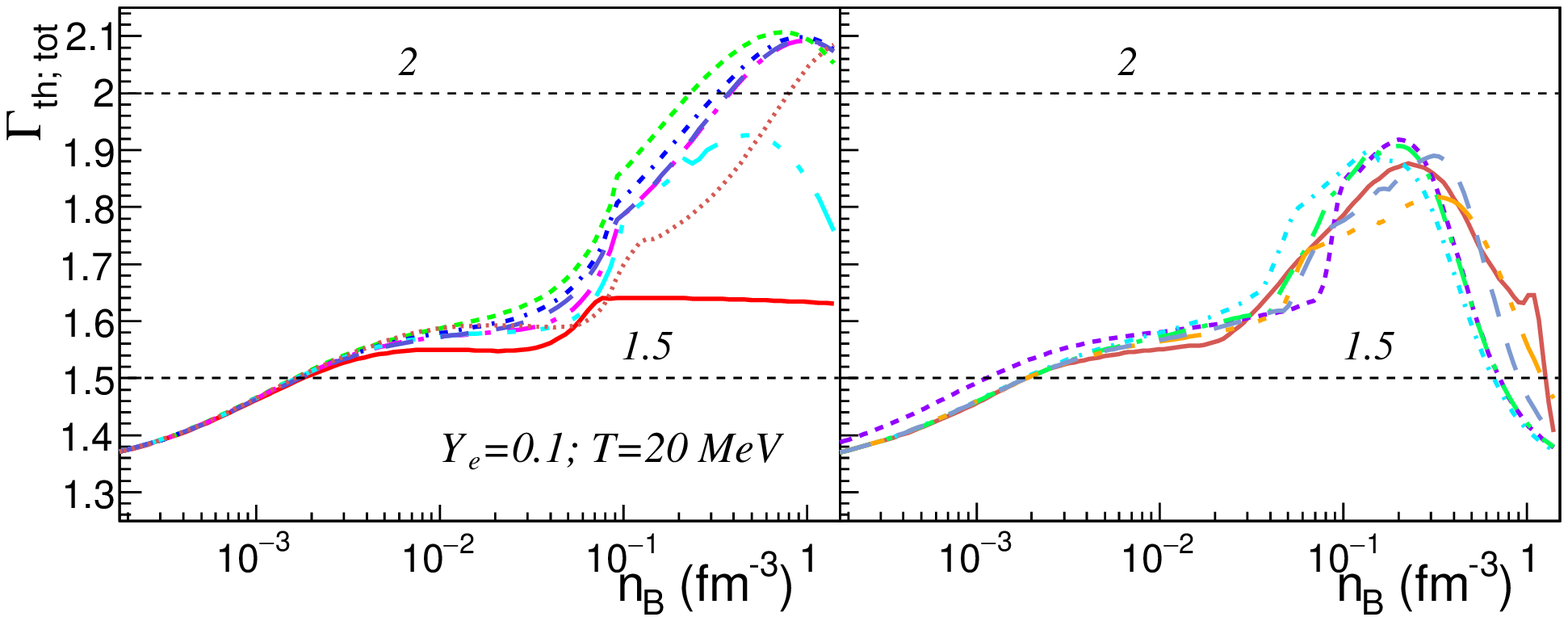}
  \includegraphics[width=0.49\linewidth]{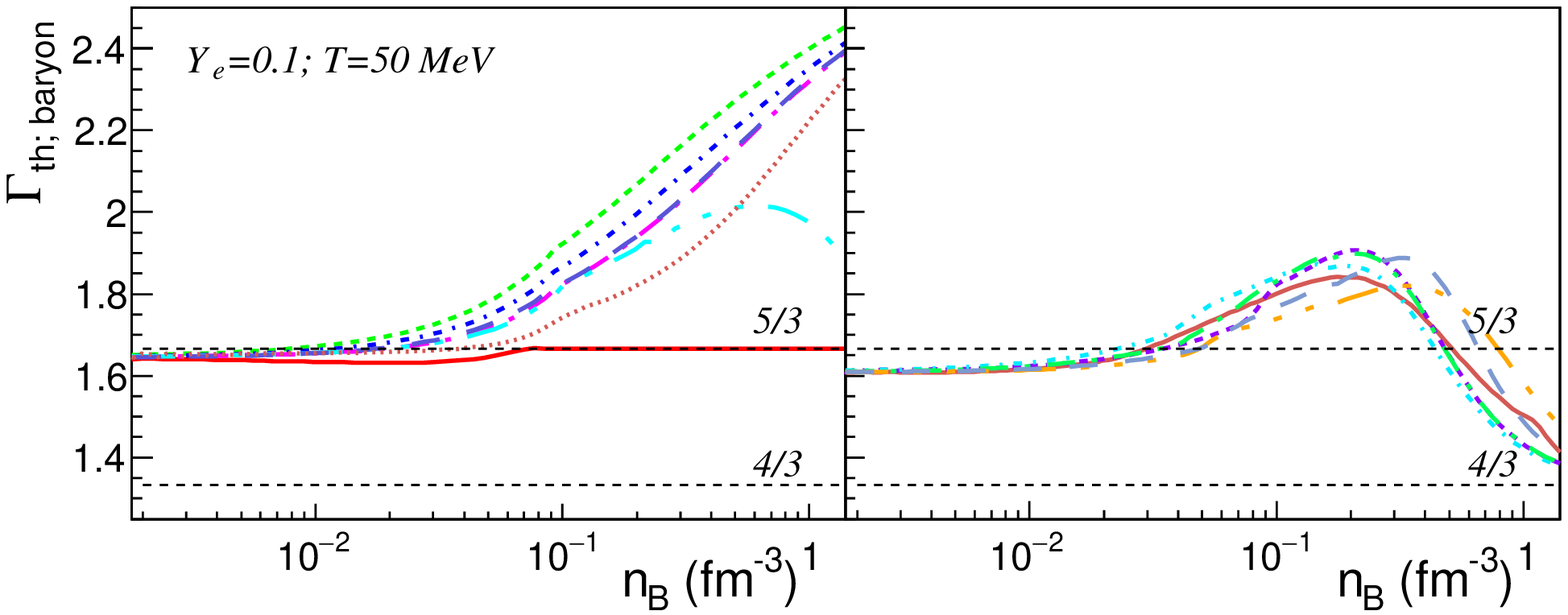}
  \hspace*{0.2cm}\includegraphics[width=0.49\linewidth]{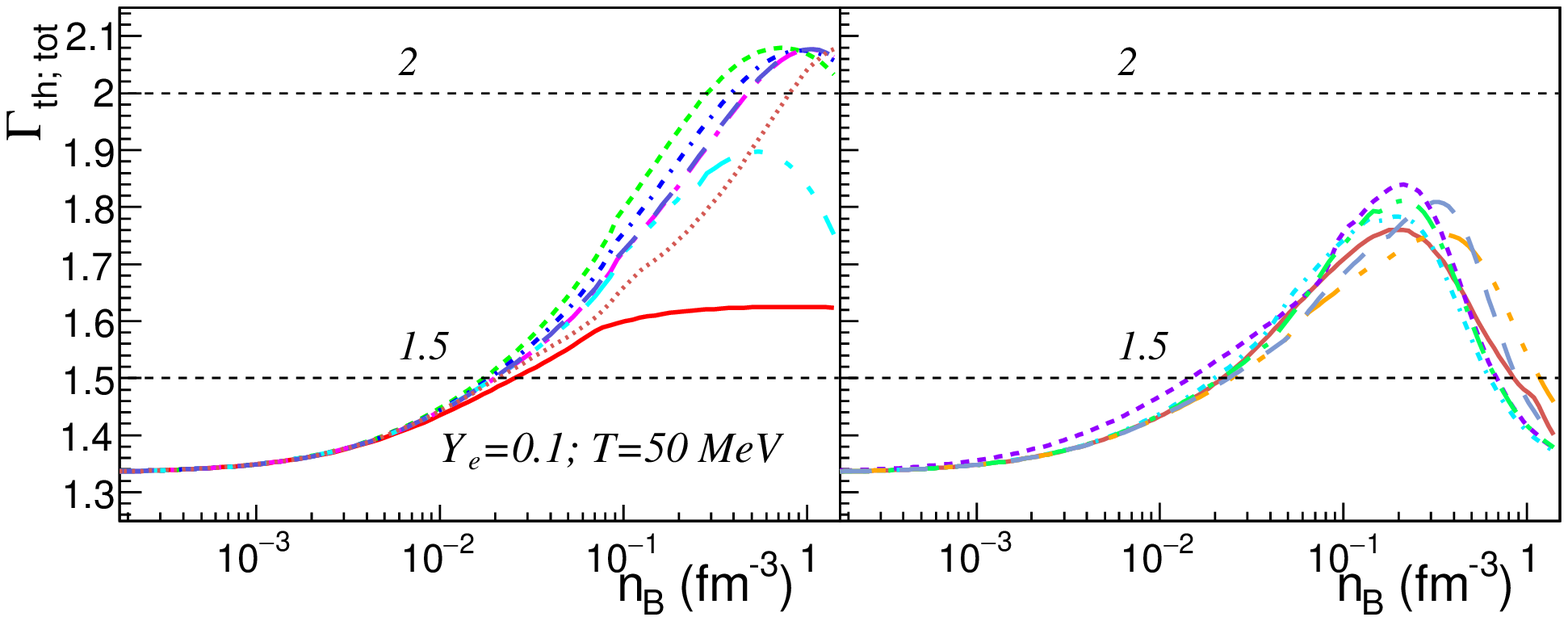}
  \caption{$\Gamma_{\mathit{th}}$ as a function of baryon number density at
    ($T$=5 MeV, $Y_e=0.3$), ($T$=20 MeV, $Y_e=0.1$), ($T$=50 MeV, $Y_e=0.1$).
    Left panels:
    results corresponding to the baryonic sector;
    right panels: 
    contributions of lepton and photon gases are also included. 
    The same EoS models as in fig. \ref{fig:EsymperA_T=0} are considered.
    Dashed horizontal lines mark the values $\Gamma_{\mathit{th}}=4/3$ and 5/3 (left panels)
    and, respectively, $\Gamma_{\mathit{th}}=1.5$ and 2 (right panels).
}
\label{fig:Gammath} 
\end{figure*}

\subsection{Thermal index}
\label{ssec:Gammath}

To complete the discussion in the preceding section, we can compute
the $\Gamma$-factor from our different EoS models as
\begin{equation}
  \Gamma_{\mathit{th}}=1+\frac{p_{\mathit{th}}}{e_{\mathit{th}}}
   \label{eq:Gammath}
\end{equation}
and compare it with the range of values usually assumed. In addition,
we will be able to see the importance of its density, temperature and
charge fraction dependence and thus check the validity of this
approximation of thermal effects.

The left panels in fig. \ref{fig:Gammath} illustrate the evolution of
$\Gamma_{\mathit{th}}$ as function of $n_B$ at ($T = 5$, $Y_e = 0.3$)
(top), ($T = 20$, $Y_e = 0.1$) (middle) and ($T = 50$ MeV, $Y_e =0.1$)
(bottom).  The same EoS models as in Figs. \ref{fig:eth} and
\ref{fig:pth} are considered.  Again, as in Figs. \ref{fig:eth} and
\ref{fig:pth}, we focus on the baryonic sector.  The low
density behavior is common for all models.  For the highest temperature
considered $\lim_{n \to 0} \Gamma_{th;B} \to 5/3$ while progressively
lower values are obtained for $\lim_{n \to 0} \Gamma_{th;B}$ when the
temperature decreases. The value $5/3$ corresponds to the
classical limit for a dilute gas. Deviations from this value at
lower temperatures are attributable to the clusterized component
and show that in these situations the plotted densities are not low enough
for the classical limit to be reached. We nevertheless note that the curves
show little sensitivity to the EoS model. At
high densities the various EoS models behave differently: i) passing
from inhomogeneous to homogeneous matter, in LS220 it smoothly goes
from the low density value to a constant value of $5/3$; ii) models
based on Skyrme-like effective interactions and, for temperatures
exceeding a certain value, also TNTYST(KOST2) provide
$\Gamma_{\mathit{th}}$ increasing monotonically with $n_B$; iii)
SRO(APR) provides a complex and non-monotonic evolution of
$\Gamma_{\mathit{th}}(n_B)$, with a maximum value at $n_B \approx
3n_{\mathit{sat}}$; iv) $\Gamma_{\mathit{th}}(n_B)$ predicted by CDFT
models has a maximum at $n_B \approx 0.2-1$ fm$^{-3}$ and then
decreases smoothly; some models manifest a sudden drop of
$\Gamma_{\mathit{th}}$ at $n_t$ which arise due to the particular way
in which the transition from inhomogeneous to homogeneous matter
has been constructed and is non physical. We note that the
high density limit of $\Gamma_{\mathit{th}}$ in CDFT,
4/3, corresponds to ultra-relativistic gases. This can be understood
from the small Dirac effective masses at high densities within
these models.
We attribute the qualitatively different behaviors of
$\Gamma_{\mathit{th}}(n_B)$ in non-relativistic versus CDFT models
to the functionally different dependence of single particle energies
on Landau/Dirac masses.
The occurrence of a maximum in $\Gamma_{\mathit{th}}(n_B)$ within CDFT
was previously linked to the minimum in
$m_{\mathit{eff};{\rm Landau}}(n_B)$, for a derivation in
the limit of low temperatures, see \cite{Constantinou_PRC_2015}.

Each of these behaviors corresponds to a certain density-dependence of
the effective masses $m_{\mathit{eff}}(n_B)$ which can qualitatively
be understood as follows.  In the limiting case of degenerate matter
without electrons, $\Gamma_{\mathit{th}}$
writes~\cite{Constantinou_PRC_2015}:
\begin{equation}
  \Gamma_{\mathit{th}}(n_B)=\frac53 -\frac{n_B}{m_{\mathit{eff};{\rm Landau}}} \frac{dm_{\mathit{eff};{\rm Landau}}}{dn_B}~.
  \label{eq:meffG}
\end{equation}
Models with constant values of Landau effective masses (LS220) will
provide $\Gamma_{\mathit{th}}(n_B)=5/3$; models with decreasing
$m_{\mathit{eff}}(n_B)$ will provide $\Gamma_{\mathit{th}}(n_B)$
increasing with $n_B$ (Skyrme models and TNTYST); models with
non-monotonic evolution of $m_{\mathit{eff}}(n_B)$ will provide
non-monotonic evolution of $\Gamma_{\mathit{th}}(n_B)$ (APR and CDFT).
Finite-temperature calculations of nuclear matter within the many-body
self-consistent Green's function method \cite{Carbone_PRC_2019} showed
that eq. (\ref{eq:meffG}) is satisfied even beyond the degenerate
regime.
Our CDFT calculations at low temperatures indicate that eq.
(\ref{eq:meffG}) is valid also for this category of models and
for any $n_B$.
This implies that, in principle, estimations of the nucleon Landau
effective mass in PNM and SNM may be obtained by integrating
eq. (\ref{eq:meffG}) for any EoS model.
The limiting condition for this integration is
$\lim_{n_B \to 0} m_{\mathit{eff};{\rm Landau}}=m_N$.

Concerning the temperature dependence, we note that for low temperatures,
$\Gamma_{\mathit{th}}(n_B)$ shows strong dependence on $T$, which becomes more
moderate with increasing $T$.

The values of $\Gamma_{\mathit{th}}$ in Figs. \ref{fig:Gammath} (left
panels) can nevertheless not be compared with those assumed when cold
EoS are phenomenologically supplemented with thermal effects $1.5 \leq
\Gamma \leq 2$ \cite{Bauswein_PRD_2010} because we have disregarded
the contributions of leptons and photons.  Therefore in the right
panels in fig. \ref{fig:Gammath} we display the total thermal index. It
turns out that while leptons and photons do not modify the qualitative
features of the various EoS models, they significantly alter the
values.  We note in particular that: i) at low densities, all
considered EoS models give $\Gamma_{\mathit{th;tot}}<1.5$, ii) at high
densities and high temperatures, models based on Skyrme-like
interactions and TNTYST provide $\Gamma_{\mathit{th;tot}}>2$, iii) at high
densities, CDFT models provide $\Gamma_{\mathit{th;tot}}<1.5$ irrespective
of temperature.  For $T>T_C$,
where $T_C$ represents the critical temperature of the liquid-gas phase
transition of subsaturated nuclear matter,
$\Gamma_{\mathit{th;tot}}$ is weakly-dependent on $T$.
We have found that the dependence on $Y_e$ is weak, too.

\subsection{Entropy, specific heats, adiabatic index and speed of sound}

\begin{figure}
\includegraphics[width=0.99\columnwidth]{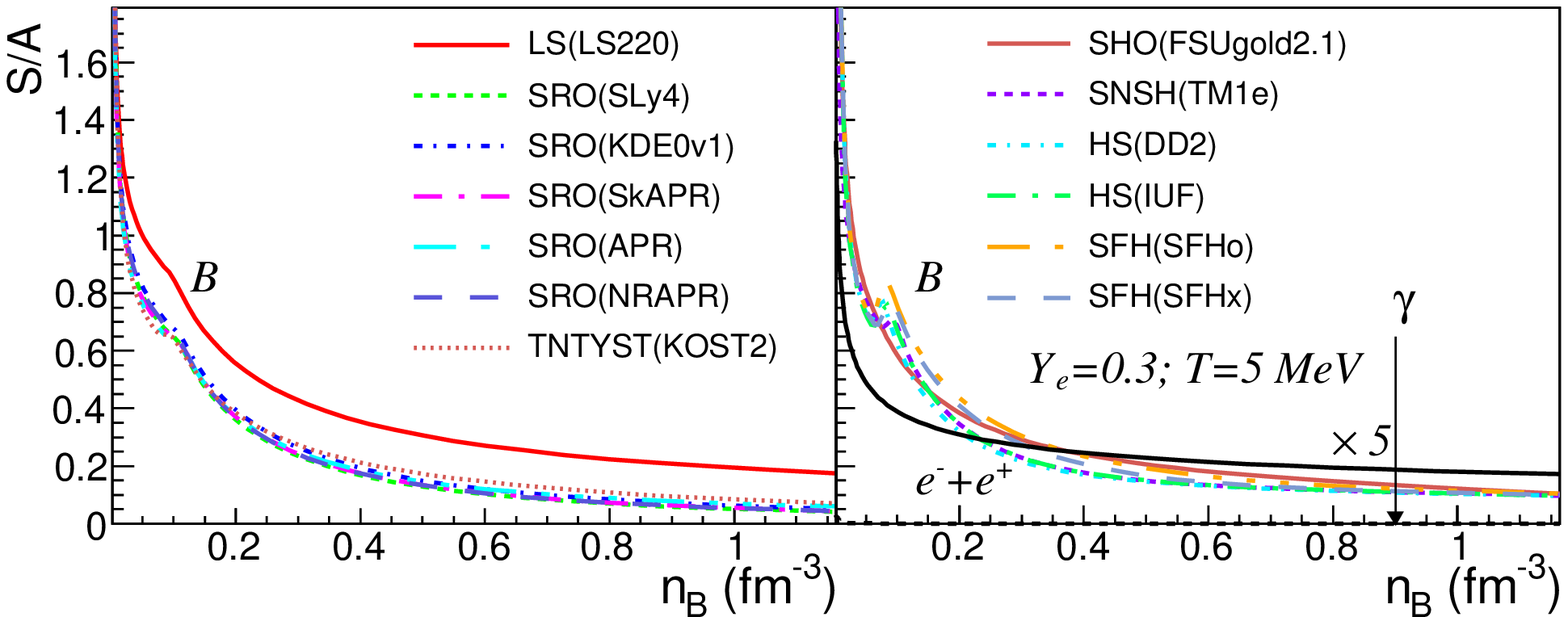}
\includegraphics[width=0.99\columnwidth]{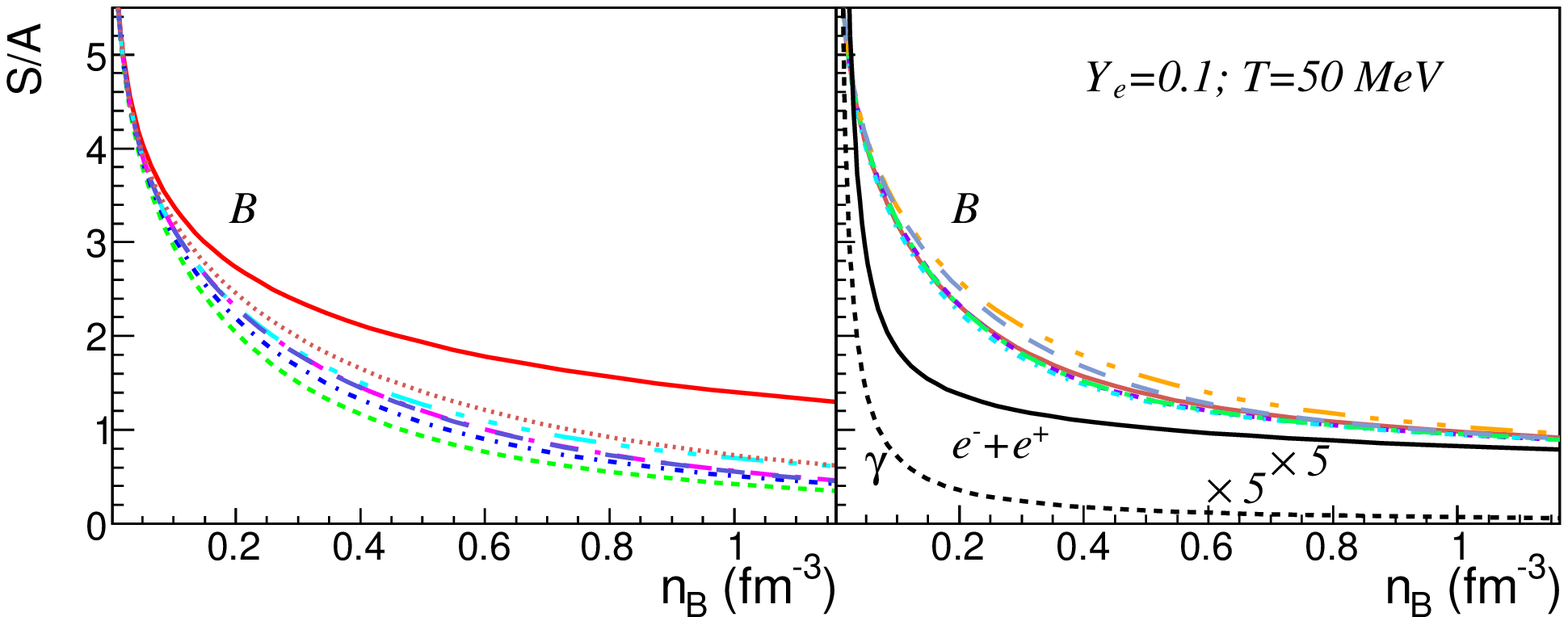}
\caption{Baryonic contribution to the entropy per baryon as function
  of baryon number density for ($T$=5 MeV, $Y_e=0.3$) and
  ($T$=50 MeV, $Y_e=0.1$).
  Contributions of leptons and photons,
  multiplied by five, are depicted in the right panel (solid and
  dashed lines, respectively).  }
\label{fig:SperA_T} 
\end{figure}

Further insight into the impact of the chosen EoS model on
thermodynamic quantities is offered in fig. \ref{fig:SperA_T}.  It
shows the baryonic contribution to the entropy per baryon for ($T$=5
MeV, $Y_e=0.3$) and ($T$=50 MeV, $Y_e=0.1$).  At fixed $T$, $S/A$
obvisouly decreases with $n_B$ within all models. However, as seen
before, quantitatively, a large dispersion is observed.
Non-relativistic models show that the larger
$m_{\mathit{eff};{\rm Landau}}$ the larger $S/A$.
This correlation among Landau effective
masses and entropy is valid at all densities, temperatures and proton
fractions and can be easily understood from the role of the Landau effective
masses in the kinetic energy.  For any $T<T_C$, all EoS models
manifest a kink or a discontinuity at $n_B \approx
n_{\mathit{sat}}/2$.  These are artifacts of the procedures employed
for the transition between clusterized and homogeneous matter and will
manifest in the behavior of most thermodynamic quantities,
{\em e.g.} the $\Gamma$ factor discussed before.

For the thermodynamic conditions considered here, contributions of
leptons and photons are quantitatively important only for low
densities, with increasing weight for increasing temperature.

\begin{figure}
\includegraphics[width=0.99\columnwidth]{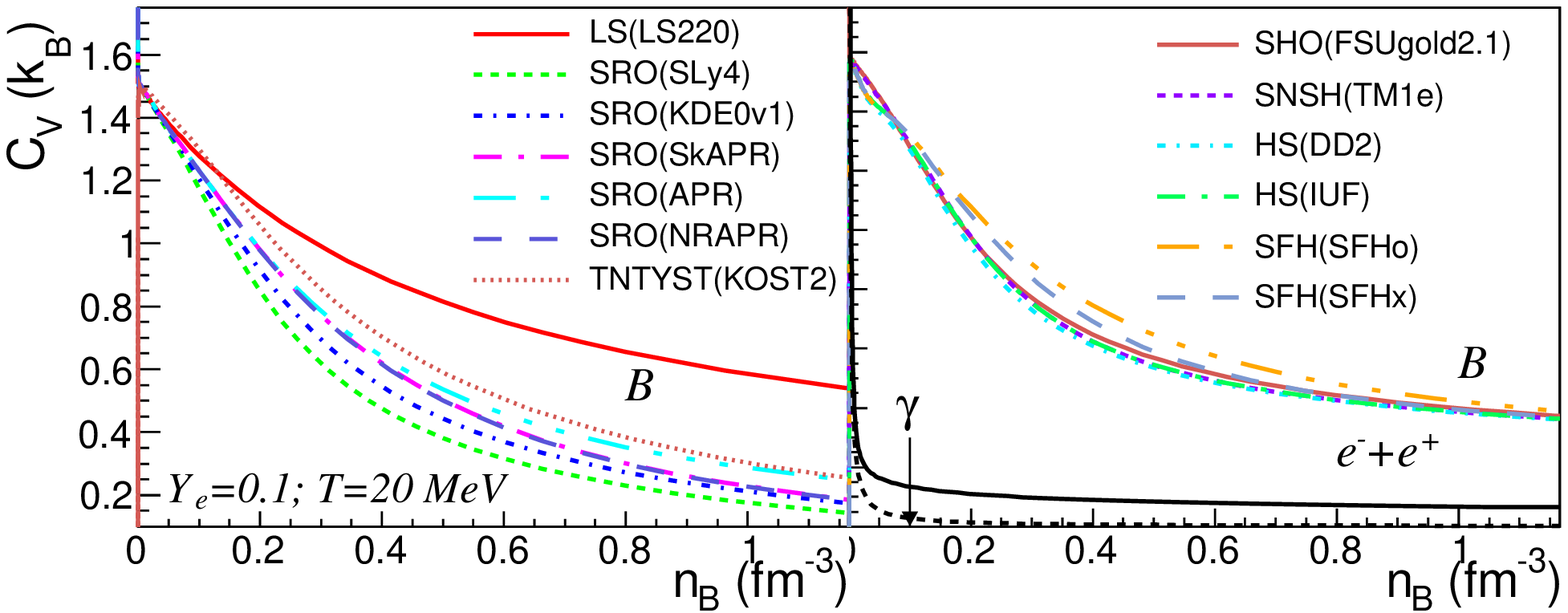}
\includegraphics[width=0.99\columnwidth]{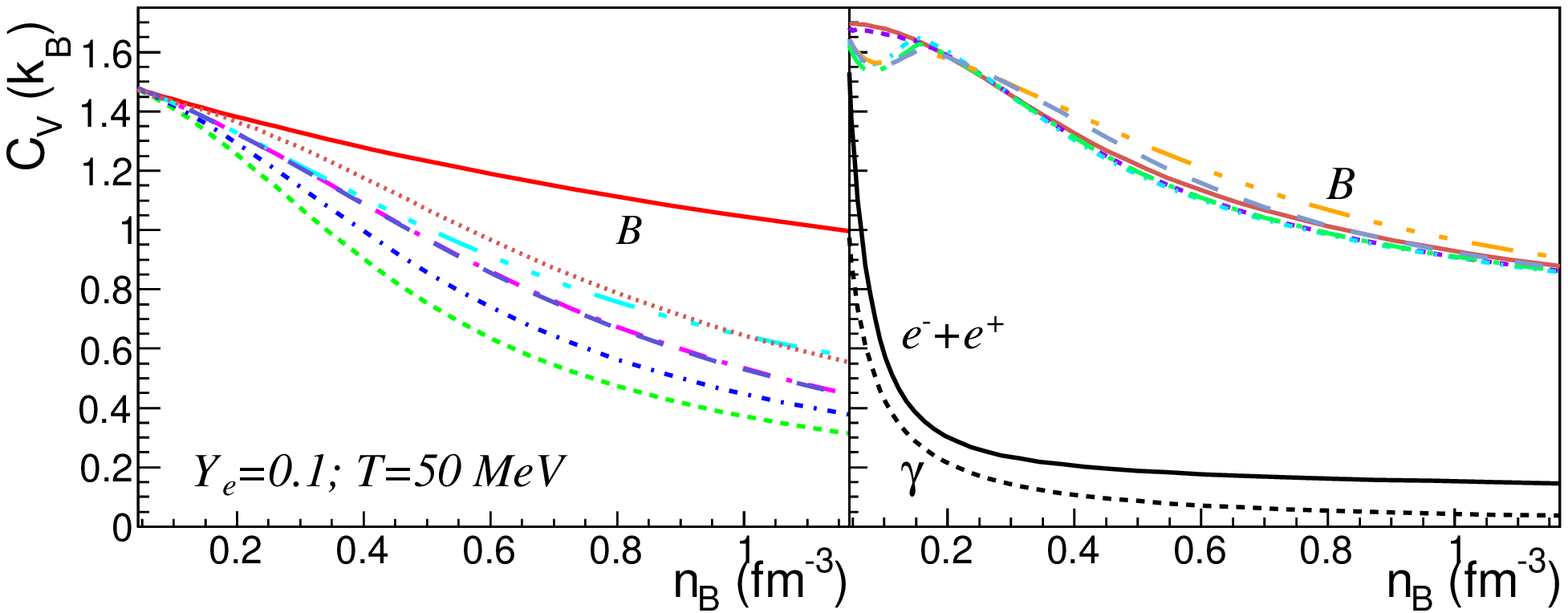}
\caption{Specific heat at constant volume, $C_V$,  as function of baryon
  number density for baryonic matter with $T$=20 MeV and $T$=50 MeV
  at $Y_e=0.1$.
  In the right panels for comparison $C_V$ of the corresponding
  lepton and photon gases are shown, too.
  }
\label{fig:Cv} 
\end{figure}

\begin{figure}
\includegraphics[width=0.99\columnwidth]{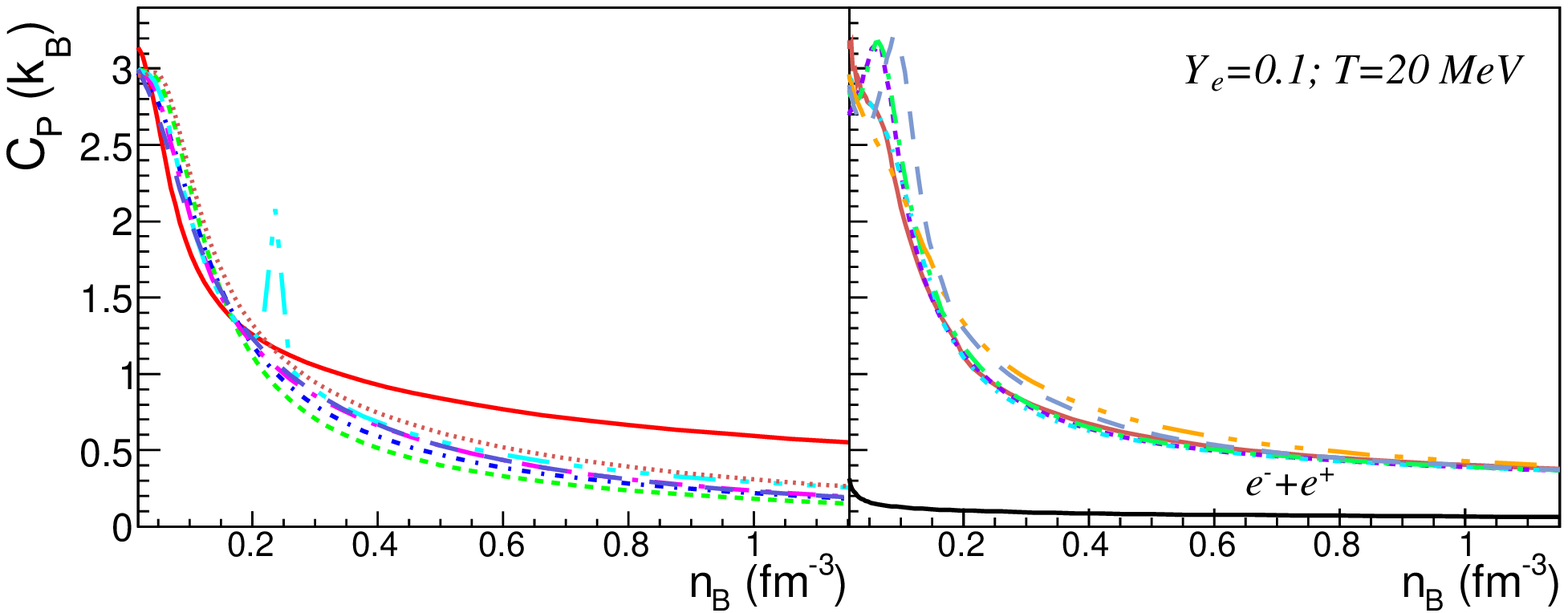}
\includegraphics[width=0.99\columnwidth]{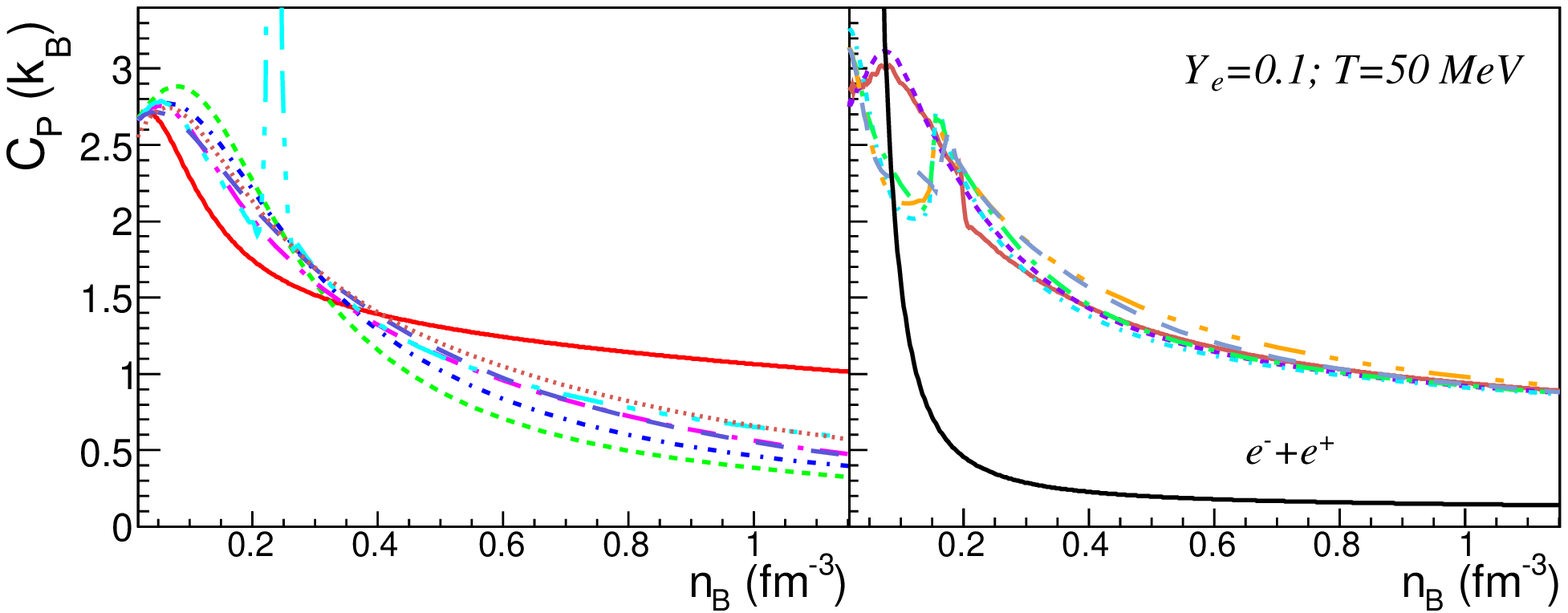}
\caption{Same as in fig. \ref{fig:Cv} for the
  specific heat at constant pressure, $C_P$.
  Line-types are the same as in fig. \ref{fig:Cv}.
  The photon contribution is not shown here as it is ill defined. 
}
\label{fig:Cp} 
\end{figure}

Specific heats at constant volume and constant pressure are defined as
\begin{equation}
  C_V=T \frac{\partial \left( S/A \right)}{\partial T}|_{V,\{N_i\}}~,
  \label{eq:cv}
\end{equation}
and, respectively,
\begin{equation}
  C_P=T\frac{\partial \left( S/A \right)}{\partial T}|_{P,\{N_i\}}=
  C_V+\frac{T}{n_B^2} \frac{\left(\frac{\partial P}{\partial T}|_n \right)^2}
  {\frac{\partial P}{\partial n}|_T}~,
  \label{eq:cp}
\end{equation}
where $A=\sum_i N_i$ is the total number of particles of the system.

Figs. \ref{fig:Cv} and \ref{fig:Cp} illustrate the evolution of $C_V$
and $C_P$ as function of baryon number density for $T$=20 MeV and
$T$=50 MeV at $Y_e$=0.1.  Contributions of baryons, leptons and, in
the case of $C_V$, also photons are reported separately.

Concerning $C_V$, except the CFDT models at $T=50$ MeV and
low densities, all models behave similarly. More precisely
they replicate the behavior seen in fig. \ref{fig:SperA_T} for
$S/A(n_B)$. The maximum value is reached in the limit of low densities
and the higher the density the smaller $C_V$.  $\lim_{n_B \to
  0}C_V=3/2$, which corresponds to the classical value of a Boltzmann
gas.  According to \cite{Constantinou_PRC_2014,Constantinou_PRC_2015}
all dilute homogeneous Fermi gases with local interactions
comply with this limit. Our results show that this is also the case of
dilute inhomogeneous nuclear matter, though for the considered
temperature values the amount of matter bound in clusters is negligible,
see Sec. \ref{sec:compo}.

At $T$=50 MeV and low densities the CFDT models present one or two
maxima in $C_V (n_B)$ and their values exceed 3/2.  SHO(FSUgold2.1)
and SNSH(TM1e) models have a maximum at $n_B \approx 0.1~{\rm
  fm}^{-3}$; models belonging to the HS~\cite{Hempel_NPA_2010} family
reach the highest $C_V$ value in the limit $n_B \to 0$ and have a
second maximum at $n_B \approx 0.2~{\rm fm}^{-3}$.
According to \cite{Constantinou_PRC_2015} the maxima in $C_V$
at non-vanishing densities are typical for non-local interactions.  
The second maximum, present only in models based on \cite{Hempel_NPA_2010},
is likely a numerical artifact with still some nuclear clusters
present.  In the high density range, the dispersion in the values
of $C_V(n_B)$ as well as the ordering of the models replicate the
behavior of the entropy per baryon, see fig. \ref{fig:SperA_T}. It is
again the impact of the effective masses which is recovered here.

Except some specific and local structures that appear
for APR and the HS~\cite{Hempel_NPA_2010} CDFT models, the qualitative
behavior of the specific heat at constant pressure as a function of
density is the same for all models.  In the limit $n_B \to 0$ all
curves converge to a unique, but temperature dependent, value. At high
temperatures $\lim_{n_B \to 0 }C_P \to 5/2$, which corresponds to
Boltzmann gases.  At low densities $C_P(n_B)$ features a short peak;
its structure is better shaped at high temperatures.  For the analytic
derivation of the onset of this peak in the case of homogeneous matter
in the non-degenerate limit, applicable to low densities and high
temperatures, see \cite{Constantinou_PRC_2014}.  Finally, at large
densities $C_P(n_B)$ decreases.  Dependence of pressure on both
effective masses and their derivatives with respect to density
\cite{Constantinou_PRC_2014} leads to a more complex evolution of
$C_P(n_B)$ than for $C_V(n_B)$.  The specific feature of APR resides
in the spike at $\approx 0.2~{\rm fm}^{-3}$, which signals the
transition to the pion condensed state.  The models based on
\cite{Hempel_NPA_2010} present a double peaked structure, as it was
also the case with $C_V(n_B)$, which is again probably due to the
presence of some nuclear clusters in this region.

Finally, for the thermodynamic conditions considered here, lepton and
photon contributions are dominant at low densities and negligible at
supranuclear densities.

\begin{figure}
\includegraphics[width=0.99\columnwidth]{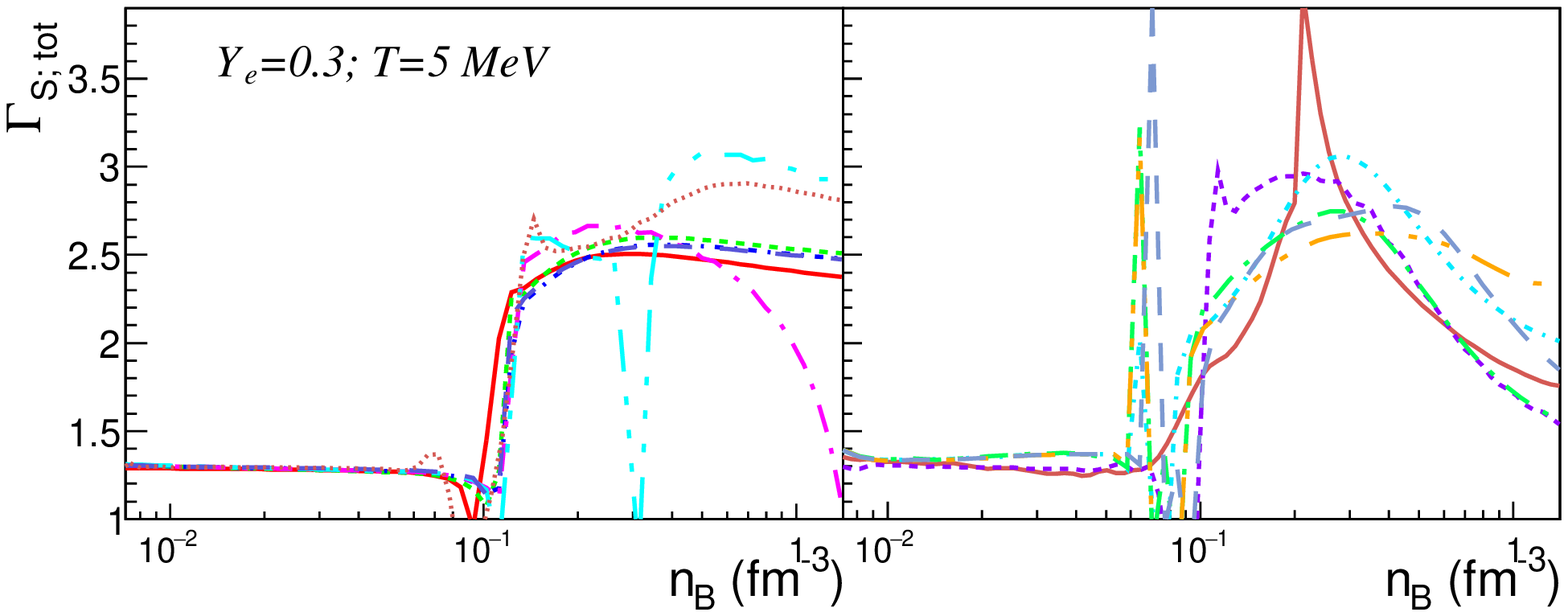}
\includegraphics[width=0.99\columnwidth]{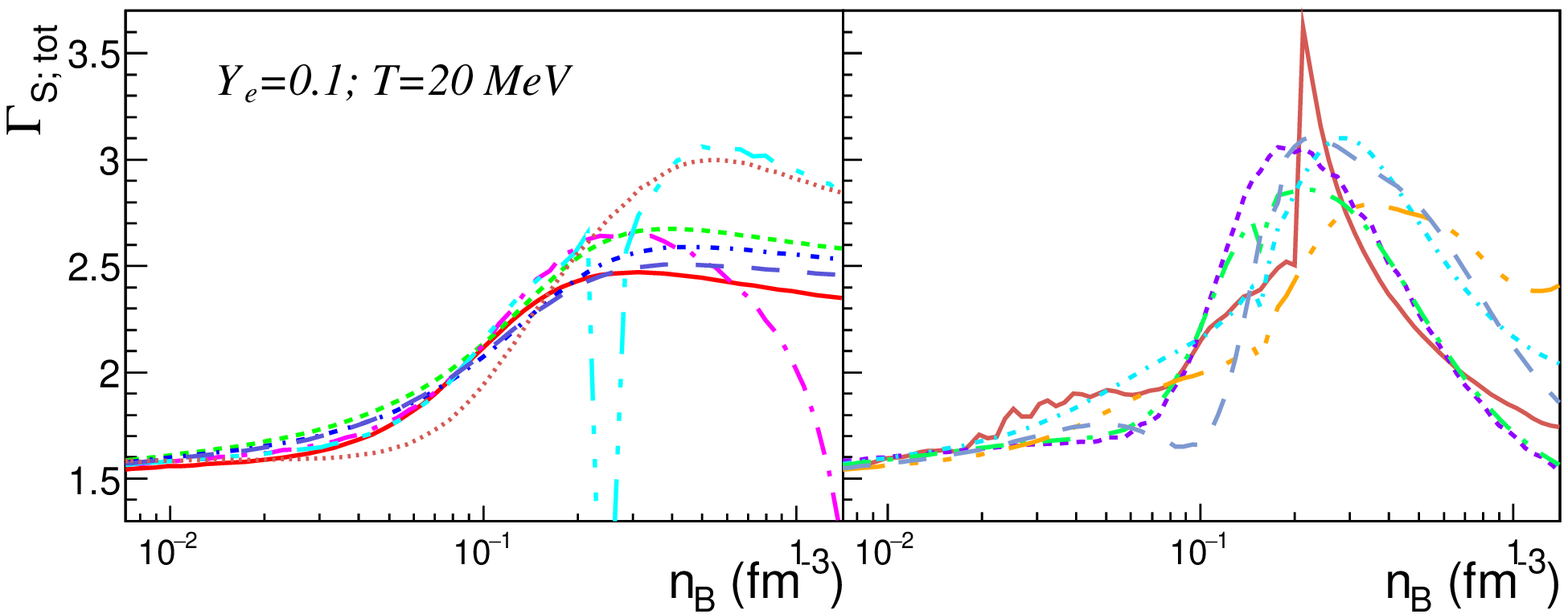}
\includegraphics[width=0.99\columnwidth]{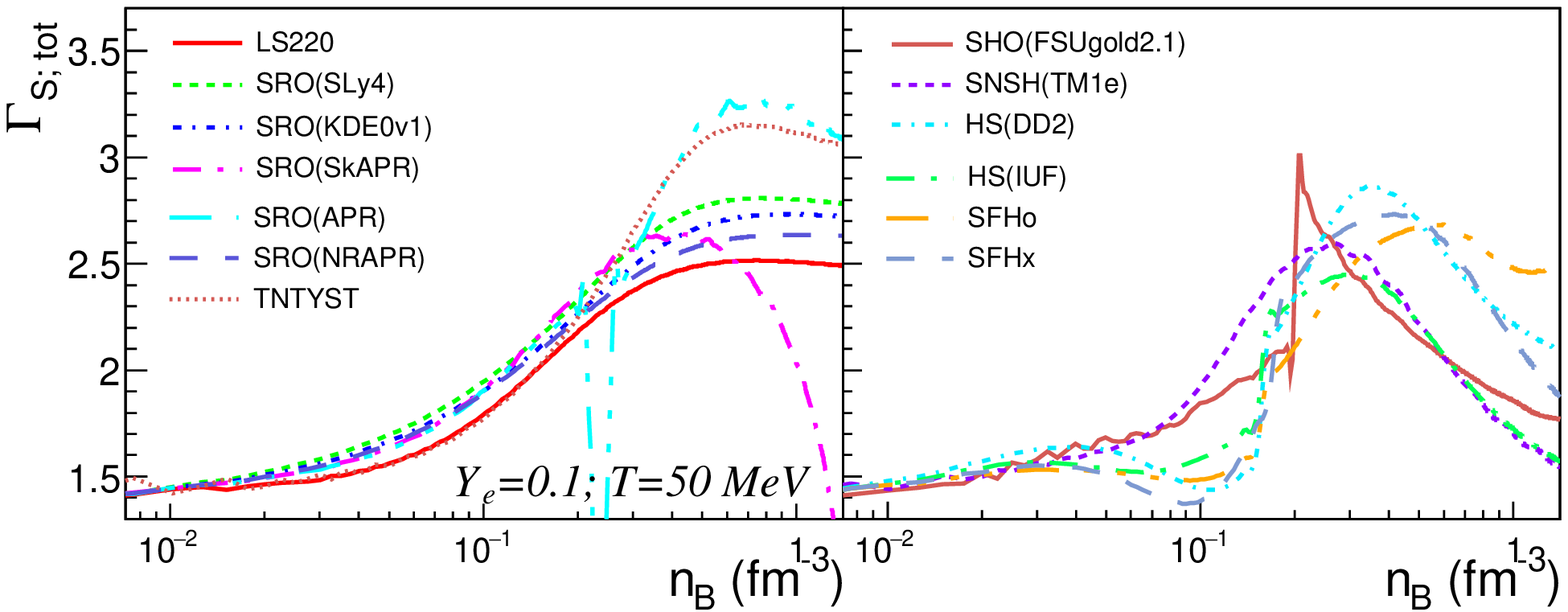}
\caption{$\Gamma_S$, eq.~(\ref{eq:GS}), as function of baryon number density
  for ($T$=5 MeV, $Y_e=0.3$),
  ($T$=20 MeV, $Y_e=0.1$) and ($T$=50 MeV, $Y_e=0.1$).
  Results corresponding to stellar matter, as predicted by various EoS models.
}
\label{fig:GammaS_T} 
\end{figure}

The behavior of the adiabatic index, defined as
\begin{equation}
  \Gamma_{S}=\frac{\partial \ln P}{\partial \ln n_B}|_{S}=
  \frac{C_P}{C_V} \frac{n_B}{P} \frac{\partial P}{\partial n_B}|_T~,
   \label{eq:GS}
\end{equation}
is illustrated in fig.~\ref{fig:GammaS_T} for stellar matter,
\textit{i.e.} including baryons, leptons and photons.  The predictions
by various EoS models at different thermodynamic conditions are
considered.  $\Gamma_S$ gives an indication about the stiffness of the
EoS in all processes occuring at constant entropy.

Qualitatively, all EoS models behave rather similarly: at low
densities $\Gamma_S$ tends to a constant value,
$\Gamma_S$ is then increasing with density, reaching a maximum around
a few times nuclear matter saturation density and decreasing again
at high densities. It is not straightforward to interpret the
quantitative differences between the EoS models at high densities,
for instance in terms of effective masses.
However, we note that the spike present in the curve
corresponding to SRO(APR) is due to the transition to pion condensed
phase and that the irregularities present at $T<T_C(Y_p)$ and $n_B
\approx n_t$ in the curves predicted by models based on the HS NSE
approach~\cite{Hempel_NPA_2010} are related to the transition from
inhomogeneous to homogeneous matter.  For $T=5$ MeV and $n_B<n_t$,
where the liquid-gas coexistence might manifest, $\Gamma_S \neq 0$.
Two alternative explanations can be put forward.  First, because of
the electron gas, the phase coexistence might be suppressed
\cite{Ducoin_NPA_2007}.  Alternatively, if the phase coexistence
persists, deviations from $\partial P/\partial n_B|_T=0$ may stem from
the fact that the transition from inhomogeneous to homogeneous matter
was not realized via a Maxwell construction, as it should, but by
minimizing the free energy for fixed value of $Y_p$.
Indeed, for numerical convenience many EoS models have adopted this
solution, see Sec. \ref{ssec:inhomo}.
$\Gamma_{S}$ depends strongly on the EoS model and
density. On the contrary, the dependence on $Y_e$ and
temperature is weak, especially at $T>T_C$.

\begin{figure}
\includegraphics[width=0.99\columnwidth]{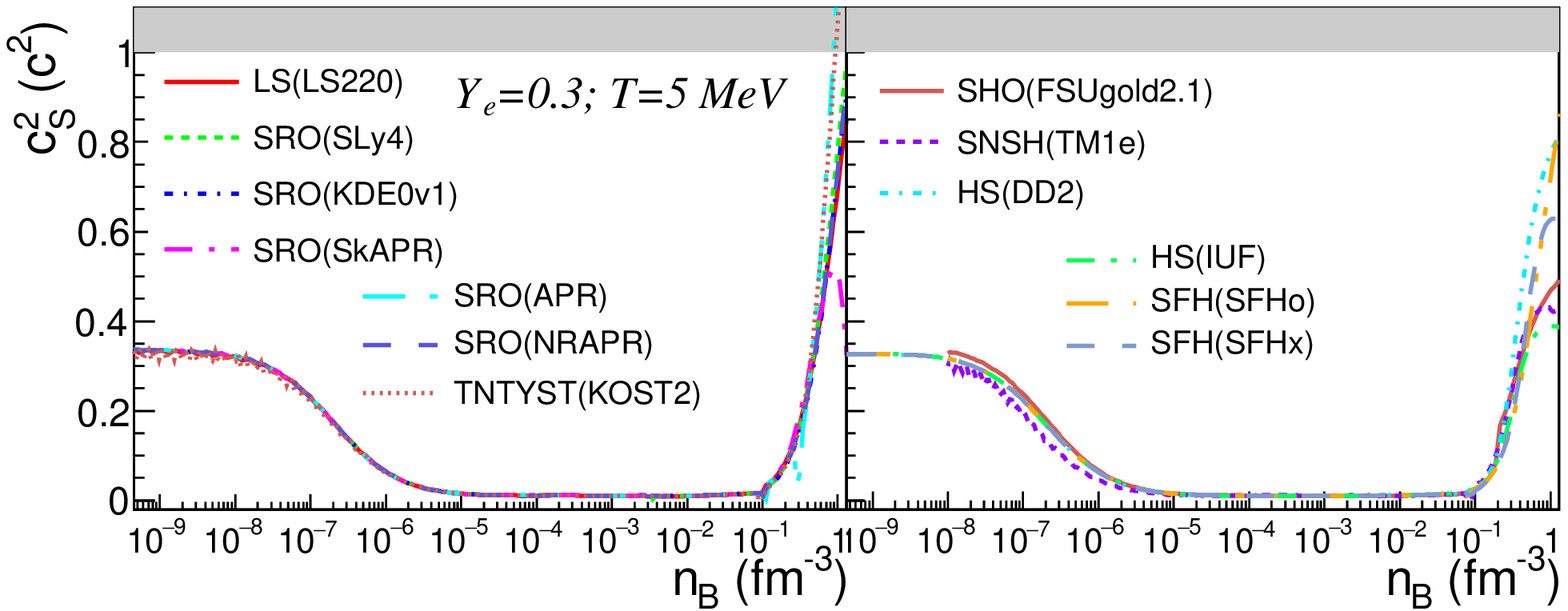}
\includegraphics[width=0.99\columnwidth]{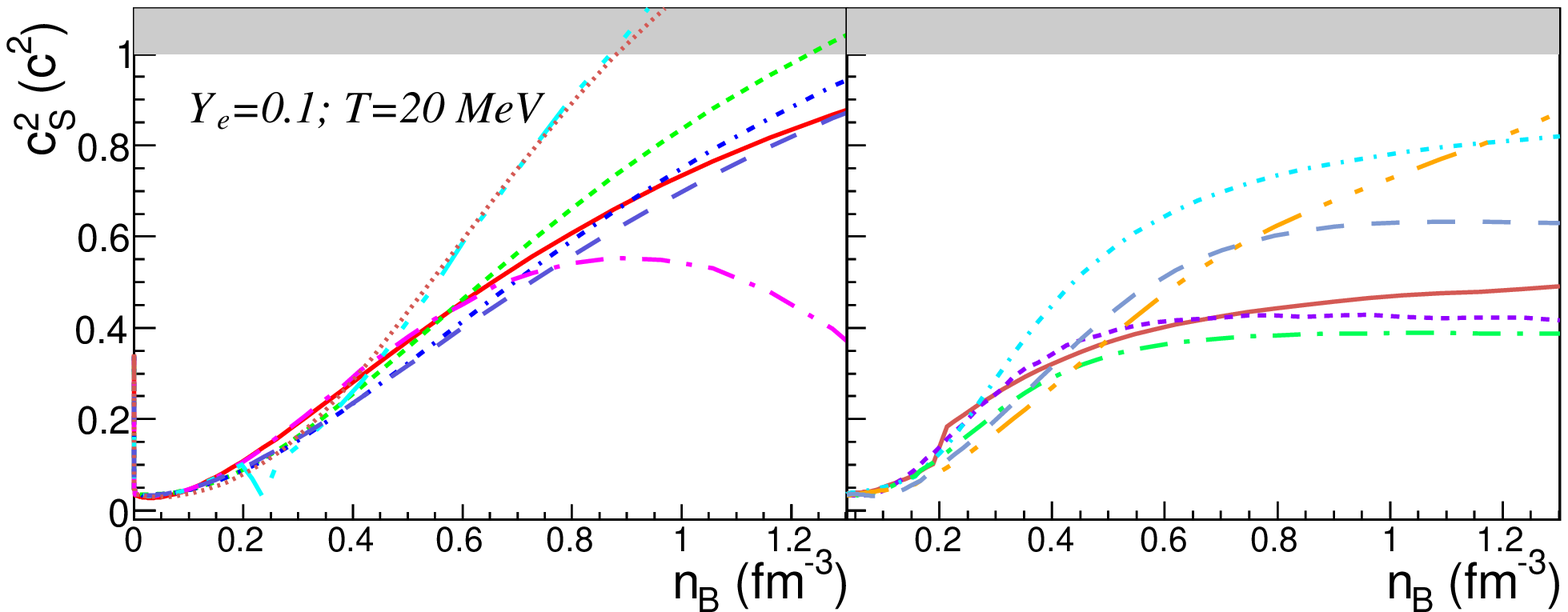}
\includegraphics[width=0.99\columnwidth]{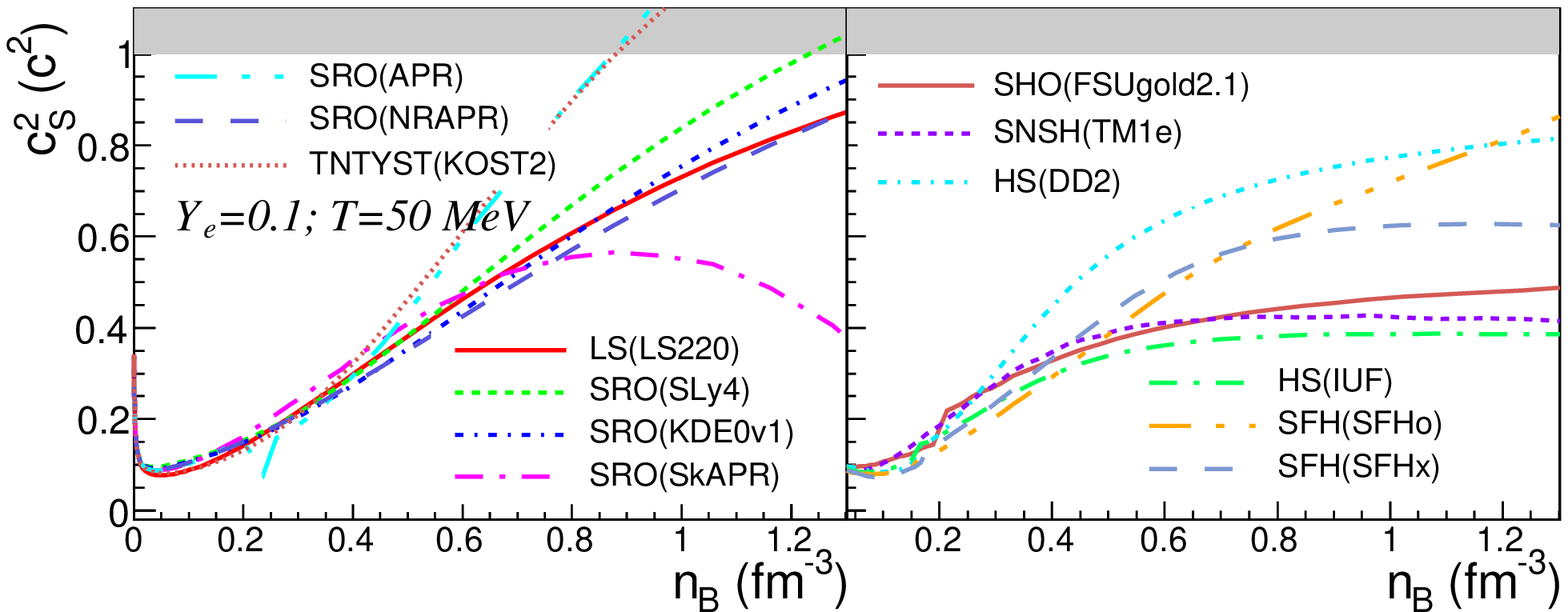}
\caption{Speed of sound squared, eq. (\ref{eq:cs2}), as function of
  baryon number density in stellar matter for various EoS models.
  The considered thermodynamic conditions are:
  ($T$=5 MeV, $Y_e=0.3$), ($T$=20 MeV, $Y_e=0.1$) and 
  ($T$=50 MeV, $Y_e=0.1$). 
  The shaded area marks the domain $c_S^2 >1$ forbidden by causality.
}
\label{fig:cs2_T} 
\end{figure}

A key quantity in dynamical numerical simulations is the speed of sound $c_S$.
In units of $c$, the speed of light, it is given by
\begin{equation}
  c_S^2=\frac{dP}{de}|_{S,A,Y_e}=\Gamma_S \frac{P}{e+P}~.
  \label{eq:cs2}
\end{equation}

Fig. \ref{fig:cs2_T} illustrates its behavior as function of baryon
number density within the different EoS models.  Different
thermodynamic conditions, indicated in the figures, are considered.
We note that for the lowest considered temperature and a
model-independent density range, $10^{-5} ~{\rm fm}^{-3} \lesssim n_B
\lesssim n_t$, $c_S^2$ has a small but non-vanishing
value. This domain roughly corresponds to the domain of coexistence
between the liquid and gas phases of pure baryonic matter.
As for $\Gamma_S$, to which $c_S^2$ is related, the non-vanishing
values may be attributed either to the quenching of the coexistence
region or, if this is not the case, to the way in which the transition
from the clusterized phase to the homogeneous one was
performed.
As such the small value of $c_S^2$ should be regarded as a consequence of
the dominance of energy density over pressure.
We also note that the transition to the pion condensate
phase in SRO(APR) is signalled by a spike.  The density where this
spike occurs ranges from $\approx 0.2$ fm$^{-3}$ to $\approx 0.32$
fm$^{-3}$, as $Y_e$ increases from 0 to 0.5 \cite{APR_PRC_1998}.
At high densities, the models
incorporating relativistic kinematics reach a saturation, whereas it
is obvious that for the non-relativistic models $c_s$ continues to
increase and some models reach the causal limit below a density of 1
fm$^{-3}$.  SkAPR singles out by $c_S$ decreasing with density over
$n_B > 0.8$ fm$^{-3}$.  This behavior is the consequence of the
condition to reproduce the pressure of SNM and PNM at $4
n_{\mathit{sat}}$ calculated by APR, imposed to SkAPR
\cite{Schneider_PRC_2019}, and the stiffer evolution of $p(n_B)$ in
SkAPR with respect to APR up to $4 n_{\mathit{sat}}$, see also
Figs. \ref{fig:EperA_SNM_T=0} and \ref{fig:EperA_PNM_T=0}.  Concerning
the temperature dependence, it turns out that for $T>T_C$, $c_S$ only
weakly depends on $T$.  $c_S$ weakly depends on $Y_e$, too, though not
shown here.

\section{Composition and thermodynamic quantities of subsaturated matter
  at finite temperature}
\label{sec:compo}
\begin{figure*}
  \begin{center}
    \includegraphics[width=0.39\textwidth]{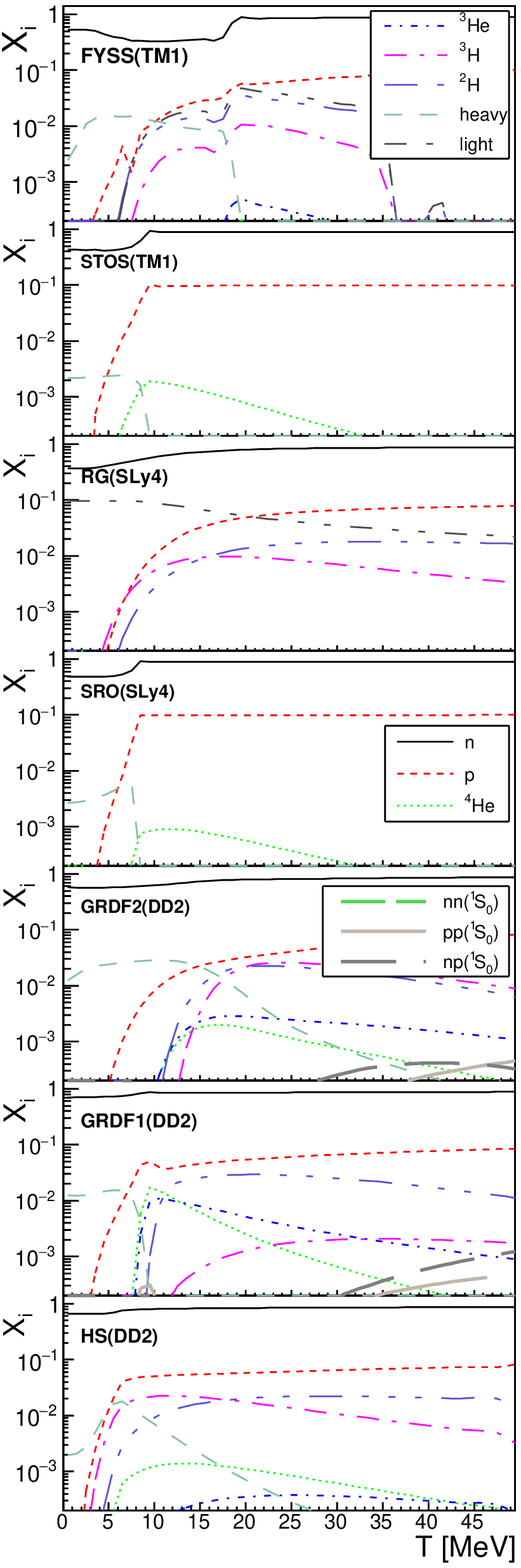}
    \includegraphics[width=0.39\textwidth]{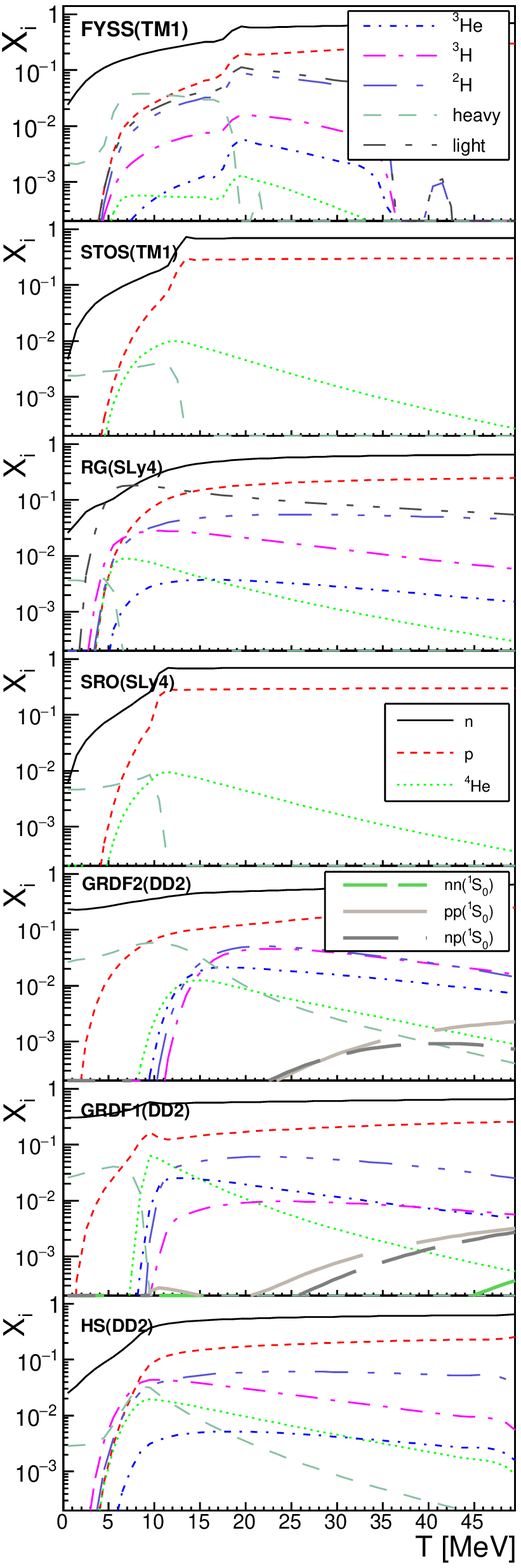}
  \end{center}
  \caption{Composition of stellar matter in terms of (average) particle
    fractions as function of temperature at $n_B=3 \cdot 10^{-2}$ fm$^{-3}$
    for $Y_e=0.1$ (left) and $Y_e=0.3$ (right) for various EoS models.
    "Light" clusters refer to nuclei with $2 < A \leq 20$ (RG(SLy4)) or
    $Z \leq 5$ (FYSS(TM1));
    "heavy" clusters refer to nuclei with $A \geq 20$ (RG(SLy4)) or
    $Z> 5$ (FYSS(TM1)) or all nuclei other than
    ($^2_1$H, $^3_1$H, $^3_2$He, $^4_2$He) (HS(DD2), GRDF1(DD2), GRDF2(DD2))
    or all nuclei other than $^4_2$He (STOS(TM1) and SRO(SLy4)).
    For GRDF1(DD2) and GRDF2(DD2) the fraction of $^2$H includes contributions
    from the deuteron bound state and continuum correlations in the np($^3S_1$)
    channel.
  }
\label{fig:compo_T} 
\end{figure*}

\begin{figure*}
  \begin{center}
    \includegraphics[width=0.39\textwidth]{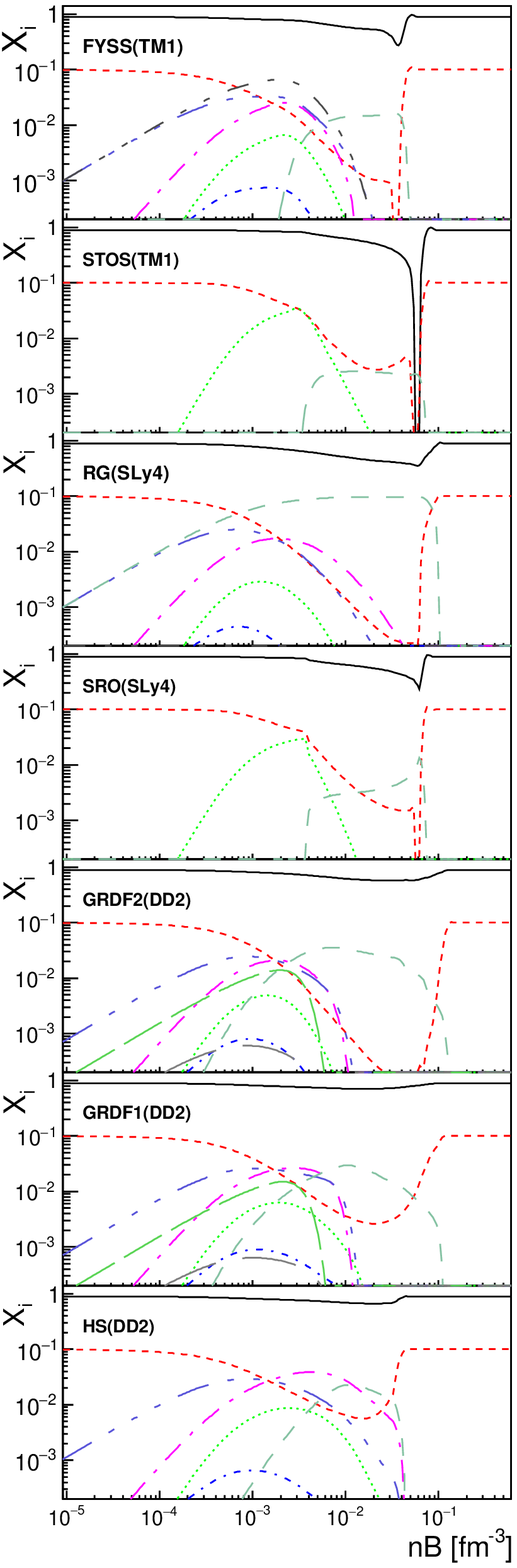}
    \includegraphics[width=0.39\textwidth]{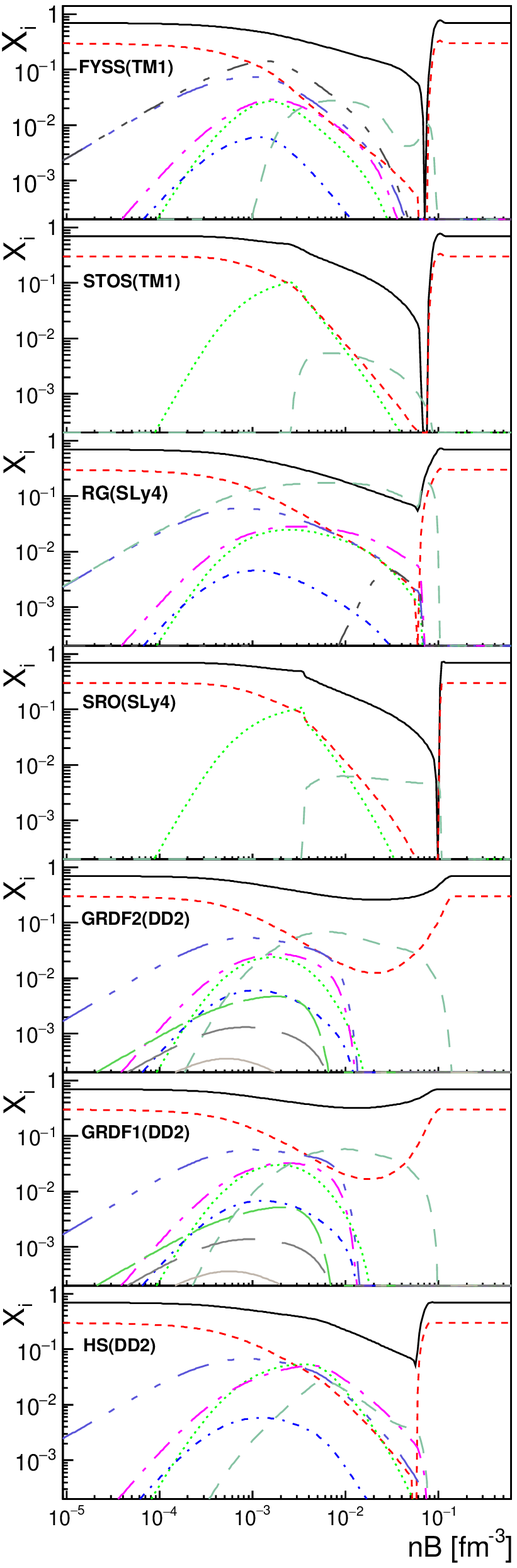}
  \end{center}
  \caption{Composition of stellar matter in terms of (average) particle fractions
    as function of baryon number density at $T$=5 MeV for
    $Y_e=0.1$ (left) and $Y_e=0.3$ (right panels) for various EoS models.
    Line types and nuclear clusters are defined as in fig. \ref{fig:compo_T}.
 }
\label{fig:compo_nB} 
\end{figure*}

For given thermodynamic conditions, the nuclear distribution in
stellar matter depends on the modelling (NSE vs. SNA, mass and isospin
asymmetry distributions of cluster degrees of freedom, shape degrees
of freedom, nuclear binding energies, excited states and level
densities, in-medium and temperature modification of nuclear energy
density, density dependence of the symmetry energy, evolution of shell
effects far from stability, etc.) and on the employed effective
nucleonic interaction
\cite{Hempel_NPA_2010,Oertel_RMP_2017,Fischer_PASA_2017,Raduta_NPA_2019,Yudin_MNRAS_2019}.

Figs. \ref{fig:compo_T} and \ref{fig:compo_nB} illustrate the
evolution with temperature and baryon number density of the (average)
particle fractions of different species (neutron, protons, $^2$H,
$^3$H, $^3$He, $^4$He, "light" and "heavy" clusters) for
$n_B=3 \cdot 10^{-2}$ fm$^{-3}$ and, respectively, $T$=5 MeV at
$Y_e=0.1$ and $Y_e=0.3$. We recall that among the considered models
SRO(SLy4) and STOS(TM1) rely on SNA while
GRDF1(DD2), GRDF2(DD2), HS(DD2), RG(SLy4) and FYSS(TM1) rely on
extended NSE. In addition to the light species whose abundances are
explicitly considered in Figs. \ref{fig:compo_T} and \ref{fig:compo_nB},
RG(SLy4) accounts for exotic isotopes of light nuclei like
$^{4 \leq A \leq 14}$H, $^{5 \leq A \leq 17}$He, $^{8 \leq A \leq 17}$Li, etc.;
for $n_B-T-Y_e$ domains where some of these neutron-rich light nuclei
are copiously produced, see \cite{Raduta_NPA_2019}.
Inclusion of unstable states in the NSE pool is motivated by the
high temperature and neutron-richness of the environment.
GRDF1(DD2) and GRDF2(DD2) account for nn($^1S_0$), np($^1S_0$),
pp($^1S_0$) as effective resonances with medium dependent properties.
Apart from differences in the considered pool of nuclei, there are
differences in modelling, see the discussion in Sec.~\ref{ssec:inhomo}.


Despite these many differences, some common features are
observed. Among others, in fig. \ref{fig:compo_T} it can be seen that
the abundances of unbound nucleons increase with temperature in all
models and light nuclei survive beyond the critical temperature of the
nuclear matter liquid-gas phase transition, for which the maximum
value $T_C \approx 16-18$ MeV, is reached in SM.  Similarly, increased
isospin symmetry of the mixture favors nuclear clusters at the cost of
unbound nucleons.  Differences in modelling mostly impact the density
domain where a significant amount of matter is bound in massive clusters.
Implementation of a temperature-dependent surface tension that
vanishes at $T_C(Y_p)$ in SRO(SLy4), STOS(TM1) and FYSS(TM1) results in the
sudden disappearance of the "heavy" nucleus at $T=T_C(Y_p)$.
Differences in the estimation of excluded volume, nuclear level
density, maximum allowed excitation energy and, in the case of
GRDF1(DD2), GRDF2(DD2) and FYSS(TM1), in-medium and/or temperature modifications
of the nuclear binding energies along with the definition of "heavy"
nuclei are reflected in a variety of behaviors for $X_{heavy}(n_B)$,
including the density where $X_{\mathit{heavy}} < 10^{-4}$, the lower
limit in the figure.

Fig. \ref{fig:compo_nB} shows that clusters, light and heavy, exist in
significant amounts only over limited domains of baryon number density and,
similarly to what we have seen in fig. \ref{fig:compo_T}, their
abundances increase with $Y_e$. In very dilute matter unbound nucleons
dominate. For $n_t(T)/100 \lesssim n_B \lesssim n_t(T)$ matter
composition is strongly model dependent.  A few other points should be
mentioned: i) in-medium modifications of binding energies in
GRDF1(DD2) and GRDF2(DD2) result in the dissolution of light clusters
at a baryonic number density one order of magnitude lower than the one
corresponding to the transition to homogeneous matter;
at variance with this heavy clusters survive up to larger densities in
GRDF models than in HS;
ii) in models where the transition to homogeneous matter is determined
by excluded volume most light clusters survive up to this transition density.

Both Figs. \ref{fig:compo_T} and \ref{fig:compo_nB} show that,
except for neutron-rich matter near the transition density, SNA
based models show the same pattern under specific thermodynamic
conditions; at moderate isospin asymmetries mass sharing in RG(SLy4)
and HS(DD2) is similar whereas it differs at large asymmetries;
predictions of these two models which do not account for in-medium
and thermal modifications of the cluster energy are very different
than those of FYSS(TM1); mass sharing is significantly different in
the two GRDF models; only a small fraction of unbound nucleons
features correlations and their importance increases with
temperature and proton fraction. 

As a remark of caution, let us remind that for $10^{-2}~{\rm fm}^{-3}
\lesssim n_B \lesssim n_t$ and temperatures of the order of a few MeV
the strongly discontinuous surface contribution of the nuclear
free energy due to nuclear shell effects results in multi-modal
cluster distributions \cite{Raduta_PRC_2016,Oertel_RMP_2017}. This
means, that the compositional information provided in \textsc{CompoSE}
in terms of one average heavy nucleus cannot correctly
describe the actual distribution. Among others, this prevents a
realistic calculation of weak interaction rates~\cite{Juodagalvis_NPA_2010,Raduta_PRC_2016}.

\begin{figure*}
  \begin{center}
    \includegraphics[width=0.3\textwidth]{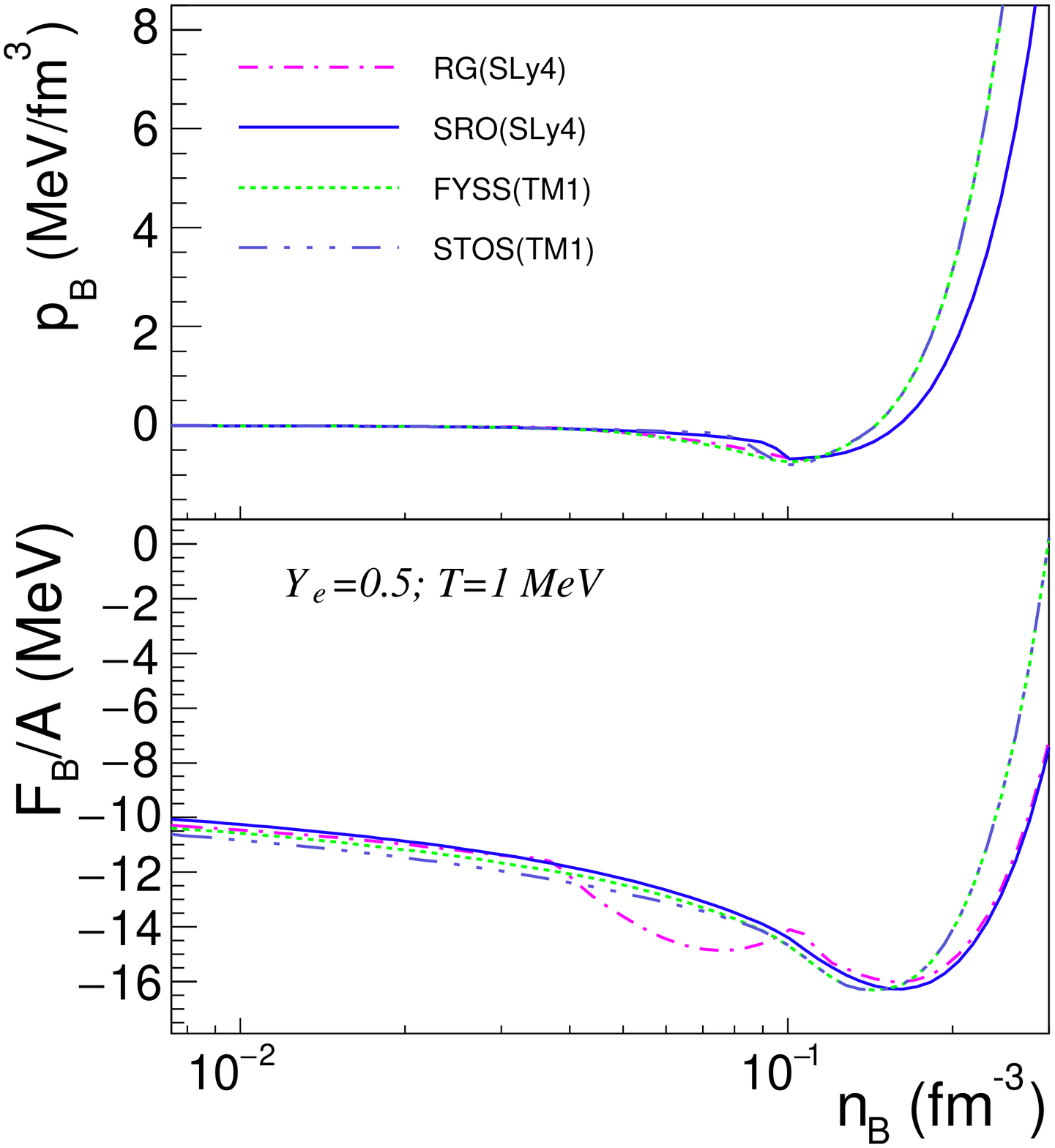}
    \includegraphics[width=0.3\textwidth]{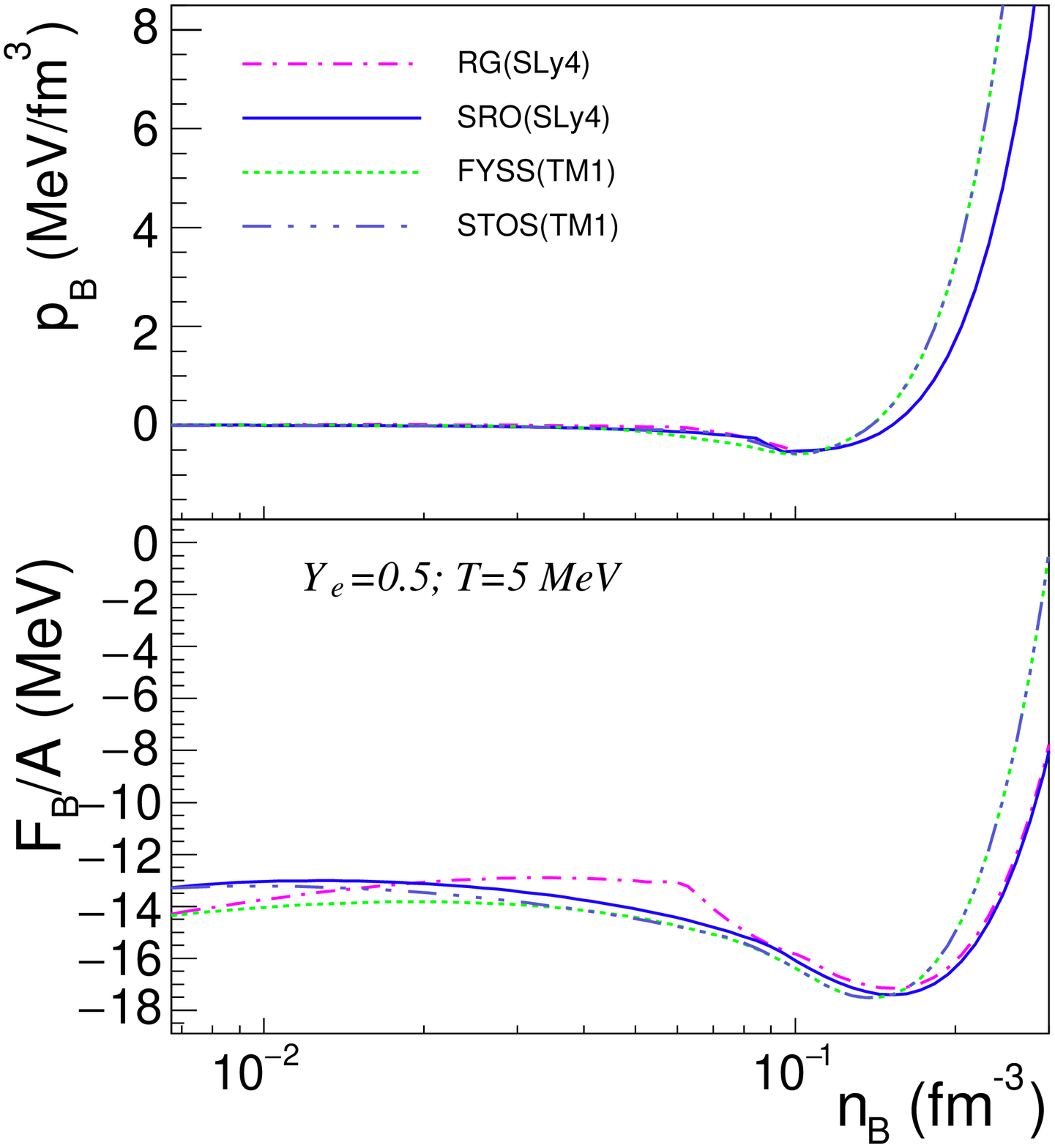}
    \includegraphics[width=0.3\textwidth]{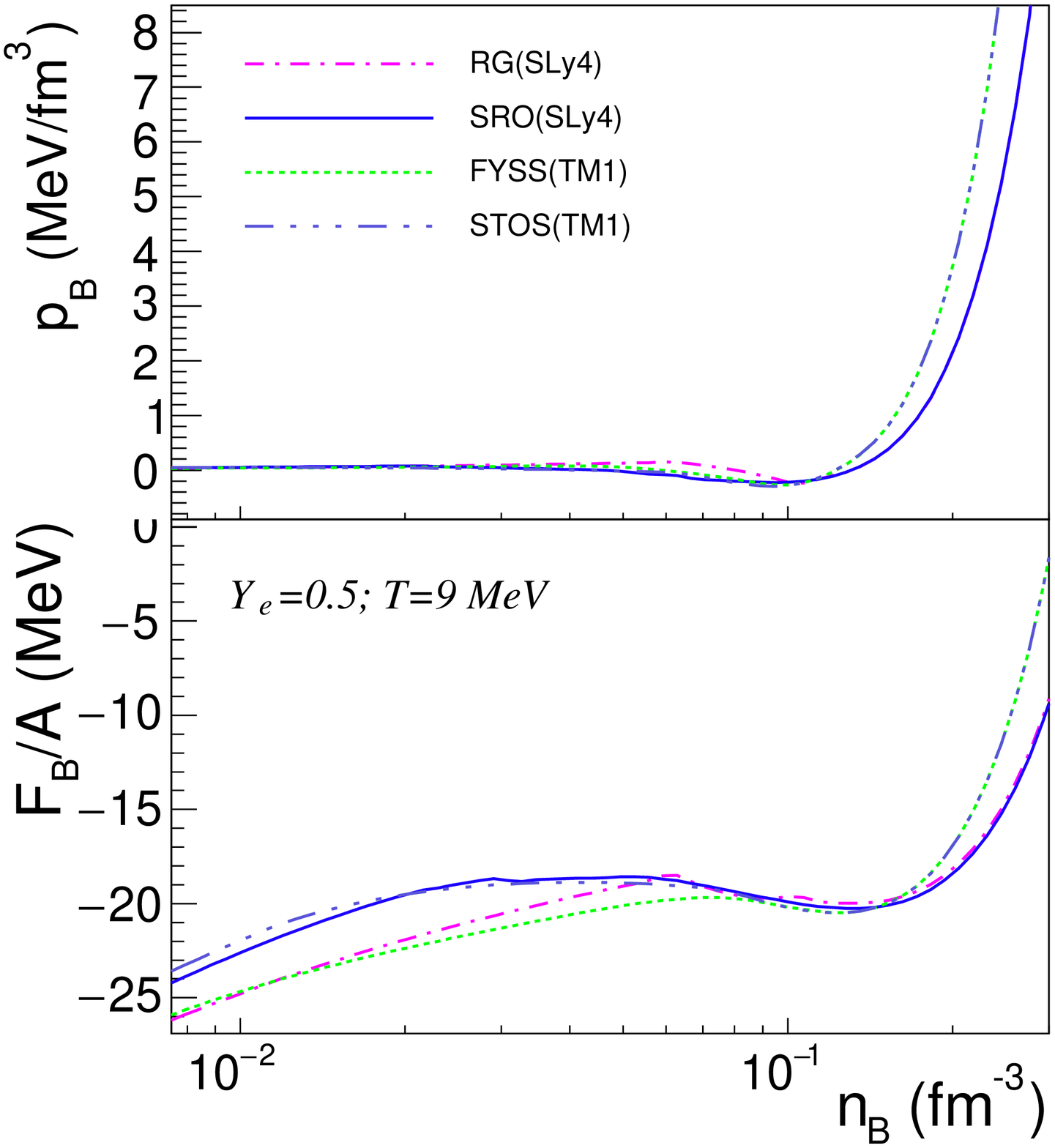}
  \end{center}
  \caption{Baryonic contribution to the free energy per nucleon
    and pressure as function of
    baryon number density for $T=1$, 5, 9 MeV and $Y_e=0.5$
    for models which employ the same effective interaction but
    adopt different models for inhomogeneous matter at sub-saturation
    densities.
    SRO(SLy4) and STOS(TM1) relay on SNA while RG(SLy4) and FYSS(TM1)
    employ extended NSE.
    }
    \label{fig:ep_SLy4TM1_nB} 
\end{figure*}

\begin{figure*}
  \begin{center}
    \includegraphics[width=0.3\textwidth]{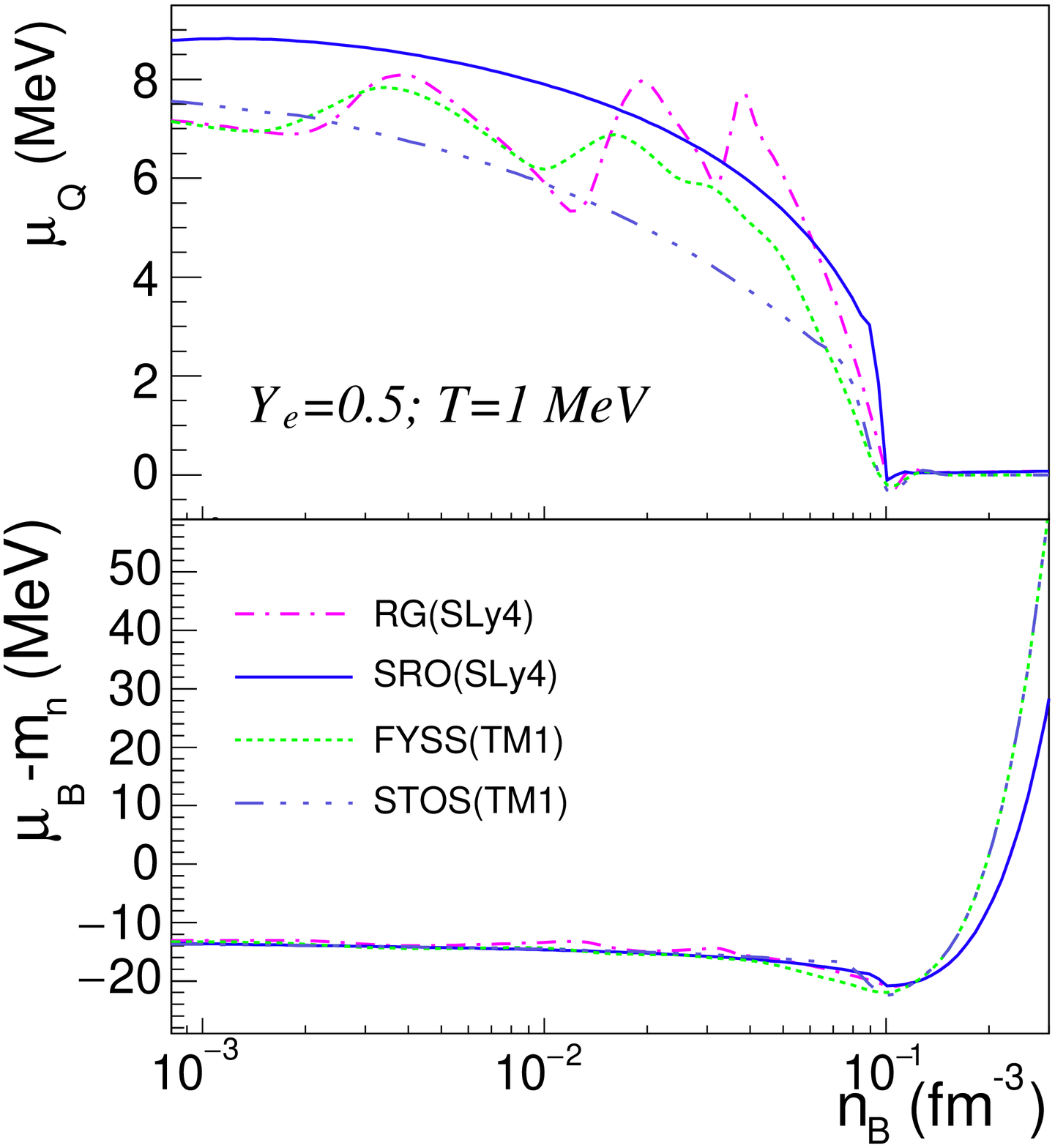}
    \includegraphics[width=0.3\textwidth]{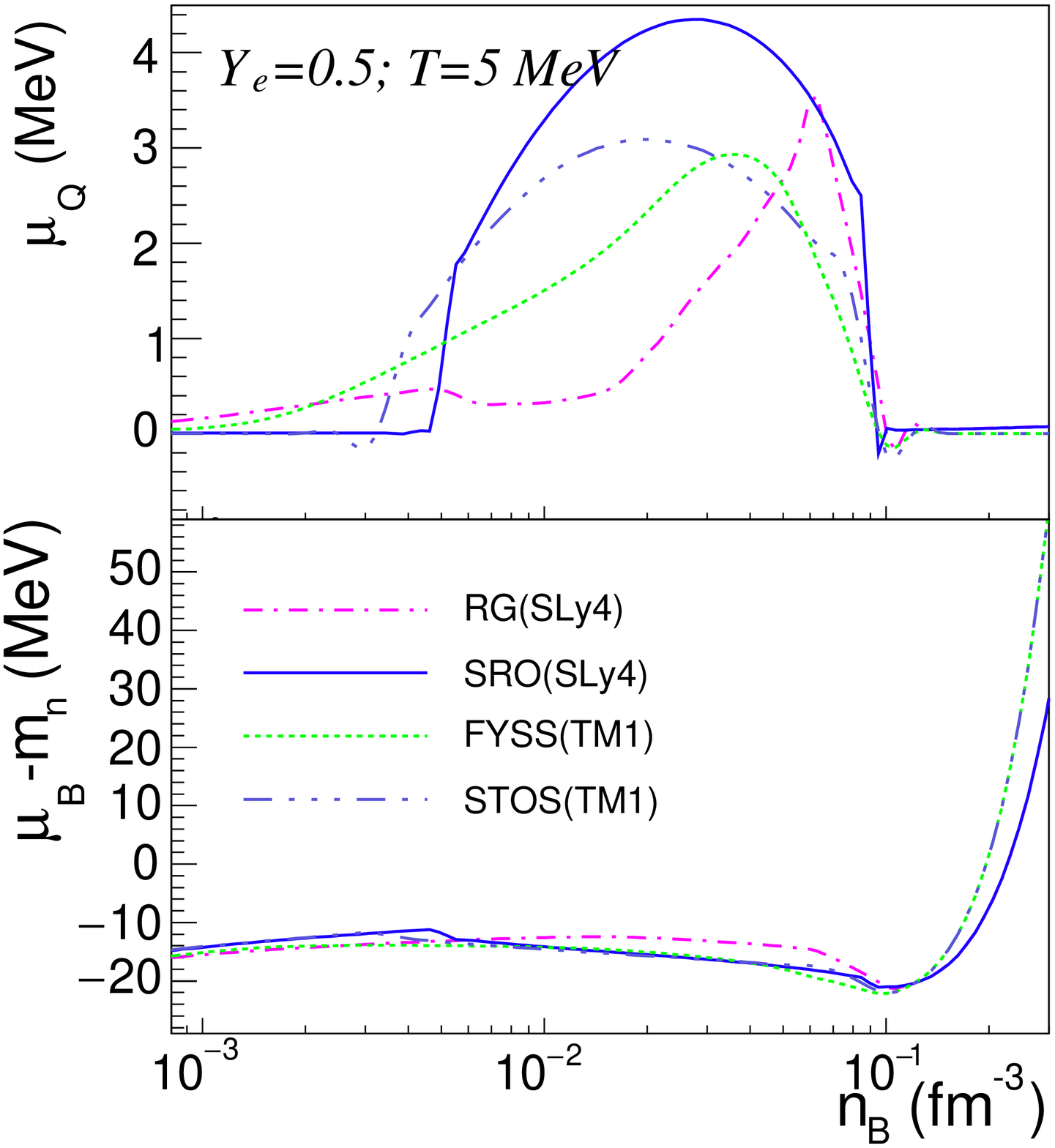}
    \includegraphics[width=0.3\textwidth]{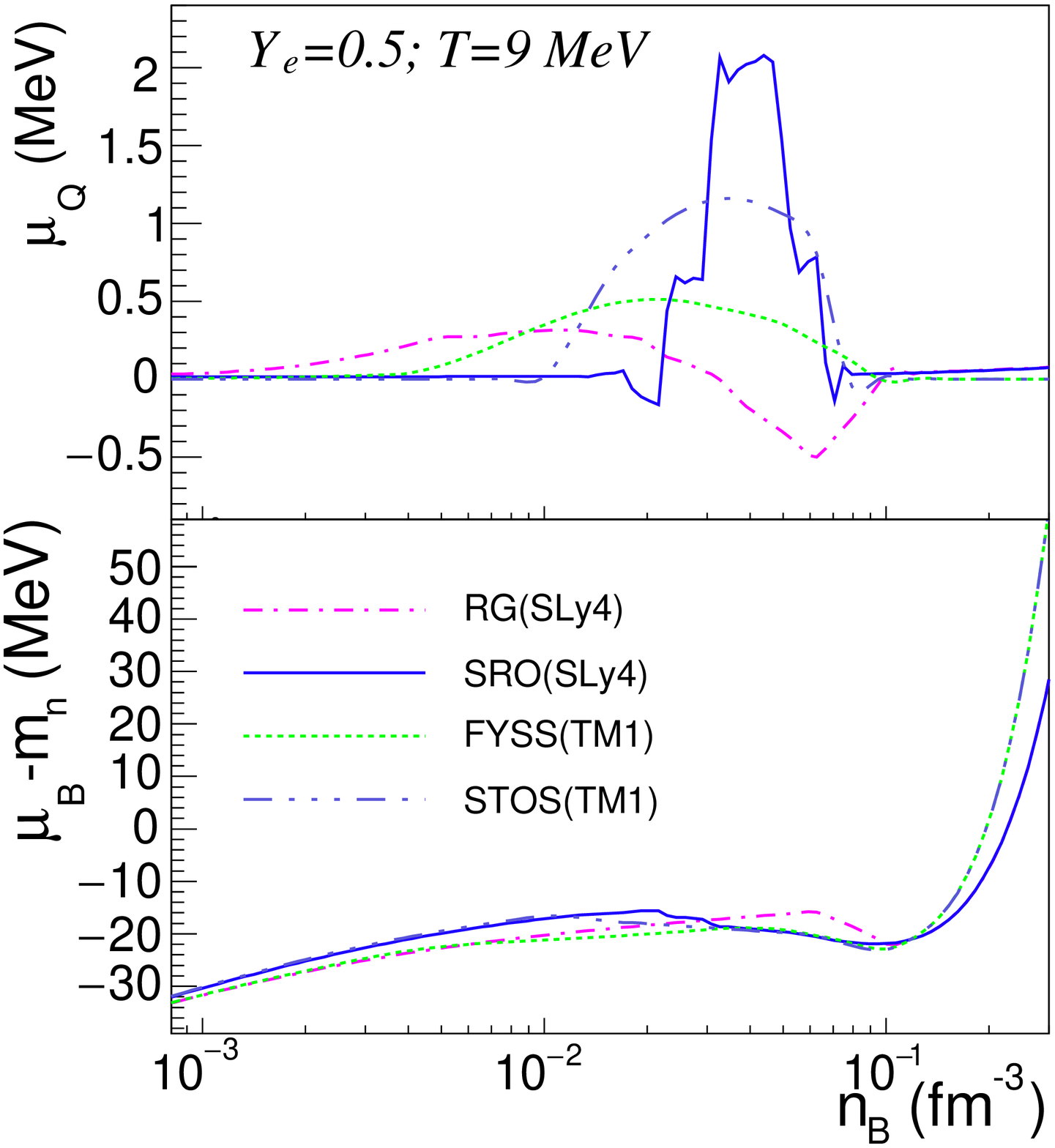}
  \end{center}
  \caption{The same as in Fig. \ref{fig:ep_SLy4TM1_nB} for baryon
    $\mu_B$ and charge chemical potentials $\mu_Q=\mu_p-\mu_n$.
    In the latter case the contribution of neutron and proton rest masses
    is not included in the corresponding chemical potentials.}
\label{fig:mu_SLy4TM1_nB} 
\end{figure*}

Despite significant effects on composition, there is consensus that
the details of the nuclear distribution at subsaturation densities has
only a small impact on the thermodynamics of the system
\cite{Lattimer_NPA_1985,Burrows_ApJ_1984,SRO_PRC_2017,Raduta_NPA_2019}.

In the following discussion we shall investigate to which 
extent this assertion is true.  We shall first confront results
corresponding to models based on the same effective interaction but
relying either on SNA or NSE.  The consequences of accounting for
the cluster's in-medium surface energy modification and particle
correlations in the continuum will be further analyzed by comparing
the results of HS(DD2) with those of GRDF1(DD2) and GRDF2(DD2).
Thermodynamic conditions, in particular the value of $Y_e$, are
chosen such as to maximize the effects.

The results of extended NSE models versus SNA approximations are
addressed in Figs. \ref{fig:ep_SLy4TM1_nB} and
\ref{fig:mu_SLy4TM1_nB}.  The models RG(SLy4) and FYSS(TM1) rely on
extended NSE while the models SRO(SLy4) and STOS(TM1) employ the SNA
approximation.  At the lowest considered temperature, $T=1$ MeV,
where the nuclear distribution is well described by a unique nucleus
embedded in a see of unbound nucleons, the discrepancies between SNA
and NSE originate from differences in the employed nuclear binding
energies and are small for all considered quantities.  The minimum
in the free energy density per nucleon as function of $n_B$
occurring at $n_{sat}/2$ within RG(SLy4) is probably an artifact of
the way in which the transition between clusterized and homogeneous
matter was realized.  RG(SLy4), which implements the experimental
mass table in \cite{Audi_2012}, and FYSS(TM1), which accounts for
shell effects, provide oscillatory behavior of the charge chemical
potential $\mu_Q$ as function of density, while the two SNA models
which employ a liquid-drop approximation (SRO(SLy4)) or the extended
Thomas-Fermi approximation (STOS(TM1)) lead to a smooth behavior.
In medium-modifications of the cluster surface energy in FYSS(TM1)
explain the damping of these oscillations as the density increases.
At $T=5$ and 9 MeV a significant dispersion is obtained between the
predictions of various models for all quantities except the baryonic
pressure.  At low densities, NSE-based models provide lower values
of the baryonic free energy density than their SNA counterparts.
Because of the chosen electron fraction value, $Y_e=0.5$, the charge
chemical potential $\mu_Q$ is small over the whole density range. At
low baryon number densities as well as for $n_B>n_t$, where unbound
nucleons are dominant or the only constituents of matter, $\mu_Q=0$.
Over the density domain where a significant amount of matter is
bound in clusters, $\mu_Q$ has non vanishing values. Most often
$\mu_Q>0$, meaning that the gas of unbound nucleons is slightly
proton rich. This is due to the slight neutron enrichment of nuclear
clusters. Among the considered cases, for $T=9$ MeV and $3 \cdot
10^{-2}~{\rm fm}^{-3} \lesssim n_B \lesssim n_t$ RG(SLy4) provides
$\mu_Q<0$.  This means that, on the contrary, clusterized matter is
proton rich and the gas neutron rich.  It turns out that for both,
chemical potentials and free energy, the results depend more on 
cluster modelling than on the effective interaction.  For all
considered conditions pressure is found to be negligibly affected by
modelling.  

\begin{figure*}
  \begin{center}
    \includegraphics[width=0.3\textwidth]{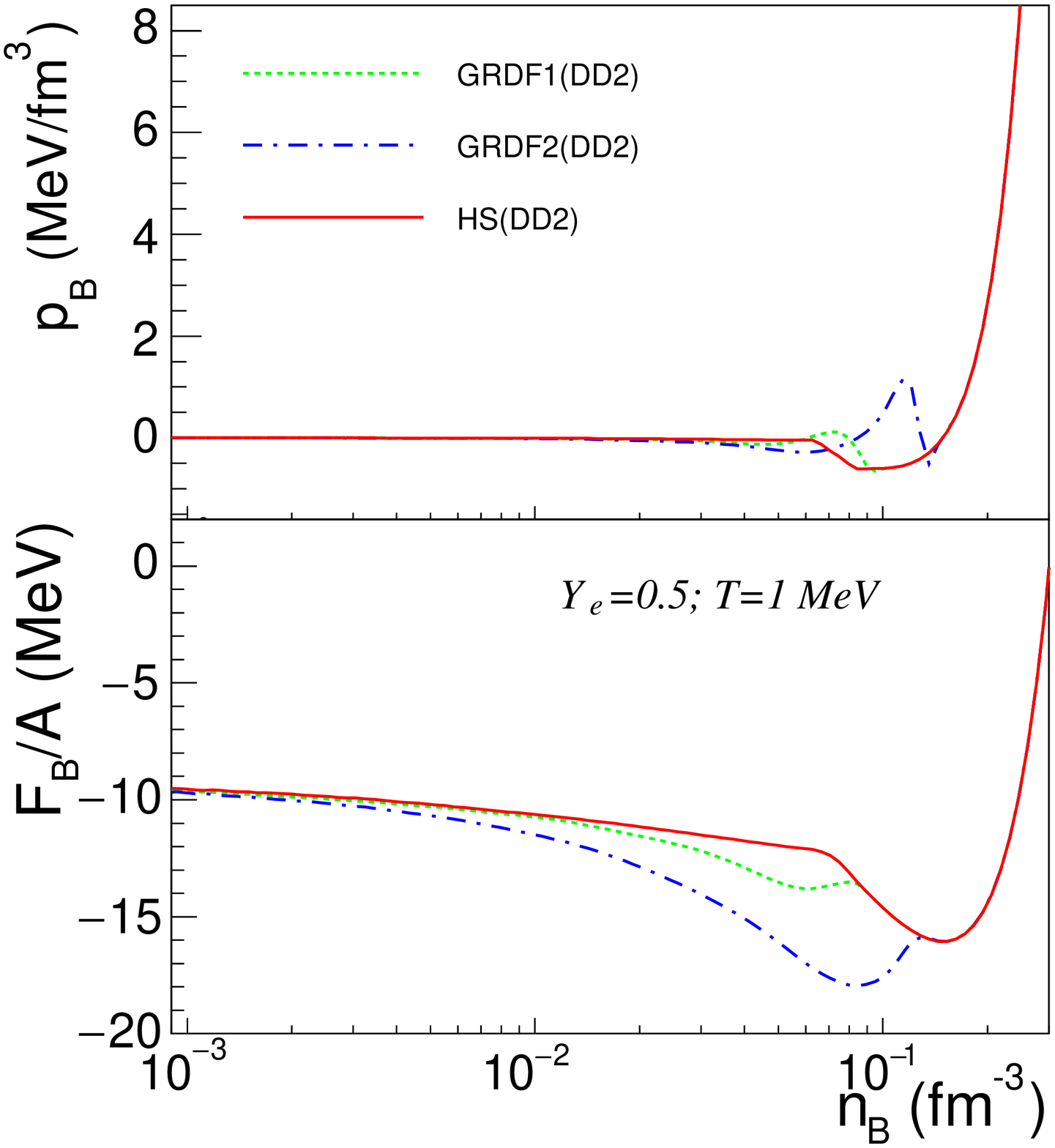}
    \includegraphics[width=0.3\textwidth]{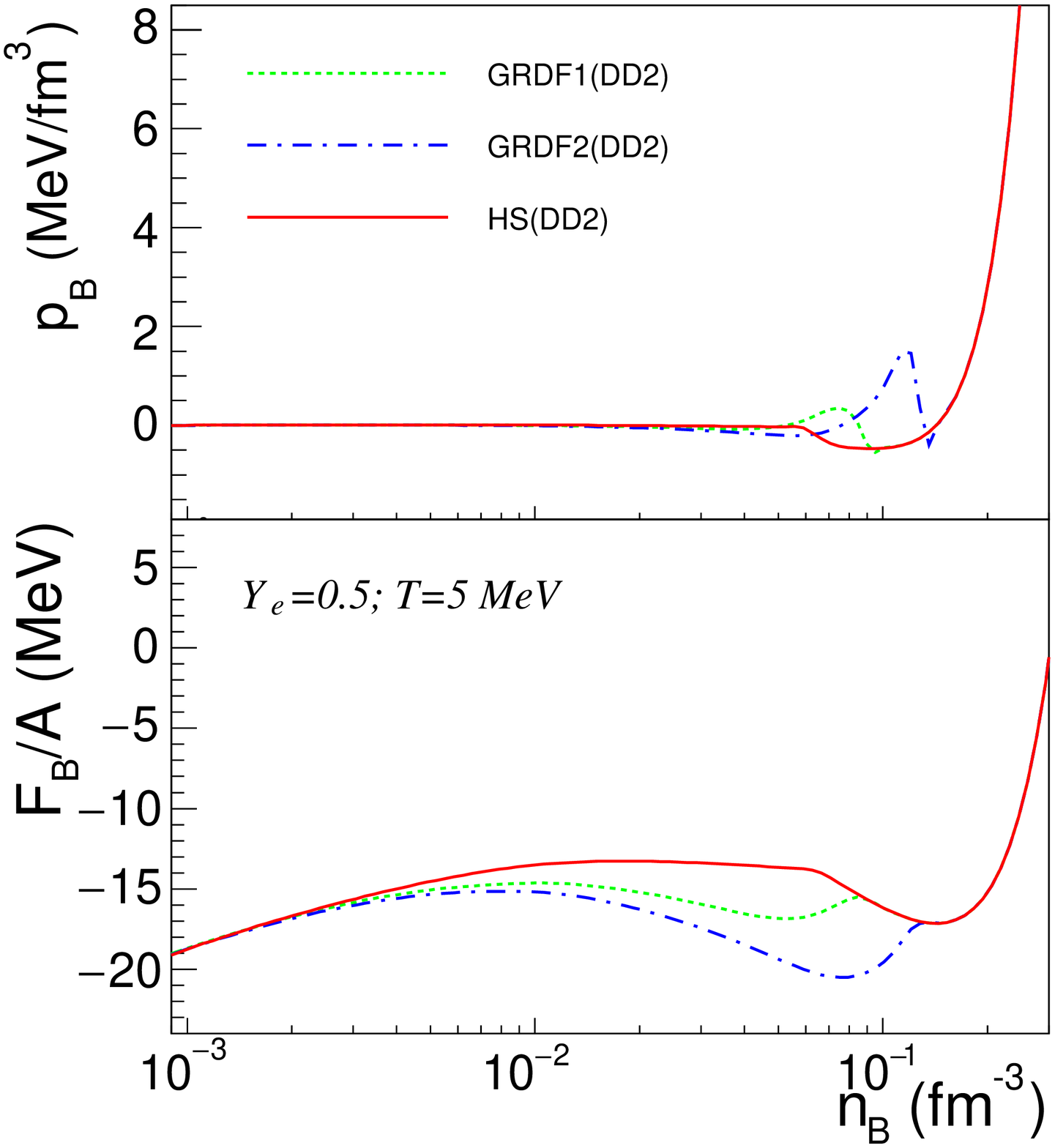}
    \includegraphics[width=0.3\textwidth]{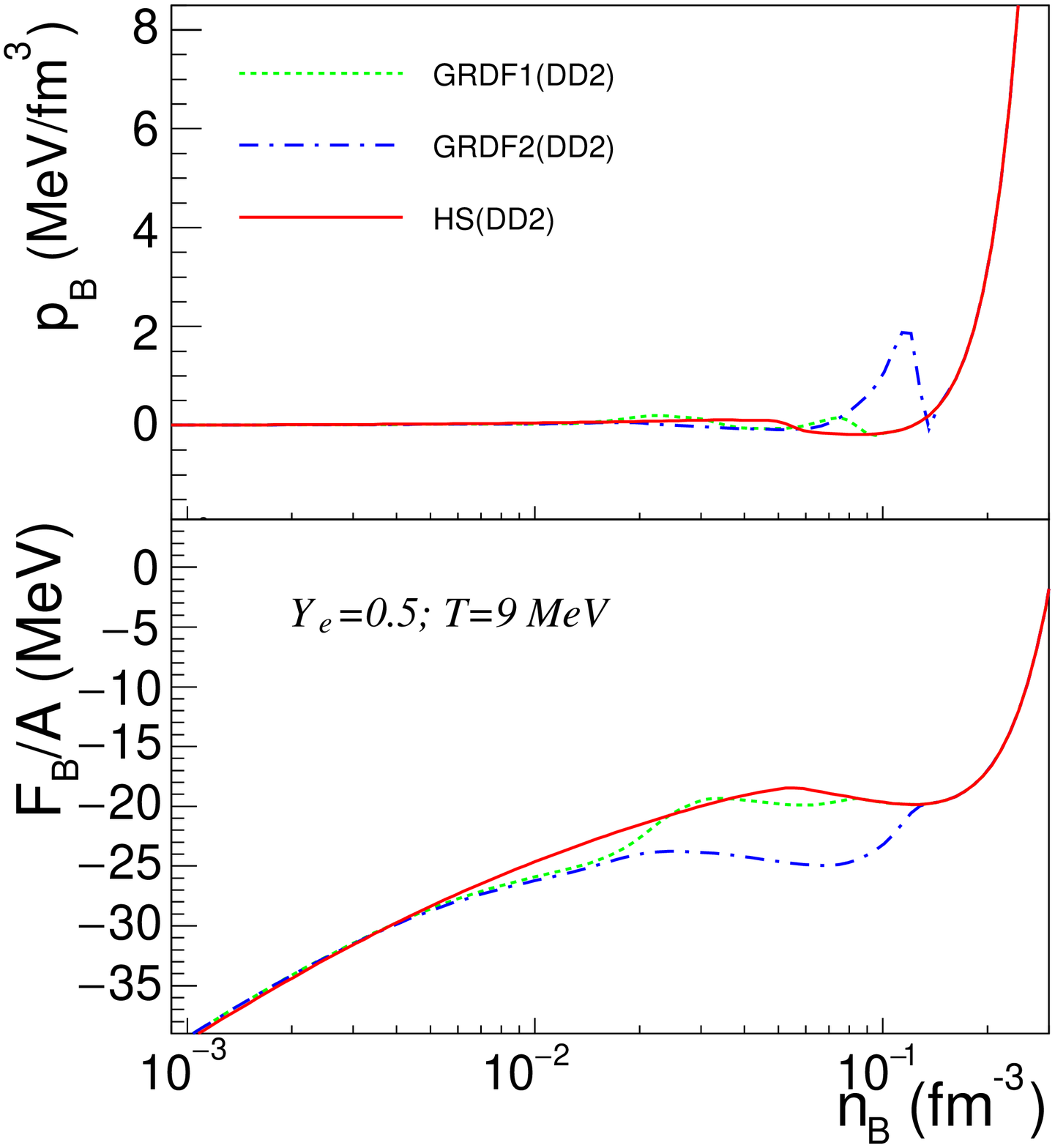}
  \end{center}
  \caption{The same as in Fig. \ref{fig:ep_SLy4TM1_nB} but for
    models which use different cluster energy functionals and,
    in the case of GRDF models, additionally account
    for particle correlations in the continuum.
    All considered models employ extended NSE and the DD2 \cite{DD2}
    effective interaction.
    }
    \label{fig:ep_DD2_nB} 
\end{figure*}

\begin{figure*}
  \begin{center}
    \includegraphics[width=0.3\textwidth]{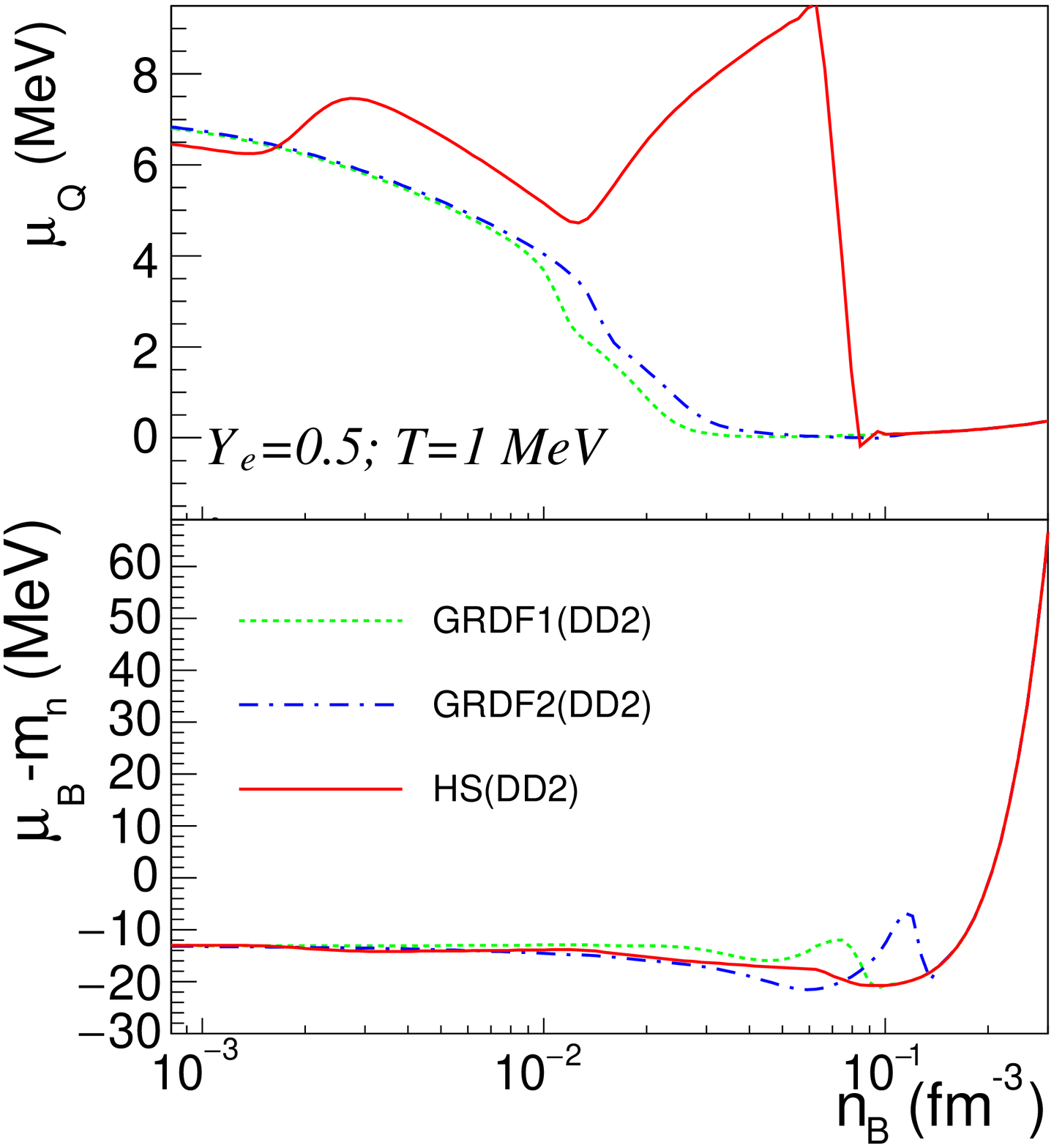}
    \includegraphics[width=0.3\textwidth]{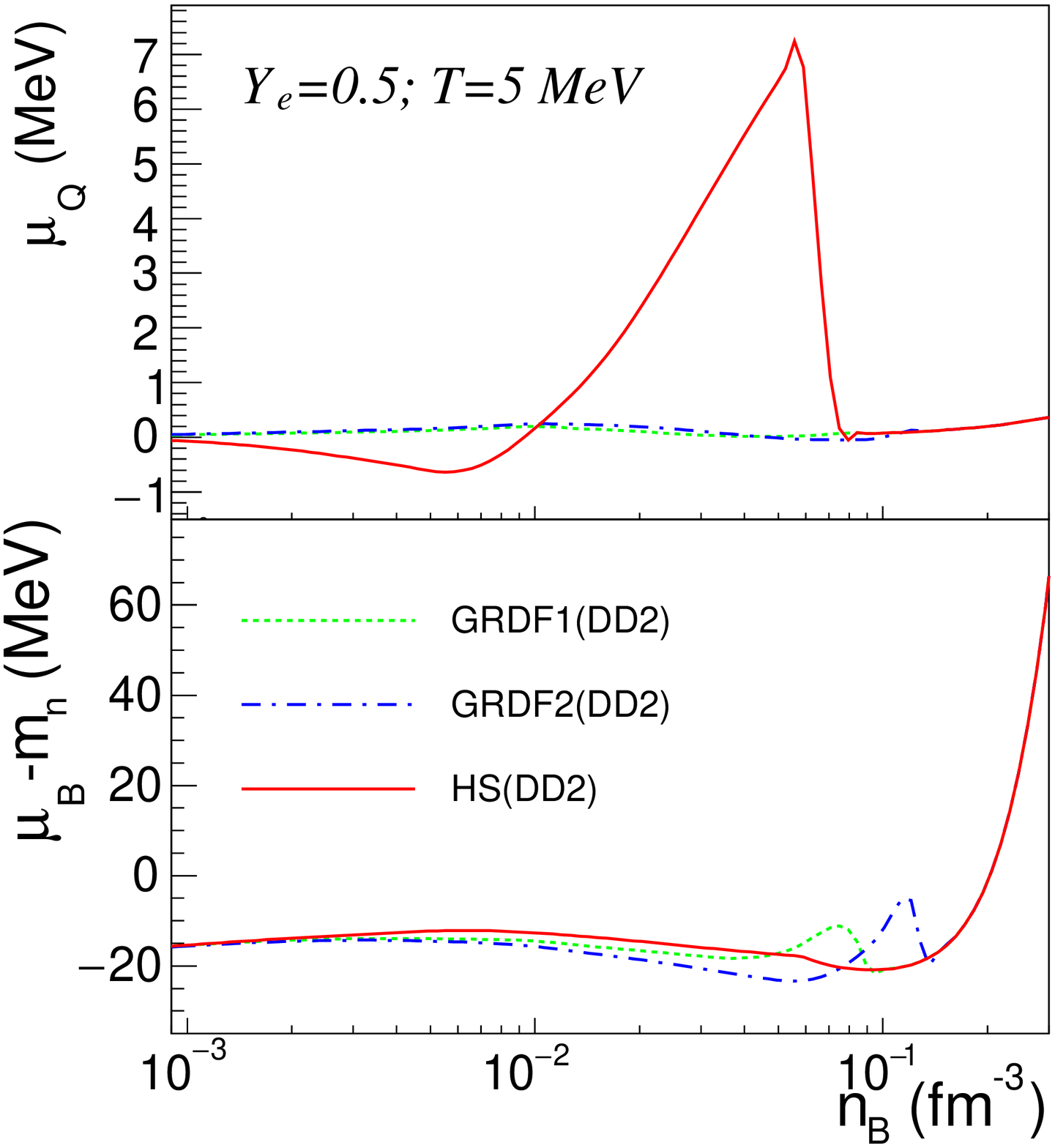}
    \includegraphics[width=0.3\textwidth]{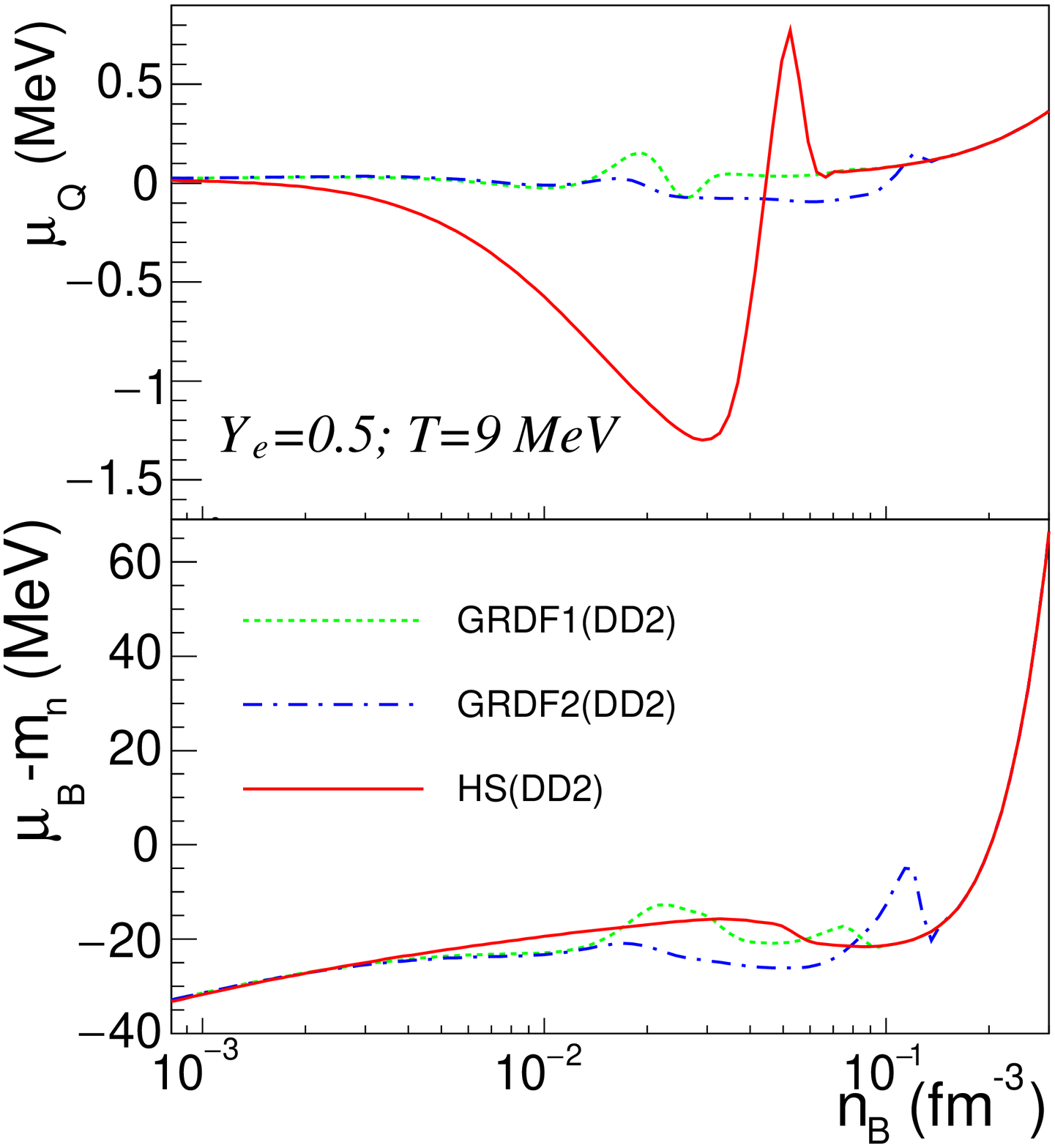}
  \end{center}
  \caption{The same as in Fig. \ref{fig:ep_DD2_nB} for baryon
    $\mu_B$ and charge $\mu_Q=\mu_p-\mu_n$ chemical potentials.
    In the latter case the contribution of neutron and proton rest masses
    is not included in the corresponding chemical potentials.}
\label{fig:mu_DD2_nB} 
\end{figure*}

Fig. \ref{fig:ep_DD2_nB} and \ref{fig:mu_DD2_nB} investigate the
modifications brought by cluster surface energy modifications and
particle correlations in the continuum by comparing the results of
HS(DD2), GRDF1(DD2) and GRDF2(DD2).  The same quantities and
thermodynamic conditions as previously discussed are considered.
For all thermodynamic quantities and conditions the effects are
stronger than those in Figs. \ref{fig:ep_SLy4TM1_nB} and
\ref{fig:mu_SLy4TM1_nB}.  Under specific model-dependent conditions
the most stable state predicted by GRDF models corresponds to $n_B
\approx n_{sat}/2$.  This situation is similar to what has been
previously observed for $T=0$ in Fig. \ref{fig:EperA_SNM_T=0}.
Similarly to RG(SLy4), the charge chemical potential in HS(DD2) has
a complex structure at low temperatures and negative values at high
temperatures.  The former feature is due to shell effects in the
binding energies while the second is due to the slight neutron enrichment
of the gas of unbound nucleons.  Information on matter composition
for these models is provided in Appendix A, see
Fig. \ref{fig:compo_T=5_DD2}. Given that, under all circumstances
considered here, particle resonances are sub-dominant deviations
between GRDF and HS can only be attributed to the 
cluster's energy functional, for which every model assumes
a different density-dependence. 

At least for SNA and NSE models which
disregard in-medium and temperature effects on
clusters energy functionals
the gross of the dynamics neither of CCSN, PNS evolution nor BNS
mergers is affected by modelling.
There is nevertheless some influence on
particular phenomena, such as those pointed out for instance for the
evolution of the shock radius, the accretion rate and the properties
of matter above the shock in
CCSN~\cite{SHF_ApJ_2013,SRO_PRC_2017}. The main influence of the
nuclear distribution is indirect via the nuclear structure dependence
of the weak interaction rates. Among others, it determines the EC
capture rate during infall in a CCSN and has thus a strong impact on
bounce properties
~\cite{Hix_PRL_2003,Sullivan_ApJ_2016,Pascal_PRC_2020} and early
post-bounce (anti)-neutrino luminosities and
spectra~\cite{Fischer_PASA_2017}.
Light and neutron-rich nuclei were also shown to play important roles in the
production of seed nuclei and $r$-process elements in supernovae \cite{Terasawa_ApJ_2001}.
According to \cite{Arcones_PRC_2008} light nuclei in the outer layers of
PNS also impact the anti-neutrino opacity and anti-neutrino average energies.
At early moments the opacity is reduced, which results in larger average energies
of the anti-neutrinos;
at late moments the appearance of light nuclei has the opposite effect
and it is expected to modify the nucleosynthesis occurring in
neutrino-driven winds.



\section{Symmetry energy at finite temperature}
\label{sec:Esym_T}

  \begin{figure*}
    \begin{center}
      \includegraphics[width=0.79\textwidth]{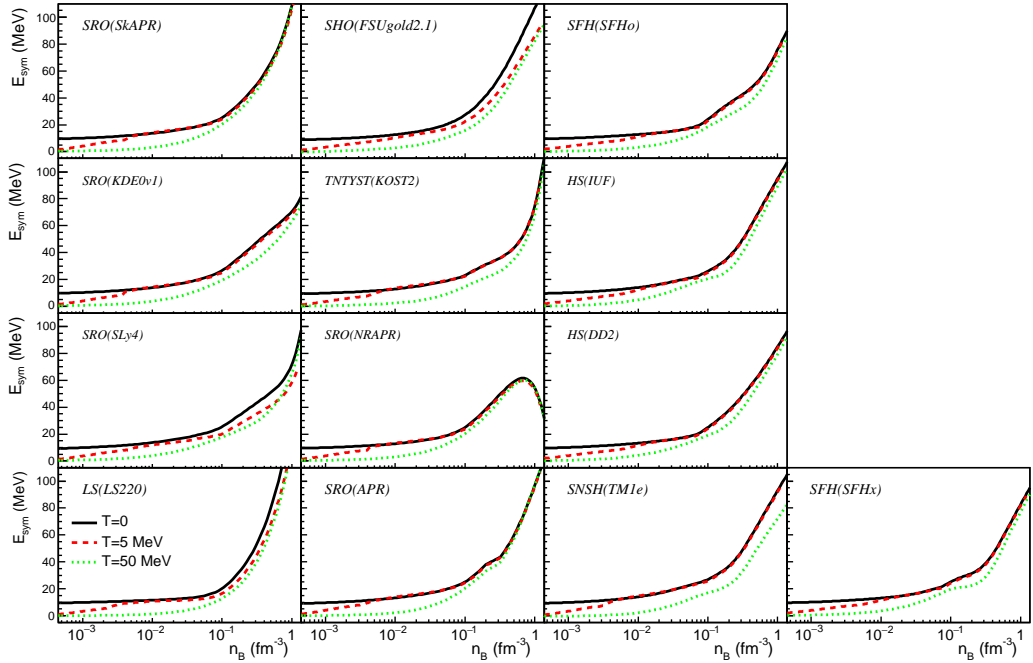}
     \end{center}
    \caption{ Symmetry energy per baryon, defined as
      $E_{sym}(n_B,T)=E_{NM}(n_B,T)/A-E_{SM}(n_B,T)/A$, as function of
      baryon number density comparing results at $T$=5 and 50 MeV with those
      at $T=0$.  The baryonic part of various EoS models is
      considered.  }
\label{fig:Esym_T} 
\end{figure*}

We have seen so far that finite temperature effects impact all
thermodynamic quantities and the nucleon effective masses play a key
role. This is expected to be the case of the symmetry energy, too.
Symmetry energy in homogeneous matter within non-relativistic
mean field models was thoroughly addressed in Ref. \cite{Ou_PLB_2011},
which highlighted a number of
correlations among the symmetry energy at finite temperature and the
density dependence of nucleonic Landau effective masses, including the
neutron-proton mass splitting.  The case of warm inhomogeneous
subsaturated matter was considered in Ref. \cite{Raduta_EPJA_2014},
who have shown that - similarly to what happens in cold matter (see
Sec. \ref{sec:T=0} ) - density dishomogeneities break the isospin
invariance of the interaction and the equivalence of the two
definitions of $E_{sym}$.  Ref. \cite{Typel_EPJA_2014} investigated
the modifications of the (free) symmetry energy due to cluster
formation, liquid-gas phase transition and Coulomb and electron
corrections for $T \leq T_C$.  Refs. \cite{Ou_PLB_2011} and
\cite{Raduta_EPJA_2014} employed non-relativistic Skyrme type models
for the nucleonic interactions; Ref. \cite{Typel_EPJA_2014} a
density-dependent CDFT model.  The widest density domain, $0<n_B \leq
1.5 n_{\mathit{sat}}$, was considered in \cite{Ou_PLB_2011}, for $5
\leq T \leq 20$ MeV.  In this section we shall address the temperature
dependence of the symmetry energy in stellar matter over wide
ranges of temperature and density considering EoS models belonging to
various classes.
  
Fig. \ref{fig:Esym_T} displays the symmetry energy as function of
baryon number density confronting the predictions of various EoS
models at $T=5$ and 50 MeV with results at $T=0$.  Two features are
worth to be noticed.  First, at variance with what happens in 
cold matter, which is inhomogeneous at sub-saturation
densities (see Sec. \ref{ssec:Esym_T=0}), sufficiently hot matter is
homogeneous at low densities, thus $\lim_{n_B \to 0} E_{sym} \to
0$~\cite{Raduta_EPJA_2014}.  This implies that, for $n_B<n_t$, the
symmetry energy diminishes as $T$ increases.  Second, the
$T$-dependence of $E_{sym}$ at high densities is not necessarily
monotonic, see for instance SRO(SLy4) which, for $n_B > 0.6~{\rm
  fm}^{-3}$, predicts $E_{sym}(T=0)>E_{sym}(T=50~{\rm
  MeV})>E_{sym}(T=5~{\rm MeV})$.  Overall, the magnitude of thermal
effects depends on density, temperature and EoS model, in agreement
with previous conclusions in \cite{Ou_PLB_2011}.

The discussion above suggests that is not straightforward to establish
correlations between the parameters characterizing the symmetry energy
in cold symmetric nuclear matter and the evolution of environments
where hot dense matter is populated. Naturally, situations
dominated by the high density part of the EoS should be more
favorable to find such correlations. Indeed, in CCSN simulations as
a consequence of the slower PNS contraction in EoS models with large
symmetry energies, lower average neutrino energies are
obtained~\cite{SHF_ApJ_2013,Schneider_PRC_2019b}.
In addition, convection in the new born PNS is sensitive to the
symmetry energy with faster stabilisation for lower symmetry
energies~\cite{Roberts2012,Nakazato2019,PascalThese}. Loose correlations between the
symmetry energy and bounce properties have been found, too~\cite{SHF_ApJ_2013,Schneider_PRC_2019b},
but it should be stressed that
the uncertainty on weak interaction
rates from electron capture on nuclei during infall is the dominant source of uncertainty in
bounce properties~\cite{Sullivan_ApJ_2016,Pascal_PRC_2020}.


Ref. \cite{Most_2021} investigates the possible signatures of
$L_{sym}$ in BNS mergers. They find that while the properties of the
merger remnant manifest a strong sensitivity on the EoS model, GW
emission and post-merger dynamics do not show any clear correlation
with $L_{sym}$, whereas large values of $L_{sym}$ lead to
systematically enhanced ejecta, which would result in increased EM
emission.

\section{Properties of hot compact stars and the role of EoS}
\label{sec:HotStars}

\begin{table*}
  \caption{Properties of the maximum mass configuration of spherically-symmetric
    non-rotating PNS with constant profiles of ($S/A$, $Y_e$).
    Predictions of different EoS models. 
    Listed are central baryon number density and temperature; baryonic and gravitational masses; radius.}
\label{tab:HotStars}       
\begin{tabular}{lccccccccccccccc}
  \noalign{\smallskip}\hline
  model &  \multicolumn{5}{c}{$S/A=1$, $Y_e=0.4$} & \multicolumn{5}{c}{$S/A=2$, $Y_e=0.2$} & \multicolumn{5}{c}{$S/A=1$, $Y_e=0.1$} \\
  & $n_{B;c}$ & $T_c$ & $M_B$ & $M_G$ & $R$ & $n_{B;c}$ & $T_c$ & $M_B$ & $M_G$ & $R$ & $n_{B;c}$ & $T_c$ & $M_B$ & $M_{G}$ & $R$ \\
  & [fm$^{-3}$] & [MeV] & [$M_{\odot}$] & [$M_{\odot}$] & [km] & [fm$^{-3}$] & [MeV]  & [$M_{\odot}$]& [$M_{\odot}$]& [km] & [fm$^{-3}$] & [MeV]  & [$M_{\odot}$] & [$M_{\odot}$] & [km] \\
  \noalign{\smallskip}\hline
  LS(LS220)       & 1.17 & 23.5 & 2.16 & 1.95 & 10.9 & 1.09 &  60.1 & 2.33 & 2.04 & 11.1 & 1.10 & 30.4 & 2.44 & 2.08 & 10.7   \\
  SRO(SLy4)       & 1.11 & 57.1 & 2.26 & 2.02 & 11.1 & 0.97 & 125.2 & 2.45 & 2.16 & 11.7 & 1.11 & 92.2 & 2.47 & 2.09 & 10.4  \\
  SRO(KDE0v1)     & 1.17 & 53.9 & 2.19 & 1.96 & 10.9 & 1.03 & 120.9 & 2.37 & 2.09 & 11.4 & 1.20 & 88.0 & 2.36 & 2.01 & 10.2 \\
  SRO(NRAPR)      & 1.13 & 57.7 & 2.23 & 2.00 & 11.1 & 1.01 & 119.7 & 2.38 & 2.10 & 11.6 & 1.20 & 81.7 & 2.32 & 1.99 & 10.2 \\
  SRO(SkAPR)      & 1.03 & 52.8 & 2.22 & 2.00 & 11.9 & 0.91 & 109.9 & 2.47 & 2.17 & 12.2 & 1.04 & 70.5 & 2.44 & 2.09 & 11.0  \\
  SRO(APR)        & 1.16 & 49.3 & 2.42 & 2.13 & 10.4 & 1.05 & 108.7 & 2.54 & 2.20 & 10.9 & 1.12 & 62.7 & 2.65 & 2.20 & 10.2 \\
  TNTYST(KOST2)   & 1.09 & 41.9 & 2.43 & 2.14 & 10.9 & 1.02 &  99.8 & 2.58 & 2.23 & 11.0 & 1.09 & 60.3 & 2.69 & 2.23 & 10.3 \\
  SHO(FSUgold2.1) & 0.89 & 31.2 & 2.30 & 2.08 & 12.9 & 0.81 &  70.6 & 2.44 & 2.17 & 13.1 & 0.88 & 40.4 & 2.48 & 2.14 & 12.0  \\
  SNSH(TM1e)      & 0.86 & 31.7 & 2.33 & 2.10 & 13.0 & 0.79 &  74.0 & 2.49 & 2.20 & 13.2 & 0.87 & 41.8 & 2.52 & 2.16 & 12.1   \\
  HS(IUF)         & 0.96 & 33.0 & 2.12 & 1.93 & 12.5 & 0.89 &  75.4 & 2.26 & 2.02 & 12.7 & 0.98 & 43.5 & 2.28 & 1.98 & 11.5 \\
  HS(DD2)         & 0.83 & 31.3 & 2.69 & 2.37 & 12.7 & 0.79 &  72.2 & 2.79 & 2.42 & 12.8 & 0.84 & 41.1 & 2.89 & 2.42 & 12.0  \\
  SFH(SFHo)       & 1.09 & 32.9 & 2.25 & 2.01 & 11.3 & 1.04 &  75.9 & 2.36 & 2.08 & 11.3 & 1.13 & 43.4 & 2.43 & 2.06 & 10.4  \\
  SFH(SFHx)       & 1.04 & 33.7 & 2.32 & 2.07 & 11.6 & 0.96 &  76.6 & 2.45 & 2.15 & 11.7 & 1.03 & 43.8 & 2.53 & 2.14 & 10.9 \\
\noalign{\smallskip}\hline
\end{tabular}
\end{table*}

\begin{figure*}
  \begin{center}
    \includegraphics[width=0.33\linewidth]{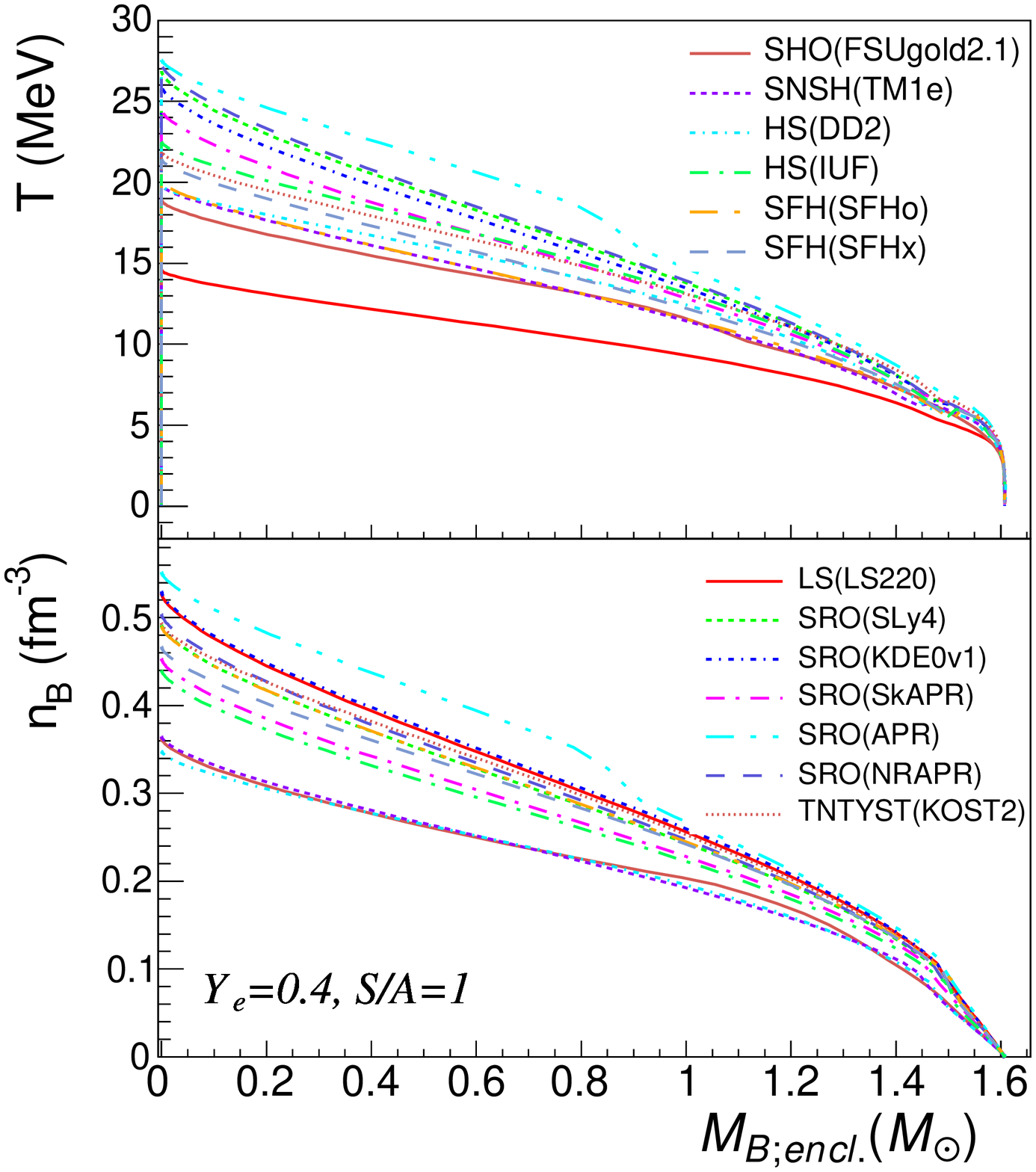}
    \includegraphics[width=0.33\linewidth]{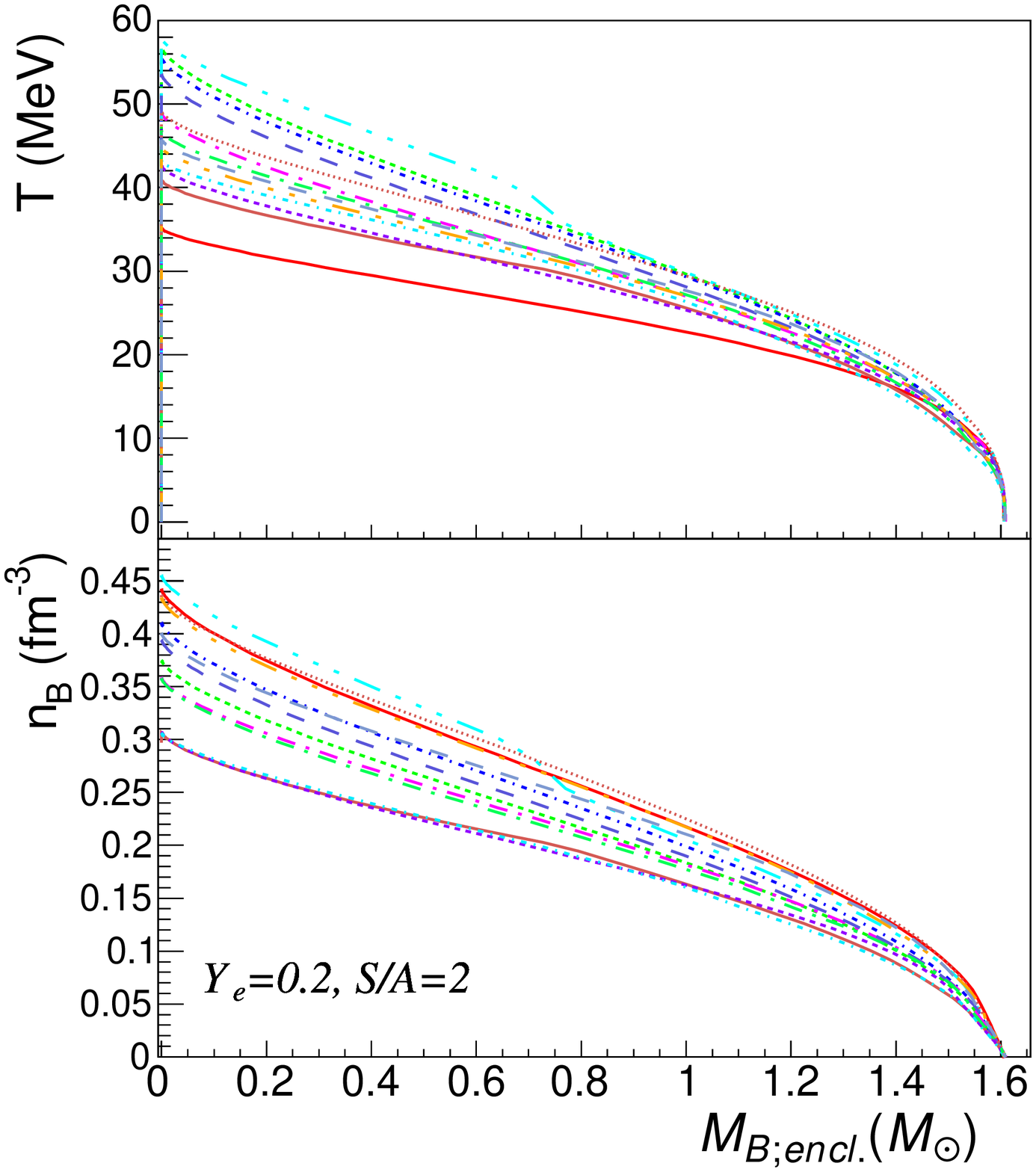}
    \includegraphics[width=0.33\linewidth]{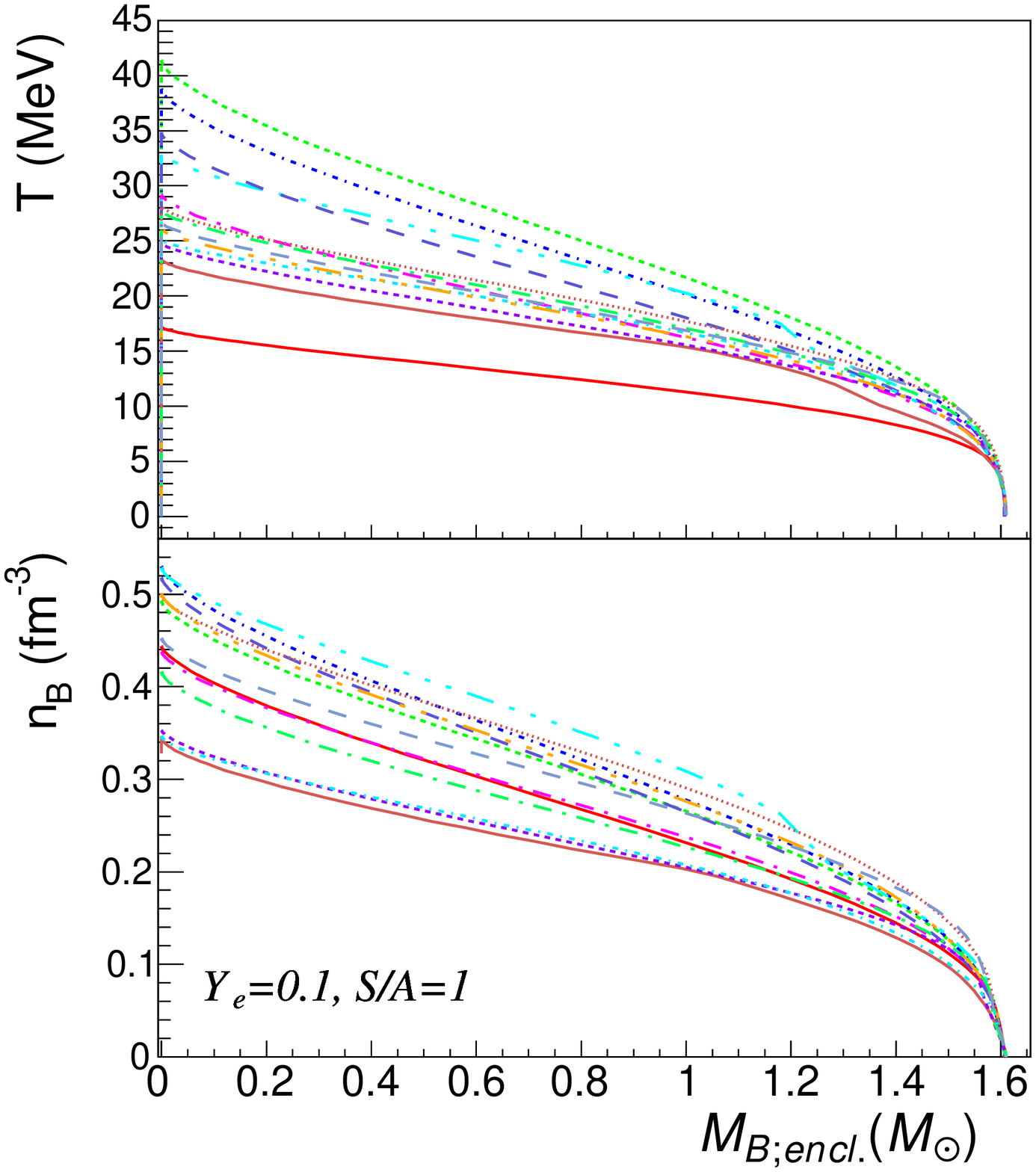}
  \end{center}
   \caption{Radial profiles of baryonic particle number density and temperature
    in spherically-symmetric non-rotating PNS with $M_B=1.6M_{\odot}$
    within different EoS models.
    The assumed constant profiles of ($S/A$,$Y_e$) are representative
    for different instances in the evolution, see the text.
  }
 \label{fig:Prof}
\end{figure*}

\begin{figure}
  \begin{center}
    \includegraphics[width=0.69\columnwidth]{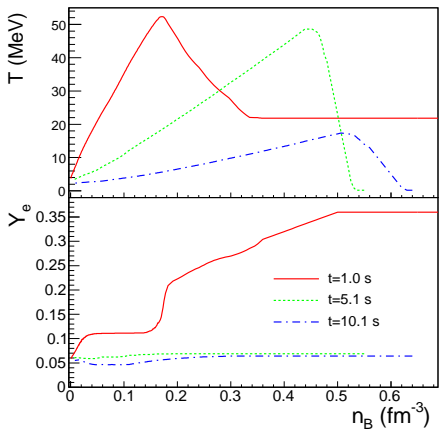}
    \end{center}
  \caption{$T(n_B)$ and $Y_e(n_B)$
    relations used for the computation of hot star models at different instances
    in PNS evolution,
    as obtained by \cite{PascalThese} for a star with a baryon mass of
    $M_B = 1.6 M_\odot$ with the RG(SLy4) EoS.
  }
\label{fig:PascalProfile} 
\end{figure}

\begin{figure}
  \begin{center}
    \includegraphics[width=0.59\columnwidth]{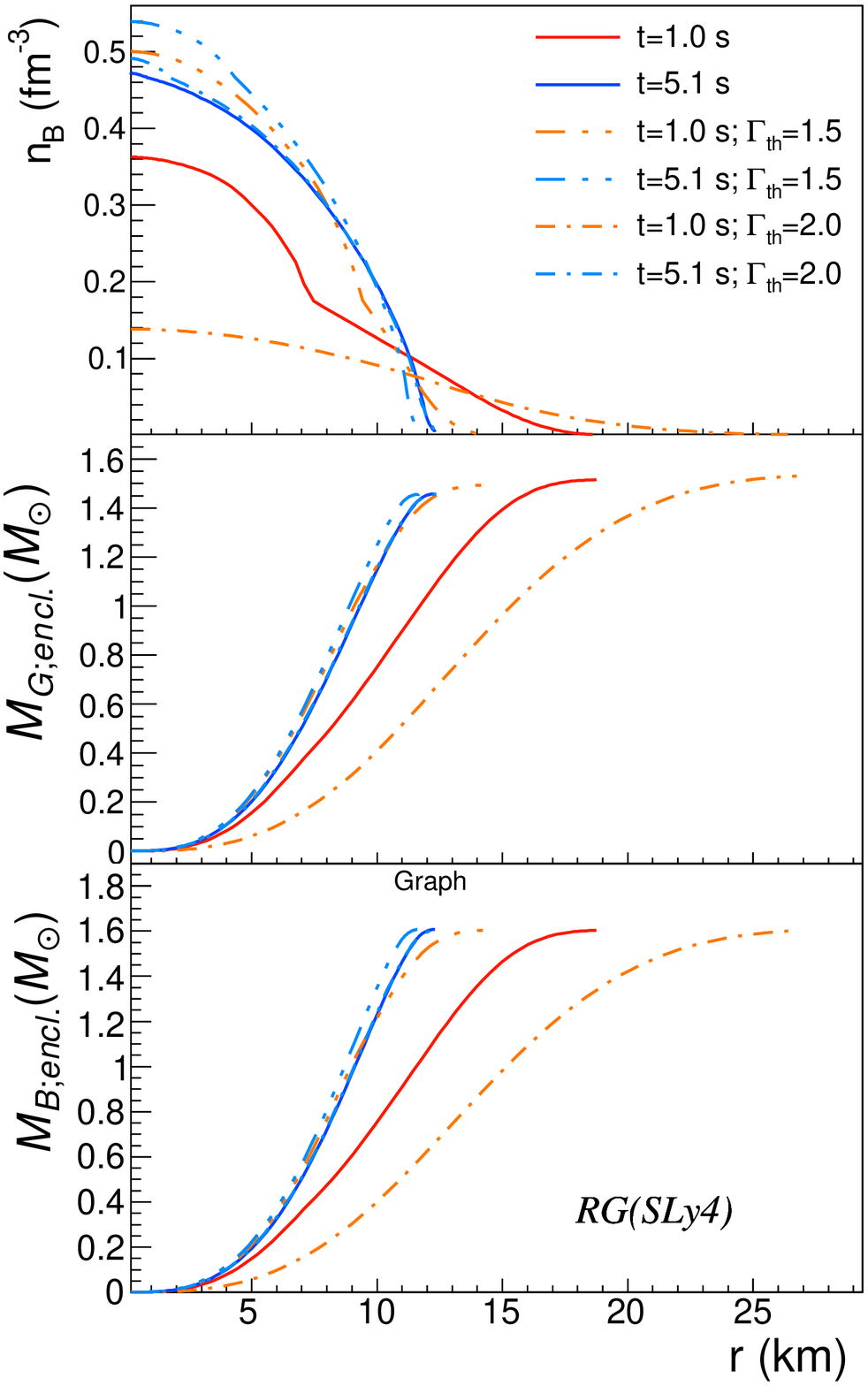}
    \end{center}
  \caption{Radial profiles of enclosed baryonic and gravitational masses
    and particle number density for a PNS with $M_B=1.6M_{\odot}$ at
    $t=1.0$ and 5.1 s.
    Exact results corresponding to RG(SLy4)
    are confronted with those obtained by the $\Gamma$-law,
    eq. (\ref{eq:GammaEoS}), for $\Gamma_{\mathit{th}}=1.5$ and 2.0.
}
\label{fig:ProfGamma} 
\end{figure}

Though significant dependence on the dense matter EoS and particle degrees
for freedom, in addition to initial conditions and weak interaction rates,
was put forward for PNS evolution since decades~\cite{Pons_ApJ_1999,Prakash_1997},
only a limited pool of EoS models has yet been considered within full numerical
simulations which require a combined solution of hydrodynamics equations and neutrino
transport, see
e.g.~\cite{Prakash_1997,Pons_ApJ_1999,Villain_AA_2004,Roberts2012code,Nakazato2019,PascalThese}.
The situation is rather similar for studies of the
post-merger remnant of a BNS merger, where full-fledged hydrodynamics
simulations are computationally very expensive. Many studies of the
EoS and composition dependence therefore focus on stationary solutions
of hot compact stars which among others can give hints on maximally
supported masses and internal structure, see
e.g.~\cite{Sumiyoshi_AASS_1999,Strobel_AA_1999,Nicotra_PRD_2006,Li_Catania_PRC_2010,Burgio_AA_2010,Martinon_PRD_2014,Camelio_PRD_2016,Marques_PRC_2017,Lu_Catania_PRC_2019,Roark_MNRAS_2019,Lenka_JPG_2019,Nunna_ApJ_2020,Raduta_MNRAS_2020}.
Profiles of entropy per baryon (or temperature) and electron/lepton
fraction are thereby obtained either directly from numerical
simulations at different evolution times or taken as constant with
typical values.

As an application of the finite-temperature EoS discussed in the
previous sections, we pursue in this direction and discuss the EoS
dependence of the structure of a hot static star.
As an indication, we assume constant profiles of $S/A$ and $Y_e$.
Three sets of values will be considered: ($S/A=1$, $Y_e=0.4$),
($S/A=2$, $Y_e=0.2$) and ($S/A=1$, $Y_e=0.1$).
Although over the considered time range, neutrinos are still
trapped, we will not consider their contribution to
the thermodynamic quantities here. The latter contribution is small
and our qualitative discussion thus remains unaffected. Given the
above profiles, we obtain static configurations of hot stars by
solving the Tolman-Oppenheimer-Volkoff equations.

Obtained radial profiles of baryonic density and temperature, expressed in terms
of enclosed baryonic mass, are plotted in fig. \ref{fig:Prof} for
PNS with $M_B=1.6M_{\odot}$ built upon various EoS models.
Values of a few key global quantities that characterize the maximum mass configurations
of hot PNS are listed in Table \ref{tab:HotStars} for each EoS model
and assumed thermodynamic conditions.

The predictions of different models vary widely.  Temperature
  profiles reflect the density dependence of nucleon effective masses.
  LS220, for which the effective masses equal the bare masses and are
  thus higher than in all other models, provides the lowest
  temperatures. HS(DD2), SNSH(TM1e) and SHO(FSUgold2.1) provide
  similar particle number density profiles and, given that their
  effective masses are similar (see fig. \ref{fig:meff_T=0}), the
  temperature profiles are similar, too. For ($S/A=1$, $Y_e=0.1$)
  SRO(SLy4), SFHo and SRO(NRAPR) provide similar density profiles
  while $T({\rm SRO(SLy4)})>T({\rm NRAPR})>T({\rm SFHo})$;
  the first inequality can be explained by the fact that
  $m_{\mathit{eff}}({\rm NRAPR})>m_{\mathit{eff}}({\rm SRO(SLy4)})$.
  For fixed values of
  $S/A$, isospin symmetric matter has lower temperatures than isospin
  asymmetric matter \cite{Raduta_MNRAS_2020}.  Isospin asymmetric
  matter allows for more massive configurations, with smaller
  radii, too.  By comparing the configurations corresponding to ($S/A=1$,
  $Y_e=0.4$) and ($S/A=2$, $Y_e=0.2$) it comes out that during the
  early deleptonization phase, when PNS gets hotter, it expands and
  maximum baryonic and gravitational masses increase.  During the late
  deleptonization phase, see ($S/A=2$, $Y_e=0.2$) and ($S/A=1$,
  $Y_e=0.1$), PNS shrinks.
  
  The maximum baryon mass is an interesting quantity in the context of
  stability against collapse to a black hole during PNS and BNS merger
  evolution. In the absence of accretion, $M_B$ is a conserved
  quantity during evolution, such that if it exceeds the maximum
  baryon mass of the cold $\beta$-equilibrated configuration, the star
  necessarily becomes unstable against collapse to a black hole at
  some point independently of the mechanism stabilising it
  temporarily~\cite{Bombaci_AA_1996}.  According to Table
  \ref{tab:HotStars} this is the case of SRO(KDE0v1), SRO(NRAPR),
  SRO(SkAPR), see also other examples in \cite{Raduta_MNRAS_2020}.
  SRO(NRAPR) and SRO(SkAPR) have almost identical nucleon effective
  masses; very similar density dependencies of the symmetry energy;
  but different behaviors in the isoscalar channels with SRO(SkAPR)
  being stiffer than SRO(NRAPR). As a consequence, the radii of
  canonical NS and the maximum mass are slightly higher for the
  former. In addition, the smaller compressibility of SRO(SkAPR) leads
  to smaller central densities in the maximum mass neutron stars, such
  that it becomes more sensitive to thermal effects.  Indeed
  fig. \ref{fig:Prof} and Table \ref{tab:HotStars} show that for all
  considered thermodynamic conditions PNS built upon SRO(SkAPR) are
  more inflated than those built upon SRO(NRAPR). We note, too, that for
  extreme isospin asymmetries thermal energy is more effective in
  increasing $M_G$ and $M_B$ in SRO(SkAPR).  Finally radial profiles
  of NS built upon SRO(APR) single out by a bump, due to the onset of the
  pion condensate.

The ability of the $\Gamma$-law to account for the structure of PNS is
investigated in fig. \ref{fig:ProfGamma}, where radial profiles of
baryonic density and enclosed baryonic and gravitational masses
obtained using eq. (\ref{eq:GammaEoS}) to account for pressure of hot
matter are confronted with results taking into account the full temperature
dependence of the EoS. Two different entropy per baryon and electron fraction
profiles are employed, corresponding to two different post-bounce
times from a simulation of PNS evolution~\cite{PascalThese},
see fig.~ \ref{fig:PascalProfile}.
The considered PNS has $M_B=1.6M_{\odot}$ and the RG(SLy4) EoS model is employed, in agreement with the underlying simulation.
Only the cases of $t=1.0~{\rm s}$ and 5.1 s are illustrated.
The results indicate that as long as stellar shells with
$n_B \lesssim n_{\mathit{sat}}$ are hot
as it is the case in the first seconds of evolution, eq. (\ref{eq:GammaEoS}) does
not provide a fair estimation of the exact results.  Moreover, in
spite of the limiting cases, $\Gamma_{\mathit{th}}=1.5$ and 2
encompassing the exact result, the domain they delineate is too broad to
allow a satisfactory description.  The performance of
eq. (\ref{eq:GammaEoS}) improves when the maximum of the temperature
distribution moves in the innermost core and the temperature in the
low density shells does not exceed 10 MeV, see the curves for
$t=5.1~{\rm s}$.  These results can be understood by considering that
thermal effects are relatively more important in dilute matter than in
dense matter, see fig. \ref{fig:Xratio}.
To quantify the error of the $\Gamma$-law on global properties of hot PNS
we specify the values of gravitational masses and radii provided by this method
together with those obtained with the full temperature dependent EoS:
full EoS at $t$=1.0 s (5.1 s):
$M_G=1.51 M_{\odot}$ ($1.45 M_{\odot}$), $R=18.8~{\rm km}$ ($12.3~{\rm km}$);
$\Gamma=1.5$ at $t$=1.0 s (5.1 s): $M_G=1.49 M_{\odot}$ ($1.45 M_{\odot}$),
$R=14.3~{\rm km}$ ($11.6~{\rm km}$);
$\Gamma=2.0$ at $t$=1.0 s (5.1 s): $M_G= 1.53 M_{\odot}$ ($1.45 M_{\odot}$),
$R=27.0~{\rm km}$ ($12.4~{\rm km}$).
As can be seen, although at first sight the $\Gamma$-law reasonably
well reproduces the star's properties and profiles at later times,
quantitatively this is indeed the case only for the mass;
radii, however are extremely sensitive to thermal effects,
as noticed earlier, see e.g.~\cite{Raduta_MNRAS_2020,Preau_MNRAS_2021}.

\section{Summary and conclusions}
\label{sec:Concl}
We have reviewed the so-called ``general purpose'' equations of state
available in tabular form on the \textsc{ComPOSE} data base. The
assumptions entering the different models concerning the treatment of
inhomogeneous matter, the effective interaction between baryons and
the particle content have been discussed. Corresponding nuclear matter
parameters and properties of cold, $\beta$-equilibrated neutron stars
have been listed for all models together with existing constraints
from observations and nuclear data.

We have then investigated in detail the properties of cold symmetric
and pure neutron matter, including a discussion of the symmetry energy
and the effective masses of nucleons within a selection of purely
nucleonic models. At finite temperature we have shown the behavior of
different thermodynamic quantities for the same models: thermal
contributions to pressure and energy density, entropy per baryon,
specific heats, sound speed, adiabic and thermal indices.
All considered quantities manifest strong sensitivity to
the underlying EoS and a strong dependence on baryon number density,
whereas for the considered ranges, temperature and electron fraction
show less impact. For homogeneous
matter, among the variables which characterize the effective
interactions and the EoS, we confirm that thermal properties strongly
depend on the nucleonic Landau (for non-relativistic models)
and Dirac (for CDFT models) effective masses. The reason lies in the role
these quantities play in the single particle energies.
For non-relativistic models, where the kinetic energy strongly depends
on the Landau effective mass, models with large values of
$m_{{\mathit eff};{\rm Landau}}$ are characterized by large values
of the thermal energy density and entropy par baryon.

Concerning inhomogeneous matter, differences in modelling -- including
the theoretical framework and the temperature and density dependence
of the cluster surface energy -- lead
to significant differences in the various species abundances.
The uncertainties in thermodynamic quantities entailed by
cluster modelling are similar in magnitude to those associated
to the effective interaction. The cluster energy density functionals
accounting for in-medium and thermal effects are found to
induce more important modifications.

The selected purely nucleonic models have been used to
investigate the EoS-dependence of hot static stars with constant
entropy per baryon and electron fraction. The considered values
correspond roughly to different instants of PNS
evolution. Naturally, EoS models with smaller values of nucleon
effective mass lead to hotter stars at a given entropy per
baryon. Comparing two models having a similar density dependence of
the effective mass, we have shown that the stiffer EoS is more
sensitive to thermal effects. The reason is that the central
densities inside the stars are lower.  

Last but not least, we have compared the description of thermal
effects via an analytic extension of the cold EoS with a $\Gamma$-law
with the full solution from the finite temperature EoS. As expected
from the fact that the thermal index $\Gamma$ is strongly density
dependent, the thermal contribution to pressure is not well reproduced
by a $\Gamma$-law with constant value. Thus, this approximation to the
full finite temperature EoS should be regarded with care in situations
where thermal effects on the EoS are important, e.g. for the post
merger BNS remnant, CCSN and the early PNS evolution. Here we should
stress that thermal effects are relatively more important in dilute
matter and that even for temperatures around 50 MeV, the
suprasaturation matter is only moderatly affected by the thermal
contribution. Therefore, as we have seen, the later time evolution of a PNS,
when the temperature maximum has moved to the central part,
can be fairly described by such a $\Gamma$-law.

\begin{acknowledgements}
  The authors gratefully acknowledge discussions with Fiorella Burgio
  and Stefan Typel. We thank Aur\'elien Pascal for providing us with the
  profiles in fig.~\ref{fig:PascalProfile}.
  We also thank the anonymous referee for constructive comments
  that significantly contributed to enhancing the manuscript's quality.
  A.R.R. and F.N. acknowledge
  support from a grant of the Ministry of Research, Innovation and
  Digitization, CNCS/CCCDI – UEFISCDI, Project
  No. PN-III-P4-ID-PCE-2020-0293, within PNCDI III.  This work has
  been partially funded by the European COST Action CA16214 PHAROS
  ''The multi-messenger physics and astrophysics of neutron stars''.
\end{acknowledgements}

{\bf Data availability statement:}
This manuscript has no associated data as that all data are already
available on the CompOSE site https://compose.obspm.fr.

\begin{appendix}{Appendix A}

\begin{figure*}
  \includegraphics[width=0.24\textwidth]{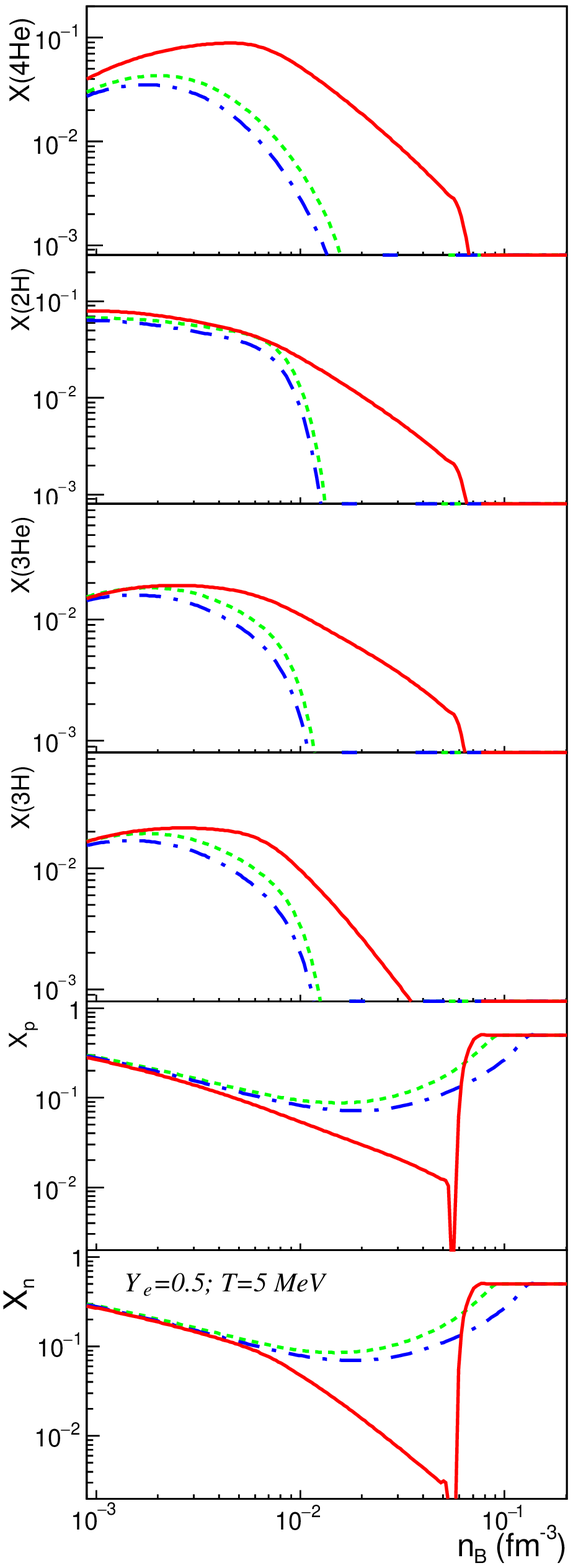}
  \includegraphics[width=0.24\textwidth]{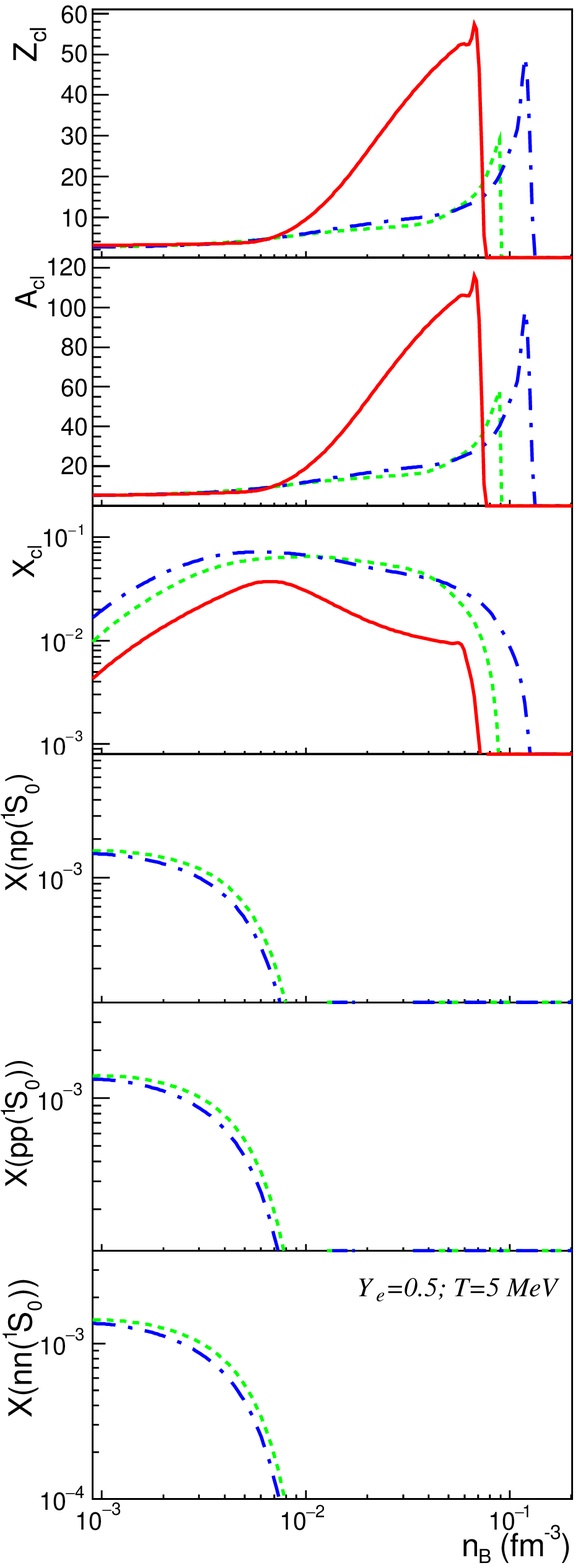}
  \includegraphics[width=0.24\textwidth]{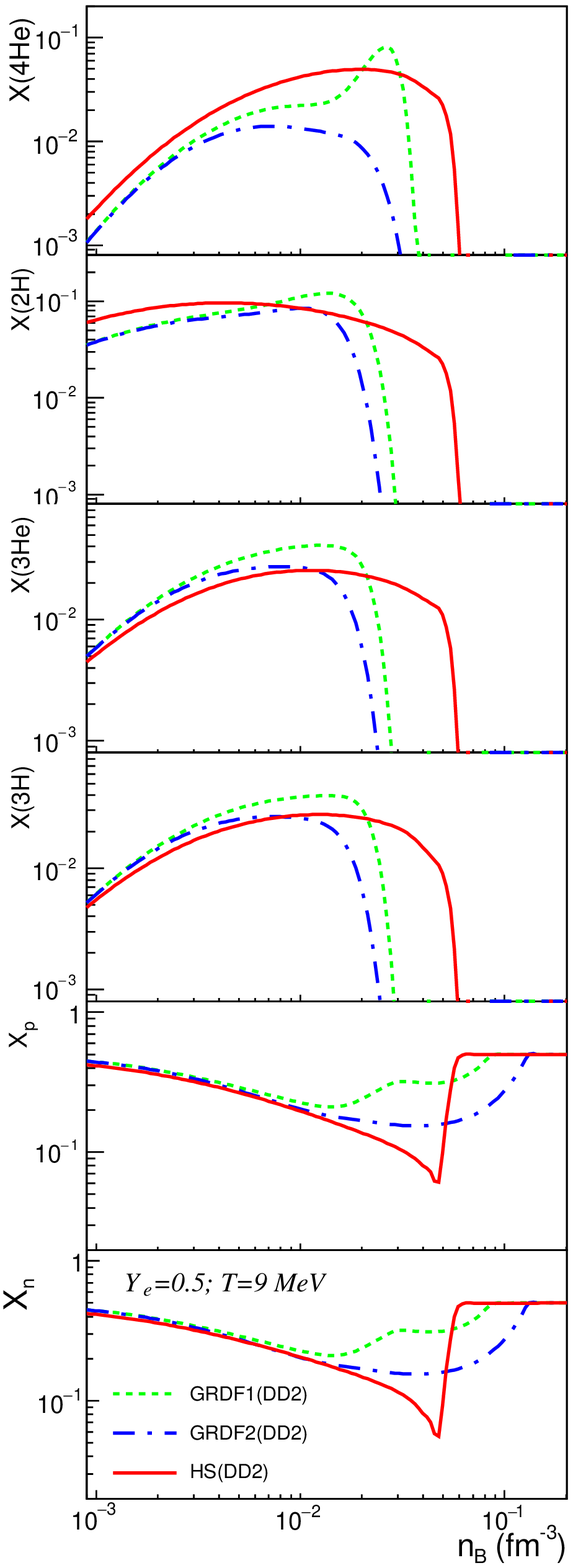}
  \includegraphics[width=0.24\textwidth]{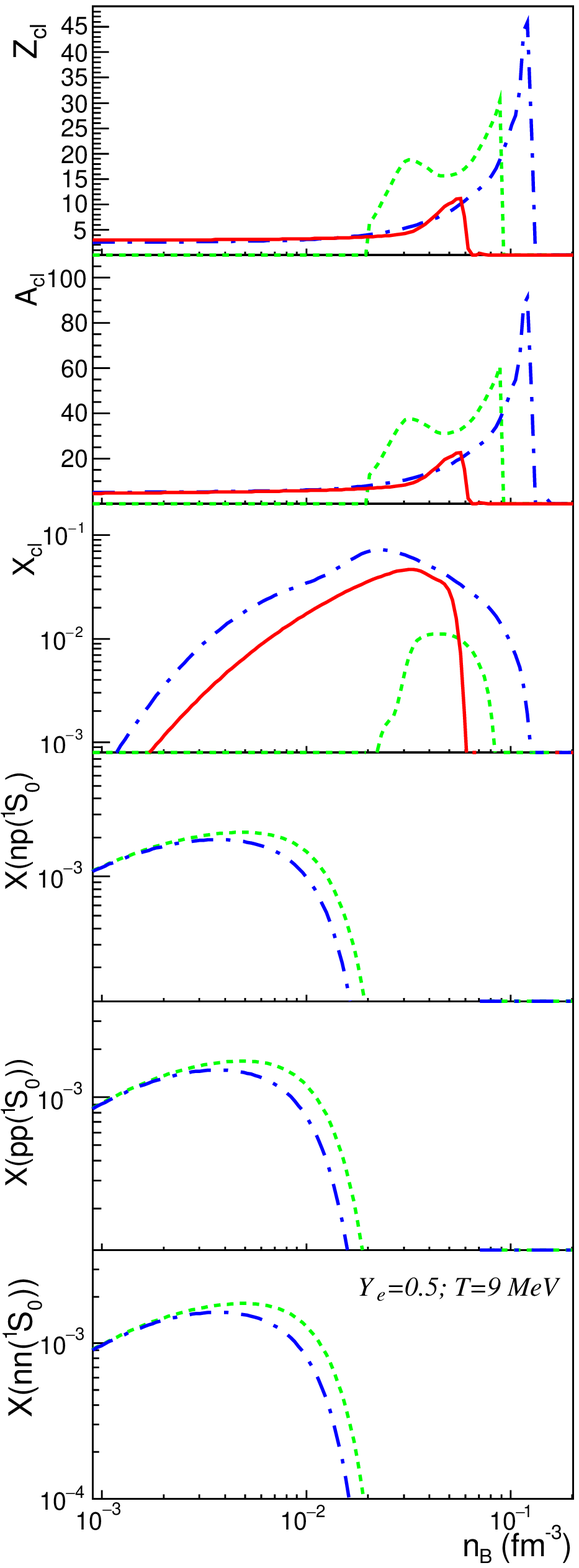}
  \caption{Matter composition in terms of particle mass fractions and average
    mass and charge numbers of nuclei as predicted by HS(DD2), GRDF1(DD2) and
    GRDF2(DD2) for $T=5$ MeV (first two columns) and $T=9$ MeV (columns 3 and 4) 
    and $Y_e=0.5$.}
\label{fig:compo_T=5_DD2} 
\end{figure*}
 
\end{appendix}

\bibliographystyle{spphys}       
\bibliography{Compose.bib}   

\end{document}